\documentclass[12pt,epsbox]{article}
\usepackage{color}
\usepackage[dvips]{graphicx}
\usepackage{amsmath}

\baselineskip=\normalbaselineskip
\renewcommand{\baselinestretch}{1.4}
\newcommand{\resection}[1]
 {\setcounter{equation}{0}\section{\large{#1}}}

\newcommand{\bel}[1]{\begin{equation}\label{#1}}
\newcommand{\bal}[1]{\begin{eqnarray}\label{#1}}

\newcommand{\be}{\begin{equation}}
\newcommand{\ee}{\end{equation}}
\newcommand{\ba}{\begin{eqnarray}}
\newcommand{\ea}{\end{eqnarray}}
\newcommand{\nn}{\nonumber \\}
\newcommand{\qq}{\qquad}

\newcommand{\n}{\nonumber}

\newcommand{\ket}[1]{\left|\,{#1}\,\right\rangle}

\newcommand{\tr}{{\rm tr}}
\newcommand{\bR}{{\bf R}}
\newcommand{\bZ}{{\bf Z}}
\newcommand{\hg}{\widehat{g}{}}
\newcommand{\hphi}{\widehat{\phi}{}}

\newcommand{\bB}{{\bf B}}

\newcommand{\eq}[1]{(\ref{#1})}
\newcommand{\cH}{{\cal H}}
\newcommand{\cL}{{\cal L}}
\newcommand{\cP}{{\cal P}}
\newcommand{\Sloc}{{S_{\rm loc}}}
\newcommand{\cLloc}{{{\cal L}_{\rm loc}}}
\newcommand{\cJ}{{\cal J}}

\newcommand{\cW}{{\cal W}}
\newcommand{\gh}{\widehat{g}{}}

\newcommand{\phib}{\bar{\phi}}
\newcommand{\pib}{\overline{\pi}}

\newcommand{\bG}{\mbox{\boldmath $G$}}
\renewcommand{\bR}{\mbox{\boldmath $R$}}
\newcommand{\bS}{\mbox{\boldmath $S$}}
\newcommand{\bGam}{\mbox{\boldmath $\Gamma$}}
\newcommand{\bcH}{\mbox{\boldmath ${\cal H}$}}
\newcommand{\bcL}{\mbox{\boldmath ${\cal L}$}}

\renewcommand{\bZ}{\mbox{\boldmath $Z$}}

\newcommand{\hS}{\widehat{S}{}}
\newcommand{\hR}{\widehat{R}{}}
\newcommand{\hnab}{\widehat{\nabla}{}}

\newcommand{\tH}{\widetilde{H}}

\newcommand{\bQ}{\overline{Q}}
\newcommand{\del}{\partial}

\newcommand{\blambda}{\overline{\lambda}}

\newcommand{\dalpha}{\dot{\alpha}}
\newcommand{\dbeta}{\dot{\beta}}
\newcommand{\half}{\frac{1}{2}}
\newcommand{\cN}{{\mathcal{N}}}
\newcommand{\hK}{\widehat{K}{}}
\newcommand{\hpi}{\widehat{\pi}{}}
\newcommand{\hL}{\widehat{L}{}}

\newcommand{\gb}{\overline{g}}

\newcommand{\cO}{{\cal O}}

\newcommand{\Diff}{{{\rm Diff}_2}}
\newcommand{\Weyl}{{{\rm Weyl}}}
\newcommand{\Conf}{{{\rm Conf}_2}}
\newcommand{\cD}{{\cal D}}
\newcommand{\gammah}{\widehat{\gamma}{}}
\newcommand{\cDh}{\widehat{\cD}{}}
\newcommand{\Th}{\widehat{T}{}}
\newcommand{\Phih}{\widehat{\Phi}{}{}}
\newcommand{\Sh}{\widehat{S}{}}
\newcommand{\phih}{\widehat{\phi}{}}
\newcommand{\epsilonh}{\widehat{\epsilon}{\,}}
\newcommand{\Nh}{\widehat{N}{}}
\newcommand{\lambdah}{\widehat{\lambda}{}}
\newcommand{\pih}{\widehat{\pi}{}}
\newcommand{\newton}{\frac{1}{2\kappa_{d+\!1}^2}}
\newcommand{\phibar}{\overline{\phi}{}}
\newcommand{\Kh}{\widehat{K}{}}

\begin{document}
\setcounter{page}{0}
\begin{flushright}
\parbox{40mm}{%
KUNS-1817 \\
YITP-02-73 \\
TAUP-2719-02 \\
{\tt hep-th/0212314} \\
December 2002}

\end{flushright}

\vfill

\begin{center}
{\Large{\bf 
Holographic Renormalization Group
}}
\end{center}

\vfill

\renewcommand{\baselinestretch}{1.0}

\begin{center}
{\sc Masafumi Fukuma}
\footnote{E-mail: {\tt fukuma@gauge.scphys.kyoto-u.ac.jp}},  
{\sc So Matsuura}
\footnote{E-mail: {\tt matsu@yukawa.kyoto-u.ac.jp}} and 
{\sc Tadakatsu Sakai}
\footnote{E-mail: {\tt tsakai@post.tau.ac.il}} 

~\\

$^1${\sl Department of Physics, Kyoto University, Kyoto 606-8502, Japan} \\
$^2${\sl Yukawa Institute for Theoretical Physics, 
      Kyoto University, Kyoto 606-8502, Japan } \\
$^3${\sl Raymond and Beverly Sackler Faculty of Exact Sciences}\\ 
\vspace{-2mm}
{\sl Schoolof Physics and Astronomy}\\
\vspace{-2mm}
{\sl Tel-Aviv University, Ramat-Aviv 69978, Israel}

\end{center}

\vfill
\begin{center}
{\bf abstract}
\end{center}

\begin{quote}

\small{%
The holographic renormalization group (RG) is reviewed 
in a self-contained manner. 
The holographic RG is based on the idea that the radial coordinate 
of a space-time with asymptotically AdS geometry can be identified 
with the RG flow parameter of the boundary field theory. 
After briefly discussing basic aspects of the AdS/CFT correspondence, 
we explain how the notion of the holographic RG comes out in the
AdS/CFT correspondence. 
We formulate the holographic RG based on the Hamilton-Jacobi equations 
for bulk systems of gravity and scalar fields, 
as was introduced by de Boer, Verlinde and Verlinde. 
We then show that the equations can be solved with  
a derivative expansion by carefully extracting 
local counterterms from the generating functional 
of the boundary field theory. 
The calculational methods to obtain the Weyl anomaly and 
scaling dimensions are presented 
and applied to the RG flow from the $\cN=4$ SYM
to an $\cN= 1$ superconformal fixed point discovered by
Leigh and Strassler. 
We further discuss a relation between the holographic RG and the 
noncritical string theory, 
and show that the structure of the holographic RG should persist 
beyond the supergravity approximation as a consequence of the 
renormalizability of the nonlinear $\sigma$ model action 
of noncritical strings. 
As a check, we investigate the holographic RG structure 
of higher-derivative gravity systems, 
and show that such systems can also be analyzed 
based on the Hamilton-Jacobi equations, 
and that the behaviour of bulk fields are determined solely 
by their boundary values. 
We also point out that higher-derivative gravity systems 
give rise to new multicritical points in the parameter space 
of the boundary field theories. 
}
\end{quote}
\vfill
\renewcommand{\baselinestretch}{1.4}

\renewcommand{\thefootnote}{\arabic{footnote}}
\setcounter{footnote}{0}
\addtocounter{page}{1}

\resection{Introduction}

The idea that there should be a close relation between gauge 
theories and string theory has a long history  
\cite{classics,hooft;73,wilson;74}. 
In a seminal work by 't Hooft \cite{hooft;73}, 
the relation is explained in terms of 
the double-line representation of gluon propagators 
in $SU(N)$ gauge theories. 
There a Feynman diagram is interpreted as a string world-sheet
by noting that each graph has the dependence on the gauge coupling 
and the number of colors as 
\begin{equation}
(g_{\rm YM}^2)^{-V+E}N^I=\lambda^{-V+E}N^{2-2g}
=(g_{\rm YM}^2)^{2g-2}\lambda^I.
\end{equation}
Here $\lambda=g_{\rm YM}^2N$ is the 't Hooft coupling, 
and $V$, $E$ and $I$ are the numbers of the vertices, propagators and index
loops of a Feynman diagram, respectively. 
We also used the Euler
relation $V-E+I=2-2g$ with $g$ a genus.
In the 't Hooft limit $N\rightarrow\infty$ 
with $\lambda$ fixed, a gauge theory can be regarded as
a string theory with the string coupling 
$g_s\propto 1/N\propto g_{\rm YM}^2$, 
and $\lambda$ is identified with some geometrical data of the string
background.
To be more precise, consider the partition function of a gauge theory
\begin{equation}
{\cal F}=\sum_{g,I}(g_{\rm YM}^2)^{2g-2}\lambda^I{\cal F}_{g,I}
=\sum_{g}(g_{\rm YM}^2)^{2g-2}{\cal F}_g(\lambda).
\end{equation}
A question is now if one can find a string theory that reproduces
in perturbation each coefficient ${\cal F}_g(\lambda)$.
In Ref.~\cite{GV;cs}, a quantitative check for this correspondence 
between Chern-Simons theory on $S^3$ and topological
$A$ model on a resolved conifold was presented.
However, it is a highly involved problem to prove such a 
correspondence in more realistic gauge theories.

The AdS/CFT correspondence is a manifestation of the idea by
't Hooft. By studying the decoupling limit of coincident D3 and 
M2/M5 branes, 
Maldacena \cite{M} argued that
superconformal field theories with the maximal amount of supersymmetry
(SUSY) are dual to string or M theory on AdS.
Soon after the ground-breaking work by Maldacena, 
this conjecture was made into a more precise statement by 
Gubser, Klebanov and Polyakov \cite{GKP} and by Witten \cite{W;holography} 
that the classical action of bulk gravity should be regarded 
as the generating functional of the boundary conformal field theory. 
Since then, the correspondence has been investigated extensively 
and a number of evidences for the conjecture have accumulated so far
(for a review, see Ref.\ \cite{review}).
As a typical example, consider the duality between the $\cN=4$ 
super Yang-Mills (SYM) theory in four dimensions and the Type IIB string
theory on AdS$_5\times S^5$.
The IIB supergravity solution of $N$ D3-branes 
reads \cite{blackp}
\begin{gather}
 ds^2=f_3^{-1/2}\left(-dt^2+dx_1^2+\cdots+dx_3^2\right)
 +f_3^{1/2}(dy_1^2+\cdots+dy_6^2) \qquad
 \left(f_3\equiv 1+\frac{\lambda l_s^4}{r^4}\right),
\end{gather}
where $r\equiv\sqrt{y_1^2+\cdots+y_6^2}$, $\lambda\equiv 4\pi Ng_s$, 
and $l_s=\sqrt{\alpha'}$ and $g_s$ are 
the string length and the string coupling, respectively. 
%
The decoupling limit is defined by $l_s\rightarrow 0$ with
$U=rl_s^{-2}={\rm fixed}$. The metric turns out to reduce to
AdS$_5\times S^5$:
\begin{equation}
l_s^{-2} ds^2={U^2\over \lambda^{1/2}}\eta_{ij}dx^idx^j
+{\lambda^{1/2}\over U^2}dU^2
 + \lambda^{1/2} d\Omega_5^2. 
\end{equation}
Or, by introducing $l=\lambda^{1/4}\,l_s$ and $z=\lambda^{1/2}\,U^{-1}$, 
this metric can be rewritten as 
\ba
 ds^2=\frac{l^2}{z^2}\,\Bigl(dz^2+\eta_{ij}\,dx^idx^j\Bigr)\,+\,
  l^2\,d\Omega_5^2,
\ea
which shows that AdS${}_5$ and $S^5$ have the same curvature radius $l$.%
\footnote{Their scalar curvatures are given 
by $R_{{\rm AdS}_5}=-20/l^2$ and $R_{S^5}=+20/l^2$, respectively.} 
On the other hand, the low energy effective theory on 
the $N$ coincident D3-branes is the ${\mathcal{N}}=4$ $SU(N)$ SYM
theory. 
From the viewpoint of open/closed string duality, it is plausible that
both the theories are dual to one another. In fact, one finds that
both have the same symmetry $SU(2,2|4)$.
Furthermore, we will find later a more stringent check of the duality
by comparing the chiral primary operators of SYM and the Kaluza-Klein (KK)
spectra of IIB supergravity compactified on $S^5$.

Recall that the IIB supergravity description is reliable only when 
the effect of both quantum gravity and 
massive excitations of a closed string is negligible.
The former condition is equivalent to%
\footnote{
The $l_{\rm Plank}$ is the ten dimensional Plank scale, 
which is given by $l_{\rm Plank}=g_s^{1/4}l_s$.
} 
\ba
l \gg l_{\rm Plank} \Leftrightarrow N\gg1,
\label{condition1}
\ea
and the latter to
\ba
l \gg l_s \Leftrightarrow g_sN\gg1.
\label{condition2}
\ea
This implies that the dual SYM is in the strong coupling regime.

One of the most significant aspects of the AdS/CFT correspondence 
is that it gives us a framework to study the 
renormalization group (RG) structure of the dual field theories  
\cite{SW}-\cite{Yamaguchi:2002pa}.
In this scheme of the  {\it holographic RG}, 
the extra radial coordinate in the bulk is 
regarded as parametrizing the RG flow of the dual field theory, 
{\em i.e.}, the evolution of bulk fields along the radial direction 
is considered as describing the RG flow of the coupling constants 
in the boundary field theory.

One of the main purposes of this article is to review various aspects of
the holographic RG using the Hamilton-Jacobi (HJ) formulation.
A systematic study of the holographic RG based on the HJ equation 
was initiated by de Boer, Verlinde and Verlinde \cite{dVV}.   
(For a review of their work see Ref.\ \cite{deBoer-rev}.)%
\footnote{The use of Hamilton-Jacobi equation was proposed by 
A.\,M.\,Polyakov sometime ago in a slightly different context  
\cite{polyakov1993}.}
In this formulation, we first perform the ADM Euclidean decomposition
of the bulk metric, regarding the normal coordinate 
to the AdS boundary, $\tau$, as an Euclidean time.
Working in the first-order formalism, we obtain two
constraints, the Hamiltonian and momentum constraints, 
which ensure the invariance of the classical action of bulk gravity 
under residual diffeomorphisms after a choice of time-slice is made.
The usual HJ procedure to these constraints 
leads to functional equations on the classical action.
These are called a {\it flow equation} and play a central role in the
study of the holographic RG.
One of the advantages of this HJ formulation is
that the HJ equation directly characterizes the classical action 
of bulk gravity without solving the equations of motion. 
In Ref.\ \cite{dVV}, a five-dimensional bulk gravity theory with scalar fields
was considered, and it was shown that the flow equation yields
the Callan-Symanzik equation of the four-dimensional 
boundary theory.
They also calculated the Weyl anomaly in four dimensions 
and found that the result agrees with those given 
in Ref.\ \cite{HS;weyl} (see also Ref.\ \cite{BK2,dSS})
For a review of the Weyl anomaly, see Ref.\ \cite{Duff;Weyl} . 

The expositions in this article are based on 
a series of work of the present authors \cite{FMS1}-\cite{FM}.
We here summarize the main results briefly.
In Ref.\ \cite{FMS1} bulk gravity systems with various scalar fields 
was investigated in arbitrary dimensionality \cite{FMS1}. 
After deriving the flow equation of this system as described above, 
we showed that the equation can be solved systematically 
with the use of a derivative expansion 
if we assign proper {\em weights} to the generating functional 
as well as to local counter terms. 
From this result, we derived the Callan-Symanzik
equation of the $d$-dimensional dual field theory. 
We also computed the Weyl anomaly and find a precise agreement 
with that given in the literature.
It was argued that the ambiguity of local counterterms
does not affect the uniqueness of the Weyl anomaly \cite{FS}.

The discussion was extended to bulk gravity with higher-derivative 
interactions in Ref.\ \cite{FMS2}. 
Higher-derivative interactions generically comes into the low-energy 
effective action of string theory by integrating out the massive modes of 
closed strings or due to the presence of orientifold planes \cite{BGN}. 
On the other hand, according to the AdS/CFT correspondence, 
these interactions are interpreted in the dual field theories 
as $1/\lambda$ corrections, 
or for orthogonal and symplectic gauge groups, 
as $1/N$ (not $1/N^2$) corrections \cite{BGN}. 
So the study of a higher-derivative
gravity theory is important in order to justify the AdS/CFT correspondence 
beyond the supergravity approximation.
We found that such evolution of classical solutions that 
maintains the holographic RG structure of boundary field theories 
can be investigated by using a Hamilton-Jacobi-like analysis, 
and that the systematic method proposed in Ref.\ \cite{FMS1} 
can also be applied in solving the flow equation.
We computed a $1/N$ correction to the Weyl anomaly of four-dimensional 
$\cN\!=\!2$ $USp(N)$ supersymmetric gauge theory, 
via higher-derivative gravity on the dual AdS 
that was proposed in Ref.\ \cite{N=2CFT}
(for an earlier work on a computation of $1/N$ corrections to Weyl
anomalies, see Refs.\ \cite{BGN,APTY}).
The result is found to be consistent with a field theoretic
computation. This implies that the AdS/CFT correspondence
is valid beyond the supergravity approximation.
In a higher-derivative gravity theory, 
new interesting phenomena of the holographic
RG develop. 
For example, one can show that adding higher-derivative interactions 
to the bulk gravity action leads to the appearance of new multicritical 
points in the parameter space of boundary field theories \cite{FM}.
For other works on the HJ formulation in the context of
the holographic RG, see Refs.\ \cite{Cor}-\cite{Martelli:2002sp}.

The expectation that the structure of the holographic RG 
should persist beyond the supergravity approximation 
can be further confirmed by formulating the string theory 
in terms of noncritical strings. 
In fact, as will be explained in \S 4, 
the Liouville field $\varphi$ of the noncritical string theory 
can be naturally identified with the RG flow parameter 
in the holographic RG. 
{}Furthermore, various settings assumed in the holographic RG 
(like the regularity of fields inside the bulk) 
have direct counterparts in the noncritical string theory. 
It will be further discussed in \S 4 that 
as a consequence of the renormalizability of the nonlinear $\sigma$ model 
action of noncritical strings, 
the behavior of bulk fields should be holographic 
in full orders of $\alpha'$ expansion, 
{\em i.e.}, it should be determined solely by their boundary values. 

The organization of this paper is the following.
In \S 2, we give a review of basic aspects of the AdS/CFT correspondence.
We outline how the notion of the holographic RG comes out in the
AdS/CFT correspondence. As an example of a holographic description
of RG flows, we consider a flow from the $\cN=4$ SYM
to an $\cN= 1$ superconformal fixed point discovered by
Leigh and Strassler \cite{LS;95}.
In \S 3, we formulate the Hamilton-Jacobi equation
of bulk gravity and derive the flow equation.
We solve it in terms of a derivative expansion by introducing the
weights.
From this solution, we derive the Callan-Symanzik equation and
the Weyl anomaly. 
\S 4 is devoted to a discussion of the relation between the holographic
RG and non-critical strings, 
and it is discussed that the structure of the holographic RG 
should persist beyond the supergravity approximation 
as a consequence of the renormalizability of the nonlinear $\sigma$ 
model action of noncritical strings. 
In \S 5, we consider the HJ formulation of a higher-derivative gravity
theory.
We first discuss a new feature of the holographic RG that appears there.
We next derive the flow equation of the higher-derivative system
and solve it by using the derivative expansion.
We show that this computation gives a consistent $1/N$ correction
to the Weyl anomaly of $\cN=2$ $USp(N)$ supersymmetric gauge theory
in four dimensions. 
In \S 6, we summarize the results of this article and discuss some
future directions in the AdS/CFT correspondence and the holographic RG.
We also make a brief comment on field redefinitions of bulk fields 
in ten-dimensional supergravity in the context of the AdS/CFT correspondence. 
In particular, we show that the holographic Weyl anomaly is 
invariant under a redefinition of the ten-dimensional metric 
of the Type IIB supergravity theory. 
In appendices, we give some useful formulae and results.

\resection{Review of the AdS/CFT correspondence}

In this section, we present a review of the AdS/CFT correspondence \cite{M} 
and the holographic renormalization group (RG). 
We first discuss 
a prescription 
given by Gubser, Klebanov and Polyakov \cite{GKP} 
and by Witten \cite{W;holography} 
to compute correlation functions of the dual CFT.
Based on these observations, we come to the idea of
the holographic RG. 
Here the IR/UV relation \cite{SW} 
in the AdS/CFT correspondence plays a central role.
As an application, 
we calculate the scaling dimensions of scaling operators of the CFT. 
We discuss in some detail a typical example of the AdS/CFT
correspondence, the duality between the four-dimensional 
${\mathcal{N}}=4~SU(N)$ 
SYM theory and 
Type IIB supergravity on AdS$_5\times S^5$. 
In order to check the duality, 
we show the one-to-one correspondence 
between the Kaluza-Klein spectra on $S^5$
and the local operators in the 
short chiral primary multiplets of the  
${\mathcal{N}}=4~SU(N)$ SYM theory.

\subsection{AdS/CFT correspondence and the IR/UV relation}

The AdS/CFT correspondence states that 
{\it a classical (super)gravity theory on a $(d+1)$-dimensional 
anti-de Sitter space-time (AdS$_{d+1}$) 
is equivalent to a conformal field theory (CFT$_d$) at the
$d$-dimensional boundary of the AdS space-time} 
\cite{M,GKP,W;holography}. 
To explain this, we first introduce some basic ingredients. 

The AdS$_{d+1}$ of curvature radius $l$ has the metric
\begin{align}
ds^2 &= \hg_{\mu\nu}^{\rm AdS}dX^{\mu}dX^{\nu} \nn
     &= \frac{l^2}{z^2}\,\left(dz^2+\eta_{ij}dx^idx^j \right) \nn
     &= d\tau^2 + e^{-2\tau/l}\eta_{ij}dx^idx^j, 
\label{ads-metric}
\end{align}
where $X^\mu=(x^i,z)$ or $X^\mu=(x^i,\tau)$ 
with $\mu=1,\cdots,d+1$ and $i=1,\cdots,d$. 
The two parametrizations for the radial coordinate, $z$ and $\tau$, 
are related as $z=l\,e^{\tau/l}$,  
and the range of $z$ (or $\tau$) is $0<z<\infty$ (or $-\infty<\tau<\infty$), 
so that the boundary is located at $z=0$ ($\tau=-\infty$). 
{}For the AdS$_{d+1}$ with Lorentzian signature, 
we take $\eta_{ij}$ to be the flat Minkowski metric 
$\eta_{ij}={\rm diag}\,[-1,+1,...,+1]$. 
In the following, we instead consider the Euclidean version 
of AdS$_{d+1}$ (the Lobachevski space) 
by taking $\eta_{ij}=\delta_{ij}$, 
which generalizes the Poincar\'{e} metric of the upper half plane. 
The AdS$_{d+1}$ has the constant negative curvature, 
$\widehat{R}=-d(d+1)/l^2$, 
and has the nonvanishing cosmological constant, 
$\Lambda=-d(d-1)/2\,l^2$.

The bosonic part of the action of $(d+1)$-dimensional supergravity 
with the metric $\gh_{\mu\nu}(X)$ and scalars $\hphi^a(X)$ 
has generically the following form:\footnote%
{We use a convention that $(d+1)$-dimensional bulk fields wear a hat 
$\widehat{~}$ whereas $d$-dimensional boundary fields do not; 
{\em e.g.}, $\widehat{\Phi}(X)=\widehat{\Phi}(x,z)$ and $\Phi(x)$. 
When bulk fields satisfy the equations of motion, 
we put bar $\overline{~}$ on the bulk fields, {\em e.g.,} 
$\overline{\Phi}(X)=\overline{\Phi}(x,z)$. 
The bulk action is written in a bold face, $\bS$, 
while the classical action (to be defined later) 
is simply written by $S$.
}
\ba
 \newton\,\bS[\gh_{\mu\nu},\hphi^a]=\newton\int d^{d+1} X\sqrt{\gh}\left[
  V\bigl(\hphi\bigr)-\widehat{R}+\frac{1}{2}\,\gh^{\mu\nu}\,
  L_{ab}(\hphi)\,\partial_\mu\hphi^a\,\partial_\nu\hphi^b\right]. 
\label{grav-action}
\ea
Throughout this article, we extract the $(d+1)$-dimensional Newton constant 
$16\pi G^{\rm N}_{d+\!1}=2\kappa_{d+\!1}^2$ from the action 
in order to simplify many of expressions in the following discussions. 
The scalar potential would be expanded as 
\ba
 V(\hphi)=2\Lambda+\sum_a\frac{1}{2}\,m_a^2\,\hphi^a\hphi^a+\cdots .
\ea
after the diagonalization of a mass-squared matrix. 
AdS gravity is obtained by substituting the AdS metric $\hg_{\rm AdS}$ 
into the bulk action $\bS$ 
with the cosmological constant $\Lambda$ set to be 
\ba
 \Lambda=-d(d-1)/2\,l^2. 
\ea
We consider classical solutions $\phibar^{\,a}(x,z)$ 
of the bulk scalar fields $\hphi^a(x,z)$ in this AdS$_{d+1}$ background. 
We impose boundary conditions on the scalar fields 
such that $\phibar^a(x,z\!=\!0)=\phi^a(x)$ 
and also that they are regular inside the bulk ($z\rightarrow+\infty$). 
The system is then completely specified solely by 
the boundary values $\phi^a(x)$, 
and thus, if we plug the classical solutions 
into the action (\ref{grav-action}), 
we obtain the classical action which is a functional of the boundary values;  
\ba
S[\phi^a(x)]\equiv\bS\left[
 \hg_{\mu\nu}(x,z)\!=\! \hg_{\mu\nu}^{\rm AdS}(x,z),\,
 \hphi^a(x,z)\!=\!\phibar^a(x,z) \right].  
\label{class.action}
\ea 
A naive form of the statement of the AdS/CFT correspondence is\footnote{%
This statement will be elaborated shortly later 
as is argued in Refs.\ \cite{GKP,W;holography}
} 
that {\em the classical action
(\ref{class.action}) is the generating functional of 
a conformal field theory living at the $d$-dimensional boundary of the
AdS space-time; }
\ba
 \exp\Bigl(-\newton\,S[\phi^a(x)]\Bigr) = \Biggl\langle \exp\left(
 \int d^dx\,\phi^a(x)\,{\mathcal{O}_a}(x)\right)
 \Biggr\rangle_{\rm CFT}, 
\ea
{\em where ${\mathcal{O}}_a(x)$'s are scaling operators of the CFT. }

This statement can be understood as a simple consequence of 
the mathematical theorem that an isometry of AdS$_{d+1}$, 
$f:~{\rm AdS}_{d+1}\rightarrow{\rm AdS}_{d+1}$, 
induces a $d$-dimensional conformal transformation at the boundary. 
In fact, if the theorem holds, 
then by using the diffeomorphism invariance of the bulk action 
\eq{grav-action}, one can easily show that 
the classical action $S[\phi^a(x)]$ is conformally invariant: 
\ba
 S[\rho^\ast\phi^a(x)]=S[\phi^a(x)], 
\ea
where $\rho\equiv f\bigr|_{\partial({\rm AdS})}$ 
is a conformal transformation on the boundary $\partial({\rm AdS})$. 
Thus, if we formally define ``connected $n$-point functions'' by 
\ba 
 \Bigl\langle{\mathcal{O}}_{a_1}(x_1)\cdots{\mathcal{O}}_{a_n}(x_n)
  \Bigr\rangle_{\rm CFT} 
  \equiv 
  \frac{\delta}{\delta\phi^{\,a_1}(x_1)}\cdots
  \frac{\delta}{\delta\phi^{\,a_n}(x_n)}
  \left(-\newton\,S[\phi^a(x)]\right)\Biggr|_{\phi^a=0}, 
\ea
then they are actually invariant under the $d$-dimensional 
conformal transformations: 
\ba
 \Bigl\langle{\rho^\ast\mathcal{O}}_{a_1}(x_1)\cdots
  \rho^\ast{\mathcal{O}}_{a_n}(x_n)
  \Bigr\rangle_{\rm CFT} 
  =\Bigl\langle{\mathcal{O}}_{a_1}(x_1)\cdots{\mathcal{O}}_{a_n}(x_n)
  \Bigr\rangle_{\rm CFT} .
\ea

We here give a proof of the theorem 
in an extended form from the above: 

\noindent{\underline{\bf Theorem}} \cite{GKP}\\
{\em Let $M_{d+1}$ be a $(d+1)$-dimensional manifold 
with boundary whose metric is asymptotically AdS near the boundary.\footnote{%
We say that a metric has an asymptotically AdS geometry 
when there exists a parametrization near the boundary ($z=0$) such that 
$\gh_{ij}=z^{-2}\,\eta_{ij}+{\mathcal O}(1)$, 
$\gh_{iz}={\mathcal O}(z)$ 
and $\gh_{zz}=z^{-2}+{\mathcal O}(1)$. 
}
Then any diffeomorphism which becomes an isometry near the boundary 
induces a $d$-dimensional conformal transformation at the boundary.}

\small{%
\noindent{\underline{\bf proof}}\\
Let us consider an infinitesimal diffeomorphism, 
$X^{\mu} \to X^{\mu}+\epsilonh^{\mu}(x,z)$. 
Since this does not move the position of the boundary off $z=0$, 
$\epsilonh^\mu(x,z)$ is expanded around $z=0$ as 
\ba
 \epsilonh^i(x,z)=\xi^i(x)+{\mathcal {O}}(z^2),\quad
  \epsilonh^z(x,z)=z\cdot \zeta(x) + {\mathcal {O}}(z^3). 
\ea
If this diffeomorphism is further an isometry near the boundary, 
the change of the metric should take the form 
\ba 
 \delta_{\epsilonh}\hg_{ij}(x,z)={\mathcal{O}}(1), \qq
 \delta_{\epsilonh}\hg_{iz}(x,z)={\mathcal{O}}(z), \qq
 \delta_{\epsilonh}\hg_{zz}(x,z)={\mathcal{O}}(1), 
 \label{isometry-condition}
\ea
around $z=0$. 
A simple calculation shows that eq.\ (\ref{isometry-condition}) 
leads to the condition that the $\epsilonh^i(x,z)$ and $\epsilonh^z(x,z)$ 
have the following expansion around $z=0$: 
\begin{align}
 \epsilonh^i(x,z)&=\xi^i(x)-\frac{z^2}{2d}\,\eta^{ij}\,\del_j\del_k\xi^k(x)
   + {\mathcal{O}}(z^4), \nn
 \epsilonh^z(x,z)&=\frac{z}{d}\,\del_i\xi^i(x) + {\mathcal{O}}(z^3),   
\end{align}
and that the $\xi^i(x)$ satisfies the $d$-dimensional 
conformal Killing equation 
\ba
 \del_i\xi_j(x)+\del_j\xi_i(x) = \frac{2}{d}\,\del_k\xi^k(x)\,\eta_{ij}. 
  \quad(\xi_i(x)\equiv\eta_{ij}\,\xi^j(x)).
\ea
This means that $\xi^i(x)$ generates a $d$-dimensional conformal 
transformation at the boundary. (Q.E.D.)
}

\normalsize

However, the naive form of the classical action \eq{class.action} 
is not defined well since the integration over $z$ generally diverges.  
This is because of the infinite volume of the AdS space-time and 
the finite cosmological constant in the Lagrangian density; 
$\bS\sim\int_{\rm AdS}d^{d+1}x \sqrt{\hg}
\left[2\Lambda+\cdots\right] \to \infty$.
Thus, we must make a proper regularization for the integration 
to make physical quantities finite.
Here we introduce an IR cutoff parameter $z_0 $ 
to restrict the bulk to the region $z_0 \leq z<\infty$,%
\footnote{
The constant in the equation below is given by 
$2\Lambda-\widehat{R}_{\rm AdS}=-d(d-1)/l^2+d(d+1)/l^2=2d/l^2$.
}  
\ba
 \newton\,\bS\bigl[\gh^{\rm AdS}_{\mu\nu}(x,z)\,\hphi^a(x,z)\bigr] 
 &=& \frac{1}{2\kappa_{d+1}^2}\,
      \int_{z_0} ^{\infty}dz \int d^dx \sqrt{\hg_{\rm AdS}}\left[
      {\rm const.}+\frac{1}{2}\,m_a^2\,\phih^a\phih^a \right.\nn
 &&~~~~~~~\left.+\,\frac{1}{2}\,\hg^{\mu\nu}_{\rm AdS}\,L_{ab}(\hphi)\,
      \del_{\mu}\hphi^a\,\del_{\nu}\hphi^b
					   \right].
\label{reg.grav-action}
\ea 
We solve the equations of motion for $\hphi^a(x,z)$ by imposing boundary
conditions at the new $d$-dimensional boundary, $z=z_0 $: 
\begin{align}
 \phibar^{\,a}(x,z\!=\!z_0) = \phi^a(x), 
\label{reg.class.sln}
\end{align}
The classical action is then properly defined by substituting the classical 
solutions $\phibar^a(x,z)$ 
into the action \eq{reg.grav-action}, 
which is also a functional of $\phi^a(x)$: 
\begin{align}
 S&=S[\phi^a(x);z_0 ] 
  \equiv \bS\left[\gh_{\mu\nu}(x,z)\!=\!\gh^{\rm AdS}_{\mu\nu}(x,z),
  \hphi^a(x,z)\!=\!\phibar^{\,a}(x,z) \right].
 \label{reg.class.action}
\end{align}
At this new boundary $z=z_0$, the conformal invariance 
disappears since this symmetry exists only at the original boundary, $z=0$. 
In fact, we will show below that 
the IR cutoff $z_0 $ in the bulk gives a UV cutoff 
$\Lambda_0=1/z_0$ of the boundary theory (the {\em IR/UV relation}). 
{}Furthermore, in order to obtain a finite classical action 
around the original conformal fixed point ($z_0\!\rightarrow\!0$), 
we need to tune the boundary values accordingly, 
$\phi^a(x)=\phi^a(x;z_0)$.
This procedure corresponds to the fine tuning of bare couplings 
encountered in usual quantum field theories. 
As we see in the next section with more general settings, 
this fine tuning exactly corresponds to the (Euclidean) time 
evolution of the classical solutions; 
$\phi^a(x;z_0)=\phibar^{\,a}(x,z_0)$. 
Thus, tracing the classical solutions as the position of the boundary $z_0$ 
changes gives a renormalization group flow of the boundary field theory. 
This is the basic idea of the {\it holographic renormalization group}  
\cite{SW}-\cite{Yamaguchi:2002pa}.

We now explain why
the cutoff parameter $z_0 $ can be regarded as a UV cutoff 
parameter, from the view point of the boundary field theory 
\cite{SW}. 
We consider a bulk scalar field $\phih(x,z)$ on (Euclidean) AdS$_{d+1}$ 
of the metric 
\ba 
 ds^2 = \frac{l^2}{z^2}\,\left(
  dz^2+\delta_{ij}\,dx^idx^j\right),
 \label{ads.sample}
\ea 
and assume that the mass $m$ of the scalar is much larger than 
the typical scale of the AdS; $m\gg l^{-1}$. 
Then, according to the AdS/CFT correspondence described above, 
the two-point function of the operator ${\mathcal{O}}$ which 
is coupled to $\hphi$ at the boundary $z=z_0 $ is evaluated as 
\begin{equation}
 \Bigl\langle
 {\mathcal{O}}(x){\mathcal{O}}(y)
 \Bigr\rangle_{z_0 } \sim 
 \sum_{{\rm paths\,connecting}\,X\,{\rm and}\,Y}
  \exp\Bigl(-m\times\bigl({\rm length\,of\,path}\bigr)\Bigr),
\end{equation}
where $X=(x^i,z\!=\!z_0)$ and $Y=(y^i,z\!=\!z_0)$. 
Under the situation $m\gg l^{-1}$, 
we can evaluate this with the geodesics and obtain 
\begin{equation}
 \Bigl\langle
 {\mathcal{O}}(x){\mathcal{O}}(y)
 \Bigr\rangle_{z_0 } \sim 
 \exp\bigl(-m\,{\mathcal D}(X,Y)\bigr),
\end{equation}
where ${\mathcal D}(X,Y)$ represents the geodesic distance 
between $X$ and $Y$ in AdS$_{d+1}$. 
For the AdS metric \eq{ads.sample}, the geodesic distance is given by 
\ba
 {\mathcal D}(X,Y)
  = l\cdot\ln\left(
  \frac{\left(|x|+\sqrt{|x|^2+z_0 ^2}\right)^2}{z_0 ^2}\right), 
\ea
where $|x|^2\equiv\delta_{ij}x^ix^j$. 
So the two-point function becomes 
\begin{align}
 \Bigl\langle
 {\mathcal{O}}(x){\mathcal{O}}(y)
 \Bigr\rangle_{z_0 } &\sim 
 \frac{z_0 ^{2m {}l}}{\left(|x-y|+\sqrt{|x-y|^2+z_0 ^2}\right)^{2m{}l}} \nn
 &\sim \frac{1}{|x-y|^{2m{}l}}\qq \text{for } |x-y| \gg z_0 . 
 \label{geodesics}
\end{align}
This means that the two-point function actually has a scaling behavior 
in the region $|x-y| \gg z_0 $ 
with scaling dimension $\Delta=ml$. 
In other words, this implies that $z_0$ gives a short-distance scale 
around which the scaling becomes broken, 
and thus $\Lambda_0=1/z_0$ can be regarded as a UV cutoff of 
the boundary field theory.

If we take into account the backreactions from bulk 
scalar fields to bulk gravity, 
we need to consider a wide class of metric 
which has an asymptotically AdS geometry near the boundary.%
\footnote{
This condition is required for gravity 
to describe a continuum theory at the boundary. 
}
This leads us to introduce the boundary conditions at the new boundary 
for the classical solutions of the induced metric of the bulk metric 
$\gh_{\mu\nu}(x,z)$,
\ba
 \gb_{ij}(x,z_0)=g_{ij}(x),
\ea
together with its regularity inside the bulk ($z\rightarrow+\infty$). 
The classical action is defined by substituting the classical solutions 
of the bulk metric and the bulk scalar fields into the bulk action,%
\footnote{
In \S 3, we prove that the classical action is independent of 
the coordinate $z_0$ of the boundary 
as a result of the diffeomorphism invariance along the radial direction. 
} 
\ba
 S\bigl[g_{ij}(x),\phi^a(x)\bigr]
  \equiv\bS\bigl[\gb_{\mu\nu}(x,z),\phib^a(x,z)\bigr]. 
\ea
The classical action can be divided into the nonlocal  
and the local parts:
\ba
 \newton S\bigl[g_{ij}(x),\phi^a(x)\bigr] 
  = -\Gamma\bigl[g_{ij}(x),\phi^a(x)\bigr]
  +\newton S_{\rm loc}\bigl[g_{ij}(x),\phi^a(x)\bigr]. 
\ea
The nonlocal part can be regarded as the generating functional 
of $d$-dimensional quantum field theory (QFT${_d}$) 
in the curved background with the metric $g_{ij}(x)$. 
The local part is the local counterterms. 
This should be actually expressed in a local form 
since singular behavior near the boundary is translated 
into the short distance singularity of QFT${_d}$.

In summary, by introducing the cutoff $z_0$ into the AdS/CFT correspondence, 
we obtain the following duality: 
\begin{equation}
 {\rm SUGRA}_{d+1}~{\rm with~IR~cutoff}~z_0~\Longleftrightarrow~
 {\rm QFT}_d~{\rm with~UV~cutoff}~\Lambda_0=z_0^{-1}. 
\end{equation}

\subsection{Calculation of scaling dimensions}

Here we calculate the scaling dimension of an operator of the 
$d$-dimensional CFT which is coupled 
to a scalar field in the background of the AdS space-time 
\cite{GKP,W;holography}.

We consider a single scalar field on the $d$-dimensional 
Euclidean AdS space-time of radius $l$.  
To determine the scaling dimension of the dual operator, 
we calculate the two-point function of the operator 
using the prescription described 
in the previous subsection. 
As the action of the scalar, we take
\begin{align}
 \newton\,& \bS\bigl[\gh^{\rm AdS}_{\mu\nu}(x,z),\phih(x,z)\bigr] \nn
 &=\frac{1}{2\kappa_{d+1}^2}\int d^{d+1}X \sqrt{\hg_{\rm AdS}}
 \Biggl[
 \frac{1}{2}\,\hg^{\mu\nu}_{\rm AdS}\,
 \del_{\mu}\hphi\,\del_{\nu}\hphi+\frac{m^2}{2}\,\hphi^2
 \Bigr] + \bigl(\phih{\rm -independent~terms}\bigr) \nn
&= \frac{l^{d-1}}{4\kappa_{d+1}^2}
 \int d^dx \int_{z_0} ^\infty \frac{dz}{z^{d-1}}
 \Biggl[
 \left(\del_z\hphi\right)^2+\left(\del_i\hphi\right)^2
 +\frac{l^2m^2}{z^2}\,\hphi^2
 \Biggr] \nn 
&= \frac{l^{d-1}}{4\kappa_{d+1}^2}\int d^dx \int_{z_0} ^\infty {dz} 
 \Biggl[
 -\hphi\Bigl(\del_z^2\hphi -\frac{d-1}{z}\,\del_z\hphi + \del_i^2\hphi
 -\frac{1}{z^{d-1}}\,\frac{l^2m^2}{z^2}\,\hphi\Bigr)  \nn 
 &~~~~~~~~~~~~~~~~~~~~~~~~~~~~~~~~~~
 +\del_z\left(\frac{1}{z^{d-1}}\,{\hphi}\,\del_z{\hphi}\right)
 +\del_i\left(\frac{1}{z^{d-1}}\,{\hphi}\,\del_i{\hphi}\right)
 \Biggr],
\label{scalar-action}
\end{align}
where $z_0 $ is the cutoff parameter to regularize the infinite volume 
of the AdS space-time. 
Using the equation of motion for $\hphi$ given by
\ba
 \del_z^2\hphi -\frac{d-1}{z}\,\del_z\hphi + \del_i^2\hphi
 -\frac{l^2m^2}{z^2}\,\hphi =0, 
 \label{eom-exam}
\ea
the classical action reads
\ba
 S=l^{d-1}\,\int d^dx
 \Bigl[
 \frac{1}{z^{d-1}}\,\phibar\,\del_z\phibar 
 \Bigr]_{z=z_0 }^{z=\infty}, 
 \label{class-action-exam}
\ea
where $\phibar$ is the solution of \eq{eom-exam}. 

To solve the equation of motion \eq{eom-exam}, we Fourier-expand 
the field $\phibar(x,z)$ as 
\ba
 \phibar(x,z) = \int \frac{d^dk}{(2\pi)^d} \,\lambda_k\,e^{ik_ix^i}\,
 \phibar_k(z)
 \qquad \left(\phibar_k(z\!\!=\!\!z_0 )=1\right).  
\ea
It turns out that $\phibar_k(z)$ is expressed by a 
modified Bessel function;%
\footnote{
Another modified Bessel function $I_{\nu}\left(kz\right)$ is not suitable 
because we require the classical solution to be regular in the limit 
$z\to\infty$.
} 
\ba
 \phibar_k(z)=\frac{z^{d/2}K_{\nu}\left(kz\right)}
 {z_0^{d/2}K_{\nu}\left(kz_0 \right)}
 \qquad \left(\nu\equiv\sqrt{l^2m^2+d^2/4}\right), 
 \label{sln-exam}
\ea
where $k\equiv\sqrt{k_1^2+\cdots k_d^2}$. 
By substituting \eq{sln-exam} into \eq{class-action-exam}, 
we obtain the classical action 
\ba
 \newton\,S\left[\lambda_k\right]
 ={2l^{d-1}\over4\kappa_{d+1}^2}\int 
  \frac{d^dk}{(2\pi)^d}\,\frac{d^dq}{(2\pi)^d}\, 
 \lambda_k\,\lambda_q\,(2\pi)^d\,\delta^{d}(k+q)\,{\mathcal{F}}(k), 
\ea
where%
\footnote{
Here we have used $\phibar_k(z=z_0 )=1$. 
} 
\begin{align}
 {\mathcal{F}}(k) &\equiv\Bigl[
 \phibar_k(z)\frac{1}{z^{d-1}}\,\del_z\phibar_k(z)
 \Bigr]_{z=z_0 }^{z=\infty} \nn 
 &= -\left.\Bigl(
 \frac{1}{z^{d-1}}\,\del_z\ln\phibar_k(z)
 \Bigr)\right|_{z=z_0 }. 
 \label{flux}
\end{align}

Writing the boundary value of the scalar as
$\phibar(x,z_0)=\int \frac{d^dk}{(2\pi)^d}\,\lambda_k\,e^{ikx}$,
the Fourier transform of the two-point function 
$\bigl\langle{\mathcal O}(x){\mathcal O}(y)\bigr\rangle_{\rm CFT}$ 
is given by%
\footnote{
The analytic terms in ${\mathcal{F}}$ give contact terms 
that only yields a contribution with a $\delta$-function-like support 
to the two-point functions. 
} 
\begin{align} 
 \Bigl\langle{\mathcal{O}}_k\,{\mathcal{O}}_q\Bigr\rangle_{\rm CFT}
 &\equiv
 \int d^d x\,d^d y\,e^{-ikx-iqy}\,
 \Bigl\langle{\mathcal{O}}(x)\,{\mathcal{O}}(y)\Bigr\rangle_{\rm CFT}
 \nn
 &=
 \frac{\delta}{\delta\lambda_{-k}}\,\frac{\delta}{\delta\lambda_{-q}}
 \left(-\newton\,S\bigl[\lambda_k\bigr]\right)
 \Bigg|_{\text{leading non-analytic part in $k$}}\
 \nn 
 &=
 -(2\pi)^d\frac{2l^{d-1}}{2\kappa_{d+\!1}^2}\,\delta^{d}(k+q)\,{\mathcal{F}}(k)
 \Bigg|_{\text{leading non-analytic part in $k$}}\,. 
 \label{2-point-exam-pre}
\end{align}
Using the identities 
\begin{gather}
 K_{\nu}=\frac{\pi}{2\sin\pi\nu}\left(I_{-\nu}-I_{\nu}\right), \\
 I_{\nu}=\left(\frac{z}{2}\right)^{\nu}\sum_{k=0}^{\infty}
 \frac{(z/2)^{2k}}{k!\,\Gamma(k+\nu+1)}, 
\end{gather}
and \eq{sln-exam}, the leading term of \eq{flux} in $z_0 $
is evaluated as 
\ba
 {\mathcal{F}}(k)=2z_0^{-d}\frac{\Gamma(1-\nu)}{\Gamma(\nu)}
 \left(\frac{kz_0}{2}\right)^{2\nu} + \left(\text{analytic in
 $k^2$}\right). 
\ea
Thus the connected two-point function \eq{2-point-exam-pre} is given by 
\ba
 \Bigl\langle{\mathcal{O}}_k\,{\mathcal{O}}_q\Bigr\rangle_{\rm CFT}
 = {\mathcal{N}}\,\delta^{d}(k+q)\,\left|k\right|^{2\nu},  
\ea
where ${\mathcal{N}}$ is a numerical factor. 
This is equivalent to 
\begin{align}
 \Bigl\langle{\mathcal{O}}(x){\mathcal{O}}(y)\Bigr\rangle_{\rm CFT}
 &= \int \frac{d^d k}{(2\pi)^d}\,\frac{d^dq}{(2\pi)^d}\,
 e^{ik x+iq y}\,
 \Bigl\langle{\mathcal{O}}_k\,{\mathcal{O}}_q\Bigr\rangle_{\rm CFT}\nn
 &\propto \frac{1}{|x-y|^{d+2\nu}}. 
\end{align}
We thus find that the scaling dimension $\Delta$ 
of the operator $\mathcal{O}$ is given by 
\begin{align}
 \Delta = \frac{d}{2}+\nu=\frac{1}{2}\left(d+\sqrt{d^2+4m^2l^2}\right),  
 \label{scaling-dimension}
\end{align}
or
\ba
 \Delta\left(\Delta-d\right)=m^2l^2. 
 \label{mass-formula}
\ea
Note that eq.\ \eq{scaling-dimension} gives $\Delta\sim ml$ 
in the limit $m\gg l^{-1}$, 
which is consistent with the expression \eq{geodesics}. 

\subsection{Example}

As discussed in the introduction, 
the duality between Type IIB supergravity on AdS$_5\times S^5$ 
and the four-dimensional ${\mathcal{N}}=4~SU(N)$ SYM theory
is one of the typical examples of the AdS/CFT correspondence. 
As an evidence for this duality,  
we make a review of the one-to-one
correspondence between the chiral primary operators of  
the four-dimensional 
${\mathcal{N}}=4~SU(N)$ SYM theory and the Kaluza-Klein modes 
of IIB supergravity compactified on $S^5$  
\cite{W;holography,review}\cite{FFZ}-\cite{FLZ}.

The four-dimensional ${\mathcal{N}}=4~SU(N)$ SYM
theory is constructed from an ${\mathcal{N}}=4$ vector multiplet, 
that is, six real scalar fields $\phi^I$ ($I=1,\cdots ,6$), 
four complex Weyl spinor fields 
$\lambda_{\alpha A}$ ($A=1,\cdots,4$) and a vector field $A_{i}$,  
each field of which belongs to the adjoint representation of $SU(N)$. 
This theory has 16 real supercharges
$\left(Q_{\alpha}^A,\overline{Q}_{\dot{\alpha} A}\right)$ and 
the supersymmetry transformations for these fields are 
\cite{Sohnius} 
\begin{align}
\left[Q_{\alpha}^A,\phi^I\right] 
&= \left(\gamma^I\right)^{AB}\lambda_{\alpha B}, 
\nn
\left\{Q_{\alpha}^A,\lambda_{\beta B}\right\} 
&= -{i\over 2}\left(\sigma^{ij}\right)_{\alpha\beta}
\delta^A_{\ B}F_{ij}+2i\left(\gamma^{IJ}\right)^A_{\ B}
 \left[\phi^I,\phi^J\right],
\nn
\left\{Q_{\alpha}^A,\overline{\lambda}_{\dot{\alpha}}^B\right\} 
&= 2i\sigma^{i}_{\alpha\dot{\alpha}}
\left(\gamma^I\right)^{AB}{\mathcal{D}}_{i}\phi^I, 
\nn
\left[Q_{\alpha}^A,A_{i}\right] 
&=i\left(\sigma_{i}\right)_{\alpha\dot{\alpha}}
\overline{\lambda}_{\dot{\beta}}^A\epsilon^{\dot{\alpha}\dot{\beta}}, 
\label{susy-trans}
\end{align}
where 
\begin{equation}
\Gamma^I=
\begin{pmatrix}
 0 & (\gamma^I)^{AB} \\
 (\overline{\gamma}^I)_{AB} & 0
\end{pmatrix}
\end{equation}
are the gamma matrices for the $SO(6)$ and 
$(\gamma^{IJ})^A_{\ B}\equiv 
\half\left(\gamma^I\overline{\gamma}^J
-\gamma^J\overline{\gamma}^I\right)^A_{\ B}$. 
The operations of $\bQ_{\dot{\alpha}A}$ are similar. 

The spectra of the operators in this theory include all the gauge
invariant quantities that can be constructed from the fields described
above. 
Here we concentrate our attention on the local operators 
that can be written as a single-trace of products of the fields in 
the $\mathcal{N}=4$ vector multiplet.%
\footnote{
Although we have also multi-trace operators  
which appear in operator product expansions of
single-trace operators, 
we do not consider them here 
since they can be ignored in the large $N$ limit. 
For a discussion of multi-trace operators in 
the AdS/CFT correspondence, see, 
Refs.\ \cite{ABS,2-trace,BSS}. 
}

The four-dimensional ${\mathcal{N}}=4~SU(N)$ SYM theory 
is a superconformal field theory as a consequence of 
the large supersymmetry.  
The generators of the superconformal transformation 
consist of the supersymmetry generators 
$\left\{M_{ij},P_{i},Q^A_{\alpha}\right\}$, the dilatation 
$D$, the special conformal transformation $K_{i}$ and its superpartner 
$S^A_{\alpha}$.  
One also needs to introduce the generators $R$ of the $R$-symmetry group 
$SU(4)$. 
The algebra also contains the bosonic conformal algebra 
$\left\{M_{ij},P_{i},K_{i},D\right\}$
as a subalgebra. 
We show some part of the algebra which are necessary 
for our discussion; 
\begin{gather}
\left[D,Q\right]=-\frac{i}{2}\,Q,\qquad \left[D,S\right]
  =+\frac{i}{2}\,S, \nn
\left[D,P_i\right]=-iP_{i},\qquad \left[D,K_{i}\right]=+iK_{i},
 \nn
\left[D,M_{ij}\right]=0, \qquad \left\{Q,S\right\}\sim M+D+R. 
\label{scft-alg}
\end{gather} 
For the complete (anti-)commutation relations of the generators, 
see Ref.\ \cite{Minwalla}.

We are interested in representations of the superconformal algebra 
whose conformal dimensions are suppressed from below. 
Let us start with the bosonic conformal subalgebra $\{M_{ij},P_i,K_i,D\}$. 
From the assumption that the conformal dimensions are suppressed 
from below, there is a state 
$\ket{{\mathcal{O}}'}$ that is characterized by the property, 
\ba
 K_i\ket{{\mathcal{O}}'} = 0.   
\ea
We can generate a tower of states from the this state 
by acting on it with the generator $P_i$, 
which is called the {\it primary multiplet}. 
The state $\ket{\mathcal{O}'}$ is called the {\it primary state} and 
the other states in the multiplet are called the {\it descendants}. 
Recalling the fact that the generator $P_i$ 
raises the conformal weight by $1$ (See \eq{scft-alg}),   
the primary state is the lowest weight state in the multiplet.

There is also the same structure in an irreducible representation of the 
superconformal algebra, that is,    
there is a state that is characterized by the property, 
\ba 
 S|{\mathcal{O}}\rangle = 0,\quad K|{\mathcal{O}}\rangle = 0, 
\ea
and a tower of states is constructed from this state 
by acting with the generators $(Q,\bQ)$ and $P_{i}$, 
which raise the conformal weight by $1/2$ and $1$, respectively.  
We call the state $\ket{{\mathcal{O}}}$ the 
{\it superconformal-primary state} and other states in the multiplet 
the {\it descendants}. 
We note that the multiplet is divided into 
several primary multiplets of the bosonic conformal subalgebra  
whose primary states are obtained by acting with the supercharges 
to the superconformal-primary state.

In primary operators%
\footnote{
We do not distinguish states and local operators because,  
in a conformal field theory,
there is one-to-one correspondence between them \cite{review}. 
} 
in the $\cN=4$ $SU(N)$ SYM theory, 
we are especially interested in the {\em chiral primary operators} 
that are eliminated by some combinations of 16 supercharges, 
not only by $S$'s. 
From the way of construction of primary multiplets described above, 
we can easily see that 
the multiplet that is made from a chiral primary operator 
contains smaller number of states than a general 
superconformal-primary multiplet. 
As discussed in Ref.\ \cite{DB}, 
the last equation of \eq{scft-alg} gives a relation among 
the conformal dimension, 
the representation of the Lorentz group 
and the representation of the R-symmetry 
($SU(4)$) of a chiral primary operator.  
This means that the conformal dimension of a chiral primary
operator is determined only by the superconformal algebra, 
being independent of the coupling constant.
Thus the chiral primary operators are appropriate
in comparing their properties with those of the dual supergravity theory,  
since the description by classical supergravity is 
reliable only in the region where the 't Hooft coupling is large 
[see eq.\ \eq{condition2}], 
for which perturbative calculation of SYM is not applicable. 
For detailed discussions of the representation theory 
of extended superconformal algebras, see, for example, 
Refs.\ \cite{Minwalla}-\cite{Andrianopoli:1998ut}. 

Let us discuss the structure of the chiral primary operators  
that are represented as the single trace of the fields 
in the $\cN=4$ vector multiplet, 
following the presentation given in Ref.\ \cite{review}.
By definition, 
the lowest component of the chiral primary multiplet is 
characterized by the fact that it cannot be obtained 
by acting on any other operator with supercharges. 
The supersymmetric transformation of the ${\mathcal{N}}=4$ 
vector multiplet \eq{susy-trans} suggests 
that the requested chiral primary operators are described by 
the trace of a symmetric product of only the scalar fields.%
\footnote{
We note that the fields in the ${\mathcal{N}}=4$ 
vector multiplet is eliminated by half of the 16 supercharges by
definition.
We must symmetrize the product because the right hand side of 
\eq{susy-trans} contains the commutators of $\phi^I$'s.} 
In fact, as discussed in Ref.\ \cite{Minwalla}, 
a scalar primary operator with conformal dimension $n$ which belongs 
to the representation of $SU(4)$ with Dynkin index $(0,n,0)$ 
is eliminated by half of the 16 supercharges. 
This means that the lowest component of the chiral primary 
multiplet is given by \cite{HST,HW} 
\ba
 {\mathcal{O}}_n\equiv \tr\left(\phi^{(I_1}\cdots\phi^{I_n)}\right)
 -({\rm traces}), 
 \qquad n=2,\cdots,N \,. 
\label{lowest}
\ea
For example, ${\mathcal{O}}_2$ stands for the set of operators 
of the form $\tr\left(\phi^I\phi^J\right)
-\frac{1}{6}\delta^{IJ}\tr\left(\sum_{K=1}^6\phi^K\phi^K\right)$. 
The conformal dimension of the operator ${\mathcal{O}}_n$ is $n$ 
because we can evaluate it in the zero coupling 
limit of the SYM theory. 
The maximum value of $n$ is $N$ because the trace of 
a symmetric product of more than $N$ commuting 
matrices can always be written 
as a sum of products of ${\mathcal{O}}_n$ $(n\leq N)$.

In the following, we examine the contents of the  
chiral primary multiplet built from the ${\mathcal{O}}_n$. 
We note that any state in the multiplet is in a representation of 
both of the superconformal algebra and the $R$-symmetry $SU(4)$. 
Recalling that $D$ and $M_{ij}$ commute each other, 
it is convenient to label the state by the conformal weight, $\Delta$, 
the left and right spins, $(j_1,j_2)$, and the Dynkin index of the
$SU(4)$, $(p,q,r)$.%
\footnote{
The dimension of the irreducible representation of $SU(4)$ 
with Dynkin index $(p,q,r)$ is given by \cite{AF} 
 $d(p,q,r)\equiv\left(p+1\right)\left(q+1\right)\left(r+1\right)
 \left(1+\frac{p+q}{2}\right)
 \left(1+\frac{q+r}{2}\right)
 \left(1+\frac{p+q+r}{3}\right)$, 
which gives the degeneracy of the state. 
}
For example, ${\mathcal{O}}_n$ and supercharges are labeled as 
\begin{equation}
\begin{array}{c|c|c|c|c}
                   & \Delta & SU(2)_L\times SU(2)_R & 
SU(4) & {\rm weight} \\ \hline
 {\mathcal{O}}_n & n    & (0,0) & (0,n,0) & 0 \\ \hline 
 Q^A_{\alpha}    & \half  & (\half,0) & (0,0,1) & +\half \\ \hline
 \bQ_{\dalpha A}  & \half  & (0,\half) & (1,0,0) & -\half \\ 
\end{array}
\end{equation}
Here, in order to keep track of operation of supercharges, 
we have introduced an additive weight 
by assigning $+1/2$ to $Q_\alpha^A$ and $-1/2$ to $\bQ_{\dalpha A}$. 
The operators in the multiplet are obtained by acting 
on the ${\mathcal{O}}_n$ with $Q$ and $\bQ$, and 
their labels are determined by 
those of the fields in the ${\mathcal{N}}=4$ 
vector multiplet, 
\begin{equation}
\begin{array}{c|c|c|c}
                   & SU(2)_L\times SU(2)_R 
		   & SU(4) & {\rm weight}\\ \hline
\phi^I             & (0,0)     & (0,1,0) & 0 \\ \hline
\lambda_{\alpha A} & (\half,0) & (1,0,0) & +\half \\ \hline
{\blambda}_{\dot{\alpha}}^A & (0,\half) & (0,0,1) & -\half \\ \hline
A_{i}              & (\half,\half) & (0,0,0) & \pm 1 \\
\end{array}
\end{equation}
and the supersymmetry transformation \eq{susy-trans}.

As an example, we explicitly construct the operators with conformal weight 
$n+1/2$ and $n+1$ by operating the supercharges to the lowest operator 
${\mathcal{O}_n}$ \cite{review}. 

\begin{description}
\item 1) $\Delta = n+1/2$

The states with the conformal dimension $n+1/2$ are obtained by 
operating the supercharges once to the lowest state 
$|{\mathcal{O}}_n\rangle$, that is,  
$Q_{\alpha}|{\mathcal{O}}_n\rangle$ and 
$\bQ_{\dot{\alpha}}|{\mathcal{O}}_n\rangle$.  
Their explicit expressions are%
\footnote{
In this subsection, we assume that 
fields in a trace are always symmetrized. 
} 
\begin{equation}
 \lambda^{(1)}_{\alpha}\equiv
  \tr\left(\lambda_{\alpha A}\,\phi^{I_2}\cdots\phi^{I_n}\right)
  \quad{\rm and}\quad
 \lambda^{(1){\dagger}}_{\dalpha} = 
 \tr\left({\blambda}_{\dot{\alpha}}^A\,\phi^{I_2}\cdots\phi^{I_n}\right).  
\end{equation}
They are spinor fields and their complex conjugate,  
whose $SU(4)$ Dynkin index and labels of the superconformal algebra are 
summarized in the table, 
\begin{equation}
 \begin{array}{c|c|c|c}
  & SU(2)_L\times SU(2)_R & SU(4) & {\rm weight} \\ \hline
   {\rm complex}\ \lambda^{(1)}_{\alpha}    & (\half,0)+(0,\half) 
   & (1,n-1,0)+(0,n-1,1) & \pm\half \\ 
 \end{array}
 \label{cweight-1/2}
\end{equation}

\item 2) $\Delta = n+1$ 

These states with the conformal weight $n+1$ are obtained by 
operating two supercharges. 
When we operate the supercharges with the same chirality, 
the irreducible representations are obtained by 
either symmetrizing or antisymmetrizing the supercharges.  
In the first case, we obtain 
$Q_{(\alpha}Q_{\beta)}|{\mathcal{O}}_n\rangle$ 
and its complex conjugate, which are self-dual and anti-self-dual 
two-form fields, respectively;   
\begin{align}
 B^{(1)}_{ij}&\equiv
 \left(\sigma_{ij}\right)^{\alpha\beta}\tr\left(
 {\bigl(\sigma^{kl}\bigr)_{\alpha\beta}\,
 F_{kl}\,\phi^{I_2}\cdots\phi^{I_{n}}}\right)+\cdots,  \nn
 {B}^{(1)\dagger}_{ij}&=
 \left(\overline{\sigma}_{ij}\right)^{\dalpha\dbeta}\tr\left(
 {\bigl(\overline{\sigma}^{kl}\bigr)_{\dot{\alpha}\dot{\beta}}\,
 F_{kl}\,\phi^{I_2}\cdots\phi^{I_{n}}}\right)+\cdots.   
\end{align}
In the second case, we obtain $\epsilon^{\alpha\beta} 
Q_{\alpha}Q_{\beta}|{\mathcal{O}}_n\rangle$ 
and its complex conjugate, 
which are scalar fields and their complex conjugate, respectively; 
\begin{align}
 \varphi^{(2)}&\equiv
 \epsilon^{\alpha\beta}\,
 \tr\left(\lambda_{\alpha A}\,\lambda_{\beta B}\,
 \phi^{I_3}\cdots\phi^{I_{n}}\right)+\cdots, \nn
 \varphi^{(2)\dagger}&=
 \epsilon^{\dot\alpha\dot\beta}\,
 \tr\left(\blambda_{\dot\alpha}^A\,\blambda_{\dot\beta}^B\,
 \phi^{I_3}\cdots\phi^{I_{n}}\right)+\cdots. 
\end{align}

On the other hand, when we operate the supercharges with different 
chiralities, the obtained states, 
$Q_{\alpha}\bQ_{\dot\alpha}|{\mathcal{O}}_n\rangle$,  
are real vector fields; 
\ba 
 A^{(1)}_{i}\equiv \left(\sigma_{i}\right)^{\alpha\dalpha}\,
 \tr\left(\lambda_{\alpha A}\,\lambda_{\dot\alpha}^B\,
 \phi^{I_3}\cdots\phi^{I_n}\right)+\cdots. 
\ea

Their $SU(4)$ Dynkin index and the labels of the 
superconformal algebra are 
summarized as 
\begin{equation}
 \begin{array}{c|c|c|c}
  & SU(2)_L\times SU(2)_R & SU(4) & {\rm weight} \\ \hline 
  {\rm complex}\ B^{(1)}_{ij} & (1,0)+(0,1) & (0,n-1,0)+(0,n-1,0) & \pm 1 \\ 
  {\rm complex}\ \varphi^{(2)} & (0,0) & (2,n-2,0)+(0,n-2,2) & \pm 1 \\ 
  {\rm real}\ A^{(1)}_{i} & (\half,\half) & (1,n-2,1) & 0  
 \end{array}
 \label{cweight-1} 
\end{equation}
\end{description}

Repeating the same operation, all the states in the multiplet 
can be constructed. 
We summarize the result in the Table \ref{states}, 
where we write only the primary states of the bosonic conformal
subalgebra in the multiplet. 
For example, we do not write such states that is obtained by 
acting with more than eight supercharges 
because such states must vanish or become descendants 
of the primary multiplets of the bosonic conformal subalgebra. 
In Table \ref{states}, for $n=2$ and $3$  
the states with negative Dynkin indices should be ignored. 

\begin{table}
\caption{The primary states in the short chiral primary 
 multiplet built on the lowest state \eq{lowest}. 
 The operator ${\mathcal{O}}_n$ corresponds to the scalar operator  
 $\varphi^{(1)}$. 
 We denote the representations of the Lorentz group 
 by the symbols 
 $\varphi$,
 $\lambda_{\alpha}$, 
 $A_{i}$, 
 $B_{ij}$,
 $\psi_{i\alpha}$,
 and $h_{ij}$, 
 which correspond to 
 states with the left and right spins $(0,0)$, $(\half,0)+(0,\half)$, 
$(\half,\half)$, $(1,0)+(0,1)$, $(1,\half)+(\half,1)$ and 
$(1,1)$, respectively. 
 The $(p,q,r)$ is the Dynkin index of the $R$-symmetry group $SU(4)$. 
}
\begin{equation*}
\begin{array}{|c|c|c|c|c|}\hline
\Delta & SO(1,3) & 
SU(4) & {\rm weight} \\ \hline\hline
n       & {\rm real}   \  \varphi^{(1)} & (0,n,0) & 0 \\ \hline 
n+\half & {\rm complex}\  \lambda^{(1)}_{\alpha}  & (1,n-1,0)+(0,n-1,1) & 
 \pm\half
\\ \hline 
n+1     & 
\begin{array}{c}
{\rm complex}\ \varphi^{(2)} \\
{\rm complex}\ B^{(1)}_{ij} \\
{\rm real}\ A^{(1)}_i
\end{array} & 
\begin{array}{c}
(2,n-2,0)+(0,n-2,2) \\
(0,n-1,0)+(0,n-1,0) \\
(1,n-2,1)
\end{array} & 
\begin{array}{c}
\pm 1 \\
\pm 1 \\
0 
\end{array} \\ \hline  
n+\frac{3}{2} & 
\begin{array}{c}
{\rm complex}\ \lambda^{(2)}_\alpha \\
{\rm complex}\ \lambda^{(3)}_\alpha \\
{\rm complex}\ \psi^{(1)}_{i\alpha}  
\end{array} & 
\begin{array}{c}
(1,n-2,0)+(0,n-2,1) \\
(2,n-3,0)+(0,n-3,2) \\
(0,n-2,1)+(1,n-2,0) 
\end{array} & 
\begin{array}{c}
\pm\frac{3}{2} \\
\pm\half \\
\pm\half 
\end{array} \\ \hline  
n+2 &
\begin{array}{c}
{\rm complex}\ \varphi^{(3)} \\
{\rm complex}\ A^{(2)}_i \\
{\rm real}\ \varphi^{(4)}  \\
{\rm complex}\ B^{(2)}_{ij} \\
{\rm real}\ h_{ij} 
\end{array} & 
\begin{array}{c}
(0,n-2,0)+(0,n-2,0) \\
(1,n-3,1)+(1,n-3,1) \\
(2,n-4,2) \\
(0,n-3,2)+(2,n-3,0) \\
(0,n-2,0) 
\end{array} & 
\begin{array}{c}
\pm 2 \\
\pm 1 \\
0 \\
0 \\
0
\end{array} \\ \hline
n+\frac{5}{2} & 
\begin{array}{c}
{\rm complex}\ \lambda^{(4)}_\alpha \\
{\rm complex}\ \lambda^{(5)}_\alpha \\
{\rm complex}\ \psi^{(2)}_{i\alpha}  
\end{array} & 
\begin{array}{c}
(0,n-3,1)+(1,n-3,0) \\
(1,n-4,2)+(2,n-4,1) \\
(1,n-3,0)+(0,n-3,1) 
\end{array} & 
\begin{array}{c}
\pm\frac{3}{2} \\
\pm\half \\
\pm\half 
\end{array} \\ \hline  
n+3 &
\begin{array}{c}
{\rm complex}\ \varphi^{(5)} \\
{\rm complex}\ B^{(3)}_{ij} \\
{\rm real}\ A^{(3)}_i
\end{array} & 
\begin{array}{c}
(0,n-4,2)+(2,n-4,0) \\
(0,n-3,0)+(0,n-3,0) \\
(1,n-4,1)
\end{array} & 
\begin{array}{c}
\pm 1 \\
\pm 1 \\
0 
\end{array} \\ \hline  
n+\frac{7}{2} & 
{\rm complex}\  \lambda^{(6)}_{\alpha}  & (0,n-4,1)+(1,n-4,0) & \pm\half
\\ \hline 
n+4     & {\rm real}   \  \varphi^{(6)} & (0,n-4,0) & 0 \\ \hline
\end{array}
\end{equation*}
\label{states}
\end{table}

\newpage
On the other hand, 
the bosonic sector of ten-dimensional Type IIB supergravity consists of 
a graviton, a complex scalar, a complex two-form field  
and a real four-form field whose five-form field strength 
is self-dual, 
and the fermionic sector consists of 
a chiral complex gravitino and a chiral complex spinor 
of opposite chirality \cite{KRN}. 
The Kaluza-Klein spectra on $S^5$ are obtained by expanding the fields 
by the spherical harmonics of $S^5$. 
Here we demonstrate the simplest example of the calculation, 
that is, the harmonic expansion of a complex scalar field 
$\bB$ in a ten-dimensional space-time $M_{10}$. 
The equation of motion is given by 
\ba
 {1\over\sqrt{-\bG}}\del_M\left(\sqrt{-\bG}\bG^{MN}\del_N\bB\right)=0, 
 \label{eom-B}
\ea
where $\bG_{MN}$ is the metric of the $M_{10}$. 
We assume that the manifold $M_{10}$ has a structure AdS$_5\times
S^5$ with the same curvature radius $l$. 
By introducing the coordinates ${\bf X}^M=(X^\mu,\,y^a)$ 
and writing the metric of AdS$_5$ and {\em unit} $S^5$ as 
$\widehat{g}_{\mu\nu}$ and $h_{ab}$, respectively, 
the equation of motion \eq{eom-B} is decomposed into 
the AdS$_5$-part and the $S^5$-part as follows:
\ba
 {1\over\sqrt{-\hg(X)}}\del_\mu
 \left(\sqrt{-\hg(X)}\hg^{\mu\nu}(X)\del_\nu\bB(X,y)\right)
 +{1\over l^2}{1\over\sqrt{h(y)}}\del_a\left(\sqrt{h(y)}
 h^{ab}(y)\del_b\bB(X,y)\right)=0. \nn
 \label{eom-B-2}
\ea
Here $\del_\mu\equiv \del/\del X^\mu$ and $\del_a\equiv \del/\del y^a$.
Next we decompose the scalar field $\bB(X,y)$with the 
scalar harmonics of unit $S^5$, 
\ba
 \bB(X,y) \equiv \sum_{j=0}^{\infty}
 \sum_{m=1}^{A_j}
 \varphi_{jm}(X)Y_{jm}(y)\,,
 \qquad \left(A_j={1\over12}(j+3)(j+2)^2(j+1)\right), 
 \label{harm-exp}
\ea
where $Y_{jm}(y)$ is the eigenfunction of the Laplacian of unit $S^5$,  
\ba 
  {1\over\sqrt{h(y)}}\del_a\left(\sqrt{h(y)}
 h^{ab}(y)\del_bY_{jm}(y)\right)= -j(j+4)Y_{jm}(y). 
\ea
Substituting \eq{harm-exp} into the equation of motion \eq{eom-B-2}, 
we obtain the equation which $\varphi_{jm}(X)$ satisfies; 
\ba
 {1\over\sqrt{-\hg(X)}}\del_\mu
 \left(\sqrt{-\hg(X)}\hg^{\mu\nu}(X)\del_\nu\varphi_{jm}(X)\right)
 - j(j+4)\,l^{-2}\varphi_{jm}(X)=0.   
\ea 
Thus the Kaluza-Klein modes made from the scalar fields $\bB$ 
consist of a tower of scalar fields of mass squared 
$ m_j^2=j(j+4)\,l^{-2}$ $(j=0,1,2,\cdots)$ with multiplicity $A_j$;
\ba
 \left\{\bigl\{\varphi_{jm}\bigr\}_{m=1}^{A_j}\,; m_j^2=j(j+4)\,l^{-2}
 \big|\, j=0,1,2,\cdots \right\}. 
 \label{KK-exam}
\ea

Thus, using the formula \eq{scaling-dimension}, 
the conformal weights 
of the corresponding scaling operators reads
\begin{align}
 \Delta_j&=\frac{1}{2}\,\left(4+\sqrt{4^2+m_j^2\,l^2}\right) \nn
  &= j+4\,, \qquad (j=0,1,2,\cdots)\,,  
\end{align}
which exactly corresponds to the scalar operator $\varphi^{(3)}$ 
in Table \ref{states} by setting $n=j+2$. 
In fact, for given $j$ $(=n\!-\!2)$, the degeneracy of the complex scalar 
modes $\varphi^{(3)}$ is given by the dimension of the representation 
of $SU(4)$ with the Dynkin index $(0,j,0)$, that is, 
${1\over12}(j+3)(j+2)^2(j+1)$,  
which exactly equals the degeneracy of the Kaluza-Klein modes \eq{KK-exam}.


The complete Kaluza-Klein spectra of Type IIB supergravity 
compactified on $S^5$ are summarized 
in TABLE III of Ref.\ \cite{KRN}. 
To compare their masses 
with the conformal weights of scalar operators 
in the chiral multiplets 
of the ${\mathcal{N}}=4~SU(N)$ SYM theory, 
we show the conformal 
weights of all the scalar states in the chiral multiplets; 
\begin{equation}
\begin{array}{cc|ccc}
              & & SU(4) & &
 \text{conformal weight}   \\ \hline
 \text{real} &  \varphi^{(1)}   
 & (0,n,0)             & (n\ge2),  & \Delta=2,3,\cdots,N\,, 
\\
 \text{complex} &  \varphi^{(2)}  
 & (2,n-2,0)+(0,n-2,2) & (n\ge2), & \Delta=3,4,\cdots,N+1\,, 
\\
 \text{complex} &  \varphi^{(3)}  
 & (0,n-2,0)+(0,n-2,0) & (n\ge2), & \Delta=4,5,\cdots,N+2 \,,
\\
 \text{real} &  \varphi^{(4)} 
 & (2,n-4,2)           & (n\ge4), & \Delta=6,7,\cdots,N+2\,, 
\\
 \text{complex} &  \varphi^{(5)}  
 & (2,n-4,0)+(0,n-4,2) & (n\ge4), & \Delta=7,8,\cdots,N+3\,, 
\\
 \text{real} &  \varphi^{(6)}  
 & (0,n-4,0)           & (n\ge4), & \Delta=8,9,\cdots,N+4\,. 
\end{array}
\label{scalars}
\end{equation}
If we apply the formula \eq{scaling-dimension} to 
the conformal dimensions of the scalar operators 
in \eq{scalars}, 
one can show that 
the mass spectra of the Kaluza-Klein scalar modes 
in TABLE III of Ref.\ \cite{KRN} are reproduced.

In Ref.\ \cite{GM}, the Kaluza-Klein spectra for $S^5$ compactification 
are classified by unitary irreducible 
representations of the superalgebra $SU(2,2|4)$ which is 
the maximal supersymmetric extension of 
the isometry group of the geometry AdS$_5\times S^5$, 
$SU(2,2)\times SU(4)$. 
The result is in the Table 1 of that literature. 
One can find the one-to-one correspondence between 
the Kaluza-Klein spectra in the 
Table 1 of Ref.\ \cite{GM} and the short chiral multiplets 
in the Table \ref{states} of this article.

The fascinating coincidence of the short 
chiral primary multiplets of ${\mathcal{N}}=4~SU(N)$ SYM with 
the Kaluza-Klein spectra IIB supergravity compactified on $S^5$ 
is a strong evidence of the AdS/CFT correspondence.


\subsection{Holographic RG}

In this subsection, we will make a review of a holographic
description of RG flows via supergravity. 
As was mentioned in \S 2.1 and will be discussed elaborately 
in the next section, the basic idea is that 
the evolution of bulk fields along the radial direction can be
identified with RG flows of the dual field theories.
When our interest is in an RG flow that connects a UV and an IR fixed 
points, the dual supergravity description is given by a background that
interpolates between two different asymptotic AdS regions along the
radial direction.
As an example, we focus on the holographic RG flow
from $\cN=4~SU(N)~{\rm SYM}_4$ to the $\cN=1$ Leigh-Strassler (LS) fixed
point \cite{LS;95}, which was investigated in Ref.\ \cite{HRG}.%
\footnote{
For analogous discussions in two-dimensional field theories, 
see Refs.\ \cite{Berg1,Berg2}.
}
The contents covered in this subsection will be re-investigated 
in \S 3.6 after we develop tools to investigate the holographic RG 
based on the Hamilton-Jacobi equations.

Let us first start by recalling the field theory stuff.
The matter content of ${\cal N}=4$ SYM 
in ${\cal N}=1$ superspace formulation reads
\begin{center}
\begin{tabular}{cc}
             & $SU(3)\times U(1)_R$ \\
 $W_{\alpha}$ & \mbox{\boldmath $1_1$} \\
 $\Phi_I$ & \mbox{\boldmath $3_{2/3}$} 
\end{tabular}
\end{center}
Here $W_\alpha$ and $\Phi_I$ $(I=1,2,3)$ are, respectively,  
$\cN=1$ vector multiplet and hypermultiplets. 
The LS fixed point can be realized by adding the mass perturbation 
to ${\cal N}=4$ SYM
\begin{equation}
 {\cal W}+\Delta {\cal W}=\tr\, \Phi_1 [\Phi_2,\Phi_3 ]
 +\frac{ m}{2}\,\tr\,\Phi_3^2,
\end{equation}
and choosing the anomalous dimensions of $\Phi_I$ as 
\begin{equation}
 \gamma_1=\gamma_2=-\frac{1}{4},~~\gamma_3=\frac{1}{2}.
\end{equation}
One can then see that the theory flows to an $N=1$ IR fixed point
with  $SU(2)\times U(1)_R^{\prime}$ global symmetry, because
the exact beta function \cite{beta} turns out to vanish:
\begin{equation}
 \beta(g)=-\frac{g^3 N}{8\pi^2}~
 \frac{3-\sum_{i=1}^3 (1-2\gamma_i)}{1-g^2N/8\pi^2}.
\end{equation}
Note that $U(1)_R^{\prime}$ is different from $U(1)_R$. 
We study the UV and IR fixed points by
computing the Weyl anomalies.
It is argued in Ref.\ \cite{an;97} that ${\cal N}=1$ superconformal
invariance relates the Weyl anomaly with the $U(1)_R$ anomaly as
\begin{eqnarray}
 \langle T_{~i}^{i}\rangle_{g,\,v}
 &=&\frac{c}{16\pi^2}\,\Bigl(\frac{1}{3}R^2-2R_{ij}^2
  +R_{ijkl}^2\Bigr)
 -\frac{a}{16\pi^2}\,\bigl(R^2-4R_{ij}^2+R_{ijkl}^2\bigr) 
 +\frac{c}{6\pi^2}\,V_{ij}^2, \nn
 \\
 \langle \partial_{i}(\sqrt{g}J^{i}) \rangle_{g,\,v}
 &=&-\frac{a-c}{24\pi^2}\,\bigl(R^2-4R_{ij}^2+R_{ijkl}^2\bigr)
 +\frac{5a-3c}{9\pi^2}\,V_{ij}\widetilde{V}^{ij}.
\end{eqnarray}
Here $g_{ij}$ is a background metric and $v_{i}$ a background 
gauge field coupled to the $R$-current $J^{i}$. 
$V_{ij}$ is the field strength of $v_i$, 
$R_{ijkl}$ is the Riemann tensor and $\widetilde{V}_{ij}$ is 
the dual of $V_{ij}$. 
The Adler-Bardeen theorem guarantees that $a$ and $c$ do not 
receive higher-loop corrections. 
So the coefficients of the Weyl anomaly
can be computed exactly in terms of perturbation.
It is then straightforward to compute $a-c$ and $5a-3c$
in the UV and IR fixed points:
\begin{equation}
 \frac{a_{\rm IR}}{a_{\rm UV}}=\frac{c_{\rm IR}}{c_{\rm UV}}
 =\frac{27}{32},~~a_{\rm UV}=c_{\rm UV},~a_{\rm IR}=c_{\rm IR}
\label{c;uvir}
\end{equation}

We will now show that the dual supergravity analysis reproduces this relation.
We first recall the computation of Weyl anomalies by supergravity 
\cite{HS;weyl}. 
It is found that the Weyl anomaly
of the dual CFT$_d$ takes the form
\begin{equation}
a=c\propto l^{d-1},
 \label{central-c}
\end{equation}
where $l$ is the radius of the AdS$_{d+1}$.
The UV fixed point is dual to AdS$_5\times S^5$ so that we get
$l_{\rm UV}=(4\pi g_sN)^{1/4}$.
On the other hand, the background dual to the IR fixed point should be
such that it has eight supercharges as well as an $SU(2)\times U(1)$
gauge group. In fact, it is shown in Ref.\ \cite{newvac} that ${\cal N}=8$
gauged supergravity in five dimensions allows this solution.
Using this result, one can obtain the radius of the new AdS background,
which turns out to yield the relation (\ref{c;uvir}).
[See also \S 3.6.]

In order to keep track of the whole RG trajectory using supergravity,
we have to find a IIB background that interpolates
along the radial direction between AdS$_5\times S^5$ corresponding to
the UV fixed point and AdS$_5\times K_5$ with $K_5$ being a compact
manifold that admits the necessary symmetries mentioned above.
Such a solution was constructed in Ref.\ \cite{PW;00} up to some unknown
functions.
Because of the background being complicated, it is difficult to get
information of the dual gauge theories from it.
One of the promising methods toward a global understanding of
holographic RG flows is to take a Penrose limit.
A Penrose limit of a background is taken by considering a null
geodesic on it and then defining an appropriate coordinate
transformation that reduces to the null geodesic equations in some
limit. So the Penrose limit amounts to probing the local geometry near
the null geodesic, and the original background often gets much simplified.
In fact, it is pointed out in Ref.\ \cite{BMN} that a Penrose limit of
AdS$_5\times S^5$ yields the pp-wave background \cite{pp}
that is maximally supersymmetric and the string theory on which
is solvable in the light-cone gauge \cite{Met;pp}.
The Penrose limit of the Pilch-Warner solution \cite{PW;00} was
studied in Ref.\ \cite{penrose;pw}.
For another application of the Penrose limit to the study of 
the holographic RG flows, see {\it e.g.} Ref.\ \cite{penrose;rg}.

Another intriguing aspect of the holographic RG is that
supergravity allows one to define a ``$c$-function'' that obeys an
analog of Zamolodchikov's $c$-theorem \cite{Z;c}.
Recalling the formula of two-dimensional Weyl anomaly 
$\bigl\langle T^i_i\bigr\rangle\propto c\,R$ with central charge $c$, 
it is natural to identify the coefficient of the Weyl anomaly 
as the central charge of the conformal field theory in arbitrary dimensions. 
Together with eq.\ \eq{central-c}, we thus define the central charge 
of the CFT dual to AdS gravity of radius $l$ as \cite{HS;weyl}
\ba
 c \sim l^{d-1}. 
 \label{central-c2}
\ea
To define the $c$-function, 
we consider a five-dimensional geometry with the metric
\begin{equation}
 ds^2 = d\tau^2 + \frac{1}{a(\tau)^2}\,\eta_{ij}\,dx^idx^j. 
 \label{metric_tau}
\end{equation}
When $a(\tau)=e^{\tau/l}$, this denotes AdS$_{d+1}$ of radius $l$.
This leads us to define the $c$-function as \cite{HRG}
\begin{equation}
 c(\tau)\propto \left(\frac{-1}{\widehat{K}(\tau)}\right)^{d-1},~~~
\widehat{K}(\tau)=-d\,\frac{d}{d\tau}\,\log a(\tau).
\label{c-function}
\end{equation}
For ${\rm AdS}_{d+1}$ of radius $l$, 
this actually gives $c(\tau)\propto l^{d-1}={\rm const}$,
in agreement with the definition \eq{central-c2}.
In order to show that $c(\tau)$ is a monotonically decreasing function of
$\tau$, we employ the null energy condition:
\begin{equation}
 \widehat{R}_{\mu\nu}\,\widehat{\xi}^\mu\,\widehat{\xi}^\nu
 =-\frac{d-1}{d}\,\frac{d\widehat{K}}{d\tau}\ge 0~~~
 {\rm for~any~null~vector}~\widehat{\xi}^\mu.
\end{equation}
Note that the inequality saturates for AdS that corresponds to a fixed
point of the dual theory.
It is not easy to verify a higher-dimensional analog of the Zamolodchikov
theorem in the purely field theory context 
(for a review, see Ref.\ \cite{high;c}).
The dual supergravity description provides us with 
a powerful framework for that.

\resection{Holographic RG and Hamilton-Jacobi formulation}

In this section, we discuss the formulation of the holographic
RG based on the Hamilton-Jacobi equation \cite{dVV,FMS1}.

\subsection{Hamilton-Jacobi constraint and the flow equation}

We start by recalling the Euclidean ADM decomposition 
that parametrizes a $(d+1)$-dimensional metric as
\ba
 ds^2&=&\widehat{g}_{\mu\nu}\,dX^\mu dX^\nu \nn
 &=&\Nh(x,\tau)^2 d\tau^2
 +\widehat{g}_{ij}(x,\tau)\Bigl(dx^i+\lambdah^i(x,\tau)d\tau\Bigr) 
 \Bigl(dx^j+\lambdah^j(x,\tau)d\tau\Bigr).
\label{metric;dvv}
\ea
Here $X^\mu=(x^i,\tau)$ with $i=1,\cdots,d$, 
and $\Nh$ and $\lambdah^i$ are the lapse and the shift function, 
respectively. 
The signature of the metric $\widehat{g}_{\mu\nu}$ 
is taken to be $(+\cdots +)$. 
As we discussed in the previous sections, the Euclidean time $\tau$ 
is identified with 
the RG parameter of the $d$-dimensional boundary field theory, 
and the evolution of bulk fields in $\tau$ 
is identified with the RG flow of the coupling constants 
of the boundary theory. 
In the following discussion, we exclusively consider scalar fields 
as such bulk fields.

The Einstein-Hilbert action with bulk scalars $\hphi^a(x,\tau)$ 
on a $(d+1)$-dimensional manifold $M_{d+1}$ 
with boundary $\Sigma_d=\partial M_{d+1}$ at $\tau=\tau_0$ is given by 
\begin{eqnarray}
 &&\bS\bigl[\widehat{g}_{\mu\nu}(x,\tau),\hphi^a(x,\tau)\bigr] \nn
 &&~~~~~=\int_{M_{d+1}} d^{d+1} X \sqrt{\widehat{g}} \left( V( \hphi ) 
  -\widehat{R}+{1 \over 2}\,L_{ab}( \hphi )\, 
  \widehat{g}^{\mu\nu}\,\partial_\mu \hphi^a\,\partial_\nu \hphi^b \right)
  -2\int_{\Sigma_d} d^d x \,\sqrt{g}\,K\,,\nn
 && \label{bulk}
\end{eqnarray}
which is expressed in the ADM parametrization as 
\begin{eqnarray}
&&\bS\bigl[\widehat{g}_{ij}(x,\tau),\hphi^a(x,\tau),\Nh(x,\tau),
 \lambdah^i(x,\tau)\bigr]
\nn
 &&~~~=\int d^dx\,\int_{\tau_0}^\infty
  d\tau \,\sqrt{\widehat{g}}\, \Bigl[\, \Nh 
  \left( V(\hphi)-\widehat{R}+\hK_{ij}\hK^{ij}-\hK^2 \right) \nonumber \\
 &&~~~~~  
   +\,{1 \over 2\Nh}\,L_{ab}(\hphi) \left( 
  \bigl(\dot{\hphi}{}^a -\lambdah^{i}\partial_{i}\hphi^a\bigr) 
  \bigl(\dot{\hphi}{}^b -\lambdah^{i}\partial_{i}\hphi^b\bigr) 
  +\Nh^2\,\gh^{ij}\,\partial_{i}\hphi^a\,\partial_{j}\hphi^b\right) 
   \,\Bigr] \nn
 &&~~~\equiv 
  \int d^d x \,\int_{\tau_0}^\infty d\tau \,\sqrt{\gh} 
  \,\cL_{d+1}\bigl[\gh,\hphi,\Nh,\lambdah\bigr],
\label{eh;off-shell}
\end{eqnarray}
where $ \cdot=\partial/\partial \tau$. 
Here $\widehat{R}$ and $\widehat{\nabla}_{i}$ are the scalar curvature 
and the covariant derivative with respect to $\gh_{ij}$, respectively. 
$\hK_{ij}$ is the extrinsic curvature of each time-slice 
parametrized by $\tau$, 
\begin{equation}
 \hK_{ij}={1 \over 2\Nh}
  \left(\dot{\gh}_{ij}-\widehat{\nabla}_{i}\lambdah_{j}
  -\widehat{\nabla}_{j}\lambdah_{i}\right),
\end{equation}
and $\widehat{K}$ is its trace, $\hK=\gh^{ij}\,\hK_{ij}$.
The boundary term in Eq.\ (\ref{bulk}) needs to be introduced 
to ensure that the Dirichlet 
boundary conditions can be imposed on the system consistently \cite{GH}. 
In fact, the second derivative of $\gh_{ij}$ in $\tau$ 
appears in the first term of Eq.\ (\ref{bulk}), 
but can be written as a total derivative 
and canceled with the boundary term.

As far as classical solutions are concerned, the action \eq{eh;off-shell}
is equivalent to the following one in the first-order form: 
\begin{eqnarray}
 \bS\bigl[\gh_{ij},\hphi^a,\pih^{ij},\pih_a, \Nh,\lambdah^i\bigr]
 \equiv \int d^d x \,d\tau \,\sqrt{\gh} 
  \left[\,\pih^{ij}\dot{\gh}_{ij}+\pih_a\dot{\hphi}{}^a
  +\Nh\widehat{\cH}+\lambdah_{i}\widehat{\cP}^{i}\,\right], 
 \label{first}
\end{eqnarray}
with 
\ba
 \widehat{\cH} &=&\cH\bigl(\gh_{ij},\phih^a,\pih^{ij},\pih_a\bigr)\nn
 &\equiv& {1 \over d-1}\left(\pih_{i}^{i}\right)^2-\pih_{ij}^2
 -{1 \over 2}\,L^{ab}(\hphi)\,\pih_a\,\pih_b+V(\hphi)-\widehat{R}
 +{1 \over 2}\,L_{ab}(\hphi)\,\gh^{ij}\,
  \partial_{i}\hphi^a\,\partial_{b}\hphi^j, 
  \nn
 \widehat{\cP}^{i}&=&\cP^i\bigl(\gh_{ij},\phih^a,\pih^{ij},\pih_a\bigr)\nn
 &\equiv& 2\,\widehat{\nabla}_{j}\pih^{ij}-\pih_a\,\widehat{\nabla}^{i}\hphi^a. 
\ea
In fact, the equations of motion for $\pih^{ij}$ and 
$\pih_a$ give the relations 
\begin{equation}
 \pih^{ij}
  =\hK^{ij}-\gh^{ij}\hK, 
  \qquad \pih_a
  ={1 \over \Nh} L_{ab}(\hphi) \left( \dot{\hphi}\,{}^b -
  \lambdah^{i}\,\partial_{i}\hphi\,{}^b \right), 
\label{Pi;gphi}
\end{equation}
and by substituting this expression into Eq.\ \eq{first}, 
\eq{eh;off-shell} is obtained. 
Here $\Nh$ and $\lambdah^i$ simply behave as Lagrange multipliers, 
and thus we have the Hamiltonian and momentum constraints: 
\begin{eqnarray}
 {1 \over \sqrt{\gh}}\,{\delta \bS \over \delta \Nh}
  &=&\widehat{\cH}~=~0,
 \label{h;constraint}\\
 {1 \over \sqrt{\gh}}\,{\delta \bS \over \delta \lambdah_{i}}
  &=&\widehat{\cP}^{i}~=~0.
 \label{m;constraint}
\end{eqnarray}
Note that these constraints generate reparametrizations
of the form $\tau \rightarrow \tau+\delta \tau(x),\,
x^i \rightarrow x^i+\delta x^i(x)$ 
for each time-slice ($\tau={\rm const}$).
One can easily show that they are of the first class 
under the canonical Poisson brackets for $g_{ij}(x),\pi^{ij}(x),
\phi^a(x)$ and $\pi_a(x)$. 
Thus, up to gauge equivalent configurations generated by 
$\cH(x)$ and $\cP^i(x)$, 
the $\tau$-evolution of the bulk fields 
is uniquely determined, being independent of the values 
of the Lagrange multipliers $N$ and $\lambda^i$, 
at the initial time-slice. 


Let $\gb_{ij}(x,\tau)$ and $\phib^a(x,\tau)$ 
be the classical solutions of the bulk action 
with the boundary conditions\footnote{
One generally needs two boundary conditions for each field, 
since the equations of motion are second-order differential equations 
in $\tau$. 
Here, one of the two is assumed to be already fixed 
by demanding the regular behavior of the classical solutions 
inside $M_{d+1}$ ($\tau\rightarrow +\infty$) \cite{M,GKP,W;holography} 
(see also Ref.\ \cite{GL}).
}
\begin{equation}
 \gb_{ij}(x,\tau\!=\!\tau_0)=g_{ij}(x),\qquad
 \phib^a(x,\tau\!=\!\tau_0)=\phi^a(x).
 \end{equation}
We also define $\pib^{ij}(x,\tau)$ and $\pib_a(x,\tau)$ to be the classical 
solutions of $\pih^{ij}(x,\tau)$ and $\pih_a(x,\tau)$, respectively. 
We then substitute these classical solutions into the bulk action 
to obtain the classical action which is a functional 
of the boundary values, $g_{ij}(x)$ and $\phi^a(x)$: 
\begin{align}
 S\bigl[g_{ij}(x),\phi(x);\tau_0\bigr] &\equiv
  \bS\left[\gb_{ij}(x,\tau),\,\phib^a(x,\tau),
  \,\pib^{ij}(x,\tau),\,\pib_a(x,\tau),\,
  \overline{N}(x,\tau),\,\overline{\lambda}^i(x,\tau)\right] \nn
 &= \int d^d x \int_{\tau_0}d\tau \,\sqrt{\gb}\,\,
 \left[\,\pib^{ij}\,\dot{\gb}_{ij}\,+\,\pib_a\,\dot{\phib}{}^a
 \,\right].
\end{align}
Here we have used the Hamiltonian and momentum constraints, 
$\overline{\cH}=\overline{\cP}_i=0$. 
One can see that the variation of the action (\ref{eh;off-shell}) is 
given by
\begin{eqnarray}
 \delta S\bigl[g(x),\phi(x);\tau_0\bigr]
 &=&-\,\int d^d x\,\sqrt{g}\,\Biggl[\,\left( 
  \pib^{ij}(x,\tau_0)\,\dot{\gb}_{ij}(x,\tau_0)
  +\pib_a(x,\tau_0)\,\dot{\phib}{}^a(x,\tau_0)\right)
 \delta \tau_0 \nn
 &&~~+\,\pib^{ij}(x,\tau_0)\,\delta\gb_{ij}(x,\tau_0)
     \,+\,\pib_a(x,\tau_0)\,\delta\phib^a(x,\tau_0)\,\Biggr] \nn
 &=&-\,\int d^d x \sqrt{g}\left[\, 
  \pib^{ij}(x,\tau_0)\,\delta g_{ij}(x)
  +\pib_a(x,\tau_0)\,\delta\phi^a(x)
 \,\right],
 \label{deviation}
\end{eqnarray}
since $\delta\gb_{ij}(x,\tau_0)=\delta g_{ij}(x)
-\dot{\gb}_{ij}(x,\tau_0)\,\delta \tau_0$, {\em etc}.
It thus follows that the classical conjugate momenta evaluated 
at $\tau=\tau_0$ are given by 
\begin{equation}
 \pi^{ij}(x)\equiv\pib^{ij}(x,\tau_0)
  ={-1 \over \sqrt{g}}\,{\delta S \over \delta g_{ij}(x)},
 \qquad
 \pi_a(x)\equiv\pib_a(x,\tau_0)
  ={-1 \over \sqrt{g}}\,{\delta S \over \delta \phi^a(x)}. 
\label{momentum;hj}
\end{equation}
Since $\delta\tau_0$ disappears on the right-hand side of 
\eq{deviation}, we find that 
\begin{equation}
 {\partial \over \partial \tau_0}S\bigl[g_{ij}(x),\phi^a(x);\tau_0\bigr]=0,
\end{equation}
that is, the classical action $S$ is independent of the coordinate 
value of the boundary, $\tau_0$. 
Thus, the classical action $S=S\bigl[g(x),\phi(x)\bigr]$ 
is specified only by the constraint equations 
\ba
 \cH\bigl(g_{ij}(x),\phi^a(x),\pi^{ij}(x),\pi_a(x)\bigr)=0,\quad
  \cP^i\bigl(g_{ij}(x),\phi^a(x),\pi^{ij}(x),\pi_a(x)\bigr)=0, 
\ea
with $\pi^{ij}(x)$ and $\pi_a(x)$ given by \eq{momentum;hj}. 
{}From the first equation (the Hamiltonian constraint), 
we obtain the flow equation 
of de Boer, Verlinde and Verlinde \cite{dVV}, 
\begin{equation}
 \bigl\{S,S \bigr\}(x)=\cL_d(x),
\label{hj}
\end{equation}
with
\ba
 \bigl\{S,S \bigr\}(x)\equiv
 \left({1 \over \sqrt{g}}\right)^2\left[ \,
 -\,{1 \over d-1}
  \left( g_{ij}{\delta S \over \delta g_{ij}}\right)^2
 +\left( {\delta S \over \delta g_{ij}}\right)^2
 +{1 \over 2}L^{ab}(\phi)\,{\delta S \over \delta \phi^a}\,
  {\delta S \over \delta \phi^b}
 \,\right],  
\ea
and
\ba
 \cL_d(x)\equiv V(\phi)-R+{1 \over 2}\,L_{ab}(\phi)\,g^{ij}\,
  \partial_{i}\phi^a\,\partial_{j}\phi^b.
\ea
The second equation (the momentum constraint) 
ensures the invariance of $S$ under $d$-dimensional diffeomorphisms 
along the fixed time-slice $\tau=\tau_0$: 
\ba
 \int d^d x\left(\delta_\epsilon g_{ij}
   \,\frac{\delta S}{\delta g_{ij}}
  +\delta_\epsilon\phi^a\,\frac{\delta S}{\delta \phi^a}\right)
  =\int d^d x\, \left[\left(\nabla_i\epsilon_j+\nabla_j\epsilon_i
  \right)\frac{\delta S}{\delta g_{ij}}+\epsilon^i\,\partial_i
  \phi^a\,\frac{\delta S}{\delta\phi^a}\right]=0,  
\ea
with $\epsilon^i(x)$ an arbitrary function.

%
\subsection{Solution to the flow equation}

In this subsection, we discuss a systematic prescription for solving 
the flow equation \eq{hj}. 

As was discussed in \S 2.1, 
when the boundary is shifted to $\tau=\tau_0$ 
from the original boundary $\tau\!=\!-\infty$ (or $z\!=\!0$) of AdS space, 
the conformal symmetry disappears at the new boundary, 
and thus the boundary field theory should be regarded as a cut-off theory. 
The limit $\tau_0\!\rightarrow\!-\infty$ yields an IR divergence 
because of the infinite volume of the bulk geometry, 
and thus we need to subtract this divergence from the classical action. 
However, as was discussed in \S 2.1, 
this divergence can also be regarded as coming from the short distance 
singularity for the boundary field theory (IR/UV relation). 
{}Since we are also taking into account the back reaction 
from matter fields to gravity, 
the required counter-term should be a local functional 
of $d$-dimensional fields, $g_{ij}(x)$ and $\phi^a(x)$.  
This consideration leads us to decompose the classical action 
into the following form: 
\begin{align}
 \newton S\bigl[g(x),\phi(x)\bigr]
  =\newton S_{\rm
 loc}\bigl[g(x),\phi(x)\bigr]-\Gamma\bigl[g(x),\phi(x)\bigr]. 
\label{eh;on-shell}
\end{align}
Here $S_{\rm loc}\bigl[g(x),\phi(x)\bigr]$ is the local counter-term, 
and $\Gamma\bigl[g(x),\phi(x)\bigr]$ is now regarded 
as the generating functional with respect to the source fields 
$\phi^a(x)$ that live in a curved background with metric $g_{ij}(x)$.

We make a derivative expansion of the local counter-term 
in the following way: 
\begin{eqnarray}
  S_{\rm loc}[g(x),\phi (x)]
  =\int d^d x \,\sqrt{g(x)}\,\cLloc(x) 
  =\int d^d x\,
  \sqrt{g(x)}\sum_{w=0,2,4,\cdots}\bigl[\cLloc(x)\bigr]_w. 
\end{eqnarray}
The order of derivatives is counted with respect to 
the weight $w$ \cite{FMS1} that is defined additively 
from the following rule\footnote{
A scaling argument of this kind is often used in supersymmetric theories
to restrict the form of low energy effective actions 
(see e.g.\ Ref.\ \cite{GSW}).
}:
\begin{center}
\begin{tabular}{c|c}
       & weight \\ \hline
$g_{ij}(x), \,\phi^a(x), \,\Gamma[g,\phi]$ & $0$ \\ \hline
$\partial_{i}$ & $1$ \\ \hline
$R, \,R_{ij},\, \partial_i\phi^a\partial_j\phi^b,\,
\cdots$ & $2$ \\ \hline
$\delta \Gamma / \delta g_{ij}(x),\, 
\delta \Gamma / \delta \phi^a(x)$ & $d$ 
\end{tabular}
\end{center}
The separation of a local counter term $S_{\rm loc}$ 
from the generating functional $\Gamma$ is usually ambiguous 
for higher weight, 
and we here assign the vanishing weight to $\Gamma$ 
since this greatly simplifies the analysis of $\Gamma$ \cite{FMS1}. 
The last line of the table is a natural consequence of this assignment, 
since $\delta\Gamma=\int d^d x \bigl(\delta\phi(x)\times
\delta\Gamma/\delta\phi(x)+\cdots\bigr)$ 
and $d^d x$ gives the weight $w=-d$.
Then, substituting the above equation into the flow equation (\ref{hj}) 
and comparing terms of the same weight, 
we obtain a sequence of equations that relate the bulk action 
(\ref{eh;off-shell}) to the classical action (\ref{eh;on-shell}). 
They take the following form \cite{FMS1}:
\ba
 \cL_d&=&\Bigl[\bigl\{\Sloc,\,\Sloc\bigr\}\Bigr]_0
  +\Bigl[\bigl\{\Sloc,\,\Sloc\bigr\}\Bigr]_2\label{eqA}\,,\\
 0&=&\Bigl[\bigl\{\Sloc,\,\Sloc\bigr\}\Bigr]_w
  \quad\quad(w=4,6,\cdots,d-2), \label{eqB} \\
 0&=&2\Bigl[\bigl\{\Sloc,\,\Gamma\bigr\}\Bigr]_d
  -\newton\Bigl[\bigl\{\Sloc,\,\Sloc\bigr\}\Bigr]_d \label{eqC}\,,\\
 0&=&2\Bigl[\bigl\{\Sloc,\,\Gamma\bigr\}\Bigr]_w
  -\newton\Bigl[\bigl\{\Sloc,\,\Sloc\bigr\}\Bigr]_w 
  \quad(w=d+2,\cdots,2d-2)\label{eqD}, \\
 0&=&\Bigl[\bigl\{\Gamma,\,\Gamma\bigr\}\Bigr]_{2d}
  -\frac{2}{2\kappa_{d+1}^2}\Bigl[\bigl\{\Sloc,\,\Gamma\bigr\}\Bigr]_{2d}
  +\frac{1}{(2\kappa_{d+1})^2}
  \Bigl[\bigl\{\Sloc,\,\Sloc\bigr\}\Bigr]_{2d}\,,\\
 0&=&2\Bigl[\bigl\{\Sloc,\,\Gamma\bigr\}\Bigr]_w
  -\newton\Bigl[\bigl\{\Sloc,\,\Sloc\bigr\}\Bigr]_w 
  \quad(w=2d+2,\cdots). \label{eqE}\,
\ea
Equations \ \eq{eqA} and \eq{eqB} determine 
$\left[\cLloc\right]_w~(w=0,2,\cdots,d-2)$, 
which together with Eq.\ \eq{eqC} in turn determine the non-local 
functional $\Gamma$. 
Although $\left[\cLloc\right]_d$ enters the expression, 
we will see later that this does not give any physically relevant
effect.

By parametrizing $[\cLloc]_0$ and $[\cLloc]_2$ as 
\begin{eqnarray}
  \left[\cLloc\right]_0&=& W(\phi), \\
  \left[\cLloc\right]_2&=& -\Phi(\phi)\,R+{1 \over 2}\,M_{ab}(\phi)\,
   g^{ij}\,\partial_{i}\phi^a\,\partial_{j}\phi^b,
\end{eqnarray}
one can easily solve \eq{eqA} to obtain \cite{FMS1}\footnote{
The expression for $d=4$ can be found in Ref.\ \cite{dVV}.
}
\begin{eqnarray}
 V(\phi)&=&-{d \over 4(d-1)}\,W(\phi)^2
  +{1 \over 2}L^{ab}(\phi)\,\partial_a W(\phi)\,\partial_b W(\phi)\,, 
  \label{hj;potential} \\
 -1&=&{d-2 \over 2(d-1)}\,W(\phi)\,\Phi(\phi)
  -L^{ab}(\phi)\,\partial_a W(\phi)\,\partial_b \Phi(\phi)\,, 
  \label{hj;kineG} \\
 {1 \over 2}\,L_{ab}(\phi)&=&-{d-2 \over 4(d-1)}\,W(\phi)\,M_{ab}(\phi)
  -L^{cd}(\phi)\,\partial_c W(\phi)\,\Gamma^{(M)}_{d;ab}(\phi)\,,
 \label{hj;kinesca}\\
 0&=&W(\phi)\,\nabla^2\,\Phi(\phi)+L^{ab}(\phi)\,\partial_a W(\phi)\,
  M_{bc}(\phi)\,\nabla^2\phi^c\,.
\end{eqnarray}
Here $\partial_a=\partial/\partial\phi^a$, and 
$\Gamma^{(M)c}_{ab}(\phi)\equiv 
M^{cd}(\phi)\,\Gamma^{(M)}_{d;ab}(\phi)$ is the Christoffel symbol 
constructed from $M_{ab}(\phi)$. 
For pure gravity ($L_{ab}=0, M_{ab}=0$), for example,  
setting $V=2\Lambda=-d(d-1)/l^2$, 
we find\footnote{
The sign of $W$ is chosen to be in the branch 
where the limit $\phi\rightarrow 0$ can be taken smoothly
with $L_{ab}(\phi)$ and $M_{ab}(\phi)$ positive definite.
} 
\ba
  W=-\,\frac{2\,(d-1)}{l},\quad \Phi=\frac{l}{d-2}.
\ea 
Here $\Lambda$ is the bulk cosmological constant, 
and when the metric is asymptotically AdS, 
the parameter $l$ is identified with the radius of 
the asymptotic AdS${}_{d+1}$.

When $d\geq4$, we need to solve Eq.\ \eq{eqB}. 
{}For the pure gravity case, for example, 
by parametrizing the local term of weight 4 as 
\begin{eqnarray}
 [\cLloc]_4=XR^2+YR_{ij}R^{ij}
 +ZR_{ijkl}R^{ijkl},
\label{l4}
\end{eqnarray}
Eq.\ \eq{eqB} with $w=4$ is expressed as 
\begin{eqnarray}
 0&\equiv&\Bigl[
  \bigl\{ S_{\rm loc}, S_{\rm loc} \bigr\}\Bigr]_4 \nn
  &=&-{W \over 2(d-1)} 
   \left( (d-4)X-{d\,l^3 \over 4(d-1)(d-2)^2}\right) R^2\nn
  &&~~~~~ -{W \over 2(d-1)} \left( (d-4)Y+{l^3 \over (d-2)^2}\right)
   R_{ij}R^{ij}~-{ d-4 \over 2(d-1)}\,W Z\,
   R_{ijkl}R^{ijkl}\ \nn
  &&~~~~~ +\left( 2X+{d \over 2(d-1)}Y+{2 \over d-1}Z\right)\nabla^2R,
\label{slsl;w4}
\end{eqnarray}
from which we find 
\begin{equation} 
 X={d\,l^3 \over 4(d-1)(d-2)^2(d-4)},\quad 
 Y=-{l^3 \over (d-2)^2(d-4)},\quad
 Z=0, \label{d=6}
\end{equation}
and $\Bigl[\{\Sloc,\Sloc\}\Bigr]_6$ can be calculated easily to be
\begin{eqnarray}
 &&\Bigl[\bigl\{ S_{\rm loc}, S_{\rm loc} \bigr\}\Bigr]_6 \nn
 &&~~=
  \Phi\left[\,\left(-4X+\frac{d+2}{2(d-1)}Y\right)
    R\, R_{ij} \,R^{ij}
  +{d+2\over 2(d-1)}XR^3
  -4\,YR^{ik} R^{jl} R_{ijkl}\right.\nn
 &&~~~~\left.+(4X+2Y)R^{ij}\nabla_{i}\nabla_{j}R
  -2YR^{ij}\nabla^2R_{ij}
  +\left(2(d-3)X+\frac{d-2}{2}Y\right)R\,\nabla^2R
 \,\right]\nn
 &&~~~~+({\rm contributions~from}~[\cLloc]_6)\nn
 &&~~=
  l^4\left[\,-\,\frac{3d+2}{2(d-1)(d-2)^3(d-4)}
   \,R\,R_{ij}\,R^{ij}
  +{d(d+2)\over 8(d-1)^2(d-2)^3(d-4)}\,R^3 \right.\nn
 &&~~~~+\,{4\over (d-2)^3(d-4)}\,R^{ik}\,R^{jl}\,
    R_{ijkl}
  -\,{1\over (d-1)(d-2)^2(d-4)}\,
   R^{ij}\,\nabla_{i}\nabla_{j}R\nn
 &&~~~~\left.  +\,{2\over (d-2)^3(d-4)}\,R^{ij}\,\nabla^2R_{ij}
  -\,{1\over (d-1)(d-2)^3(d-4)}\,R\,\nabla^2R
 \,\right]\nn
 &&~~~~+\,({\rm contributions~from}~[\cLloc]_6).
\label{slsl;w6}
\end{eqnarray}

On the other hand, from the flow equation of weight $d$, \eq{eqC}, 
we find
\ba
  \frac{2}{\sqrt{g}}\,g_{ij} {\delta\Gamma \over \delta g_{ij}}
  -\beta^a(\phi)\frac{1}{\sqrt{g}}\,{\delta \Gamma \over \delta \phi^a} 
 =-\newton\,{2(d-1)\over W(\phi)}\,
  \Bigl[\{ S_{\rm loc},S_{\rm loc} \}\Bigr]_d,
\label{cs;functional}
\ea
with
\begin{equation}
 \beta^a(\phi)\equiv 
  {2(d-1) \over W(\phi)}L^{ab}(\phi)\,\partial_b W(\phi). 
\label{cs;beta}
\end{equation}

It is crucial that $\beta^a$ can be identified with the RG beta
function.
To see this, we recall that an RG flow in the boundary field theory
is described by a classical solution in the bulk. 
Here we consider the classical solutions 
$\gb_{ij}(x,\tau)$ and $\phib^a(x,\tau)$ with 
the boundary conditions 
\begin{equation}
 \gb_{ij}(x,\tau_0)=g_{ij}(x)\equiv{1\over a^2}\,\delta_{ij},
 \qquad
 \phib^a(x,\tau_0)=\phi^a(x)\equiv\phi^a. 
 \quad\bigl(a,\phi:~{\rm const.}\bigr)
\end{equation}
This represents the most generic background 
that preserves the $d$-dimensional Poincar\'{e} (or Euclidean) symmetry. 
Since we set the fields to constant values, the system 
is now perturbed finitely. 
{}Furthermore, since $a$ gives the unit length of 
the $d$-dimensional space, 
this perturbation should describe the system 
with the cutoff length $a$, which corresponds to the time 
$\tau=\tau_0$ in the RG flow.
From Eq.\ (\ref{Pi;gphi}) and the Hamilton-Jacobi equation 
(\ref{momentum;hj}), we obtain 
\begin{eqnarray}
 {d\over d\tau}\,\gb_{ij}(x,\tau)\Big|_{\tau=\tau_0}
  &=&{1\over d-1}\,W(\phi)\,{1\over a^2}\,\delta_{ij},
  \label{barG} \\
 {d\over d\tau}\,\phib^a(x,\tau)\Big|_{\tau=\tau_0}
  &=&-L^{ab}(\phi)\,\partial_bW(\phi). 
\label{barphi}
\end{eqnarray}
We then assume that the classical solutions take the following form 
for general $\tau$: 
\begin{equation}
 \gb_{ij}(x,\tau)={1\over a(\tau)^2}\delta_{ij}, \qquad
 \phib^a(x,\tau)=\phi^a(a(\tau)),
\end{equation}
with $a(\tau_0)=a$. Note that $a(\tau)$ can be identified with the cutoff 
length at $\tau$. It then follows from (\ref{barG}) and (\ref{barphi}) that 
\begin{eqnarray}
 a\,{ d\tau \over da}&=&-\,{2(d-1)\over W(\phi)}, \\
 a\,{d \over da}\,\phi^a(a)&=&{2(d-1)\over W(\phi)}\,
 L^{ab}(\phi)\,\partial_bW(\phi). 
\end{eqnarray}
Comparing the latter with Eq.\ \eq{cs;beta}, we thus conclude that 
the $\beta^a(\phi)$'s in \eq{cs;functional} 
are actually the beta functions of the holographic RG;\footnote{%
Note that $a$ increases under our RG flow which moves to IR. 
So our definition of $\beta^a$ has the opposite sign 
to the usual one. 
}
\ba
 \beta^a(\phi)=a\frac{d}{da}\phi^a(a). 
\ea

Eq.\ (\ref{cs;functional}) is one of the key ingredients
in the study of the holographic RG. 
In fact, we will show that this yields the Weyl anomalies 
and the Callan-Symanzik equation in the dual field theory.


\subsection{Holographic Weyl anomaly}

We first notice 
that $(2/\sqrt{g})\,\delta\Gamma/\delta g_{ij}(x)$ 
gives the vacuum expectation value of the energy momentum tensor 
in the background $g_{ij}(x)$ and $\phi^a(x)$; 
\ba
 \frac{2}{\sqrt{g}}\,\frac{\delta\Gamma\bigl[g,\phi\bigr]}
  {\delta g_{ij}(x)}
  =\bigl\langle T^{ij}(x) \bigr\rangle_{g,\phi}. 
\ea
Thus, if we choose the couplings $\phi^a$ such that their beta functions 
vanish, Eq.\ \eq{cs;functional} shows that 
its right-hand side gives the Weyl anomaly: 
\ba
 \cW_d(x)\equiv\Bigl\langle T^i_{~i}(x)\Bigr\rangle\Big|_{\beta(\phi)=0}
  =-\,\newton\,{2(d-1)\over W(\phi)}\,
  \Bigl[\{ S_{\rm loc},S_{\rm loc} \}\Bigr]_d\Bigg|_{\beta(\phi)=0}. 
\ea

Before turning to a computation of the holographic Weyl anomaly,
we here would like to clarify the relation between the uniqueness of Weyl
anomalies and an ambiguity of the solution of the flow equation,
that was argued in Ref.\ \cite{FS}.

Generically, the Weyl anomaly has the form 
\ba
 {\cal W}_d&=&-\,\newton\,{2(d-1)\over W(\phi)}\,\left(
  \Bigl[\bigl\{ S_{\rm loc},S_{\rm loc} \bigr\}^\prime\Bigr]_d
  +2\,\bigl\{ S_{{\rm loc};\,-d},S_{{\rm loc};\,0}\bigr\}\right)
  \Big|_{\beta(\phi)=0}, 
 \label{ambig}
\ea
where $\{S_{\rm loc},S_{\rm loc}\}^\prime$ 
is the part of $\{S_{\rm loc},S_{\rm loc}\}$ 
which does not include contributions from $[\cL_{\rm loc}]_d$, 
and we have introduced\footnote{%
The weight shifts by $-d$ after the integration 
because the weight of $d^d x$ is $-d$. 
}
\begin{equation}
S_{{\rm loc};\,\,w-d}\equiv\int d^dx\,\sqrt{g(x)}\,[{\cal L}_{\rm loc}]_w. 
\end{equation}
The first term on the right-hand side of \eq{ambig} is written 
only with $\bigl[\cLloc\bigr]_0,\cdots,\bigl[\cLloc\bigr]_{d-2}$, 
all of which can be determined by the flow equation.
On the other hand, the second term contains 
$\bigl[\cLloc\bigr]_d$ that cannot not be determined 
by the flow equation. 
However, this can be absorbed into the effective action $\Gamma$.
In fact, by using the relations 
\begin{equation}
 {\delta S_{{\rm loc};\,-d}\over\delta g_{ij}}=
 {\sqrt{g}\over 2}\,W(\phi)\,g^{ij},\qquad
 {\delta S_{{\rm loc};\,-d}\over\delta \phi^a}=
 \sqrt{g}\,\,\partial_a W(\phi), 
\end{equation}
one finds that 
\ba
 2\,{2(d-1)\over W(\phi)}\,
  \bigl\{ S_{{\rm loc};\,-d},S_{{\rm loc};\,0}\bigr\}
  =-\frac{2}{\sqrt{g}}\,g_{ij}{\delta S_{{\rm loc};\,0}\over \delta g_{ij}}
   +\beta^a(\phi)\,\frac{1}{\sqrt{g}}\,
   {\delta S_{{\rm loc};\,0}\over \delta\phi^a}, 
 \label{slocd}
\ea
and can rewrite the flow equation (\ref{cs;functional}) as 
\begin{align}
& \frac{2}{\sqrt{g}}\, g_{ij}{\delta\over\delta g_{ij}}\left(
  \Gamma-\newton\,S_{{\rm loc};\,0}\right)
  -\beta^a(\phi)\,\frac{1}{\sqrt{g}}\,{\delta\over\delta \phi^a}\left(
  \Gamma-\newton\,S_{{\rm loc};\,0}\right) \nn
& =-\,\newton\,{2(d-1)\over W(\phi)}\,
  \Bigl[\bigl\{ S_{\rm loc},S_{\rm loc} \bigr\}^\prime\Bigr]_d. 
\end{align}
Thus, we have seen that the contribution from the term 
$\bigl[\cLloc\bigr]_d$ can be absorbed into $\Gamma$ 
by redefining it as 
$\Gamma'=\Gamma-(1/2\kappa_{d+\!1}^2)\,S_{{\rm loc};\,0}$.
Note that $\Gamma'$ still has vanishing weight.

Instead of redefining $\Gamma$, one can modify the Weyl anomaly 
without making any essential change. 
To show this, we first notice that the second term in eq.\ 
\eq{slocd} can be written as a total derivative: 
\begin{equation}
 2\,g_{ij}\,{\delta S_{{\rm loc};\,0}\over\delta g_{ij}}
  =-\,\sqrt{g}\,\nabla_{i}{\cal J}^{i}_d 
 \label{totalder}
\end{equation}
with ${\cal J}_d^{i}$ some local current. 
In fact, for infinitesimal Weyl transformations 
($\sigma(x)\ll 1$: arbitrary function), 
we have  
\begin{equation}
 S_{{\rm loc};\,0}[e^{\sigma (x)}g(x),\phi(x)]
  -S_{{\rm loc};\,0}[g(x),\phi(x)]
  =\int d^dx\,\sigma (x)\,
  g_{ij}\,{\delta S_{{\rm loc};0}\over\delta g_{ij}}.
 \label{dilatation}
\end{equation}
One can easily understand that 
$S_{{\rm loc};\,0}[g(x),\phi(x)]$ is invariant 
under {\em constant} Weyl transformations 
($g_{ij}(x)\!\rightarrow\!e^\sigma g_{ij}(x)$,
$\phi^a(x)\!\rightarrow\!\phi^a(x)$ with  $\sigma$ constant), 
so that the left-hand side of eq.\ \eq{dilatation} 
can generally be written as
\ba
  \int d^d x \,\partial_i\sigma(x)\,\sqrt{g}\,\cJ_d^i
\ea
with some local function $\cJ_d^i$. 
By integrating this by parts and comparing the result with the 
right-hand side of eq.\ \eq{dilatation}, 
one obtains eq.\ \eq{totalder}. 
Thus we have shown that eq.\ \eq{cs;functional} can be rewritten 
into the following form: 
\ba
&& \frac{2}{\sqrt{g}}\,g_{ij}\,{\delta\Gamma\over\delta g_{ij}}
  -\beta^a(\phi)\frac{1}{\sqrt{g}}\,{\delta\Gamma\over\delta\phi^a}\nn
&&~~~~=-\,\newton\,{2(d-1)\over W(\phi)}\,
  \Bigl[\bigl\{ S_{\rm loc},S_{\rm loc} \bigr\}^\prime\Bigr]_d
   -\nabla_{i}{\cal J}^{i}_d 
   +\beta^a(\phi)\frac{1}{\sqrt{g}}\,
  {\delta S_{{\rm loc};\,0}\over \delta\phi^a}. 
\ea
This implies that when we take $\Gamma$ as the generating functional, 
the Weyl anomaly $\cW_d$ has an ambiguity 
which can be always made into a total derivative term 
(since we set $\beta^a(\phi)=0$).

Now that the flow equation is found to provide us with a unique
form of Weyl anomalies, 
we will consider two simple examples
to illustrate how the above prescription works.

\noindent\underline{\bf 5D dilatonic gravity} \cite{FMS1}:

We normalize the Lagrangian with a single scalar field
as follows:
\ba
  \cL_4=-\frac{12}{l^2}-R+\frac{1}{2}
  \,g^{ij}\,\partial_i\phi\,\partial_j\phi.
\ea
Then, assuming that all the functions $W(\phi), M(\phi)$ and 
$\Phi(\phi)$ are constant in $\phi$,
we can solve Eqs.\ \eq{hj;potential}--\eq{hj;kinesca} with  
$V=-d(d-1)/l^2=-12/l^2$ and $L=1$, 
and obtain 
\ba
  W=-\frac{6}{l},\quad \Phi=\frac{l}{2},\quad M=\frac{l}{2};
\ea
that is, 
\begin{equation}
 S_{{\rm loc}}[g,\phi]=\int d^4x\sqrt{g}\left( -\frac{6}{l}-{l\over 2}R
 +{l\over 2}g^{ij}\partial_{i}\phi\,\partial_{j}\phi
\right).
\label{sloc4;dvv}
\end{equation}
We can calculate $\Bigl[\{\Sloc,\Sloc\}\Bigr]_4$ easily and find 
\begin{eqnarray}
 \cW_4&=&\frac{l}{2\kappa_5^2}\,\,\Bigl[\{\Sloc,\Sloc\}\Bigr]_4\nn
 &=&\frac{l^3}{2\kappa_5^2}\left(-{1\over 12}R^2+{1\over 4}R_{ij}R^{ij}
  +{1\over 12}R\,g^{ij}\,\partial_{i}\phi\,
  \partial_{j}\phi\right. \nn
 &&\left.-{1\over 4}\,R^{ij}\,\partial_{i}\phi\,\partial_{j}\phi 
  +{1\over 24}\left(g^{ij}\,\partial_{i}\phi\,
   \partial_{j}\phi\right)^2
  +{1\over 8}\left(\nabla^2\phi\right)^2\right) .
 \label{weyl4}
\end{eqnarray}
This is in exact agreement with the result in Ref.\ \cite{dilaton}.

In the duality between IIB supergravity on AdS$_5\times S^5$ 
and the large $N~SU(N)$ SYM$_4$, 
the radii of AdS$_5$ and $S^5$ both have 
$l=(4\pi g_s N)^{1/4}\,l_s$. 
This gives the five-dimensional Newton constant 
\ba
 \frac{1}{2\kappa_5^2}=\frac{{\rm Vol}(S^5)}{2\kappa_{10}^2}\,
  =\frac{\pi^3\,l^5}{128\,\pi^7 g_s^2}. 
\ea
Thus, by setting $\phi=0$, we obtain
\ba
 {\cal W}_4&=&\frac{l^8}{128\,\pi^4 g_s^2}\,
  \left(-\,{1\over 12}\,R^2+{1\over 4}\,R_{ij}R^{ij}\right)\nn
 &=&\frac{N^2}{2\,(4\pi)^2}\left(-\,{1\over 3}\,R^2+R_{ij}R^{ij}\right), 
 \label{result;weyl}
\ea
which exactly gives the large $N$ limit of the Weyl anomaly 
of the the large $N$ $SU(N)$ SYM$_4$ \cite{Duff;Weyl}.\footnote{%
The Weyl anomaly of four-dimensional field theories 
is perturbatively calculated \cite{Duff;Weyl} as 
\ba
 {\cal W}_4=\frac{c}{(4\pi)^2}\,\Bigl(\frac{1}{3}R^2-2R_{ij}^2
  +R_{ijkl}^2\Bigr)
  -\frac{a}{(4\pi)^2}\bigl(R^2-4R_{ij}^2+R_{ijkl}^2\bigr)
\ea
with 
\ba
 a=\frac{1}{360}\bigl(n_{\rm S}+(11/2)\,n_{\rm F}+62\,n_{\rm V}\bigr),\quad
 c=\frac{1}{120}\bigl(n_{\rm S}+3\,n_{\rm F}+12\,n_{\rm V}\bigr).
\ea
Here $n_{\rm S}$, $n_{\rm F}$ and $n_{\rm V}$ are 
the number of real scalars, Majorana fermions and vectors, respectively. 
The result \eq{result;weyl} can be obtained 
by setting $n_{\rm S}=6(N^2-1)$, $n_{\rm F}=4(N^2-1)$ 
and $n_{\rm V}=N^2-1$ and taking the large $N$ limit. 
}

\noindent\underline{\bf 7D pure gravity} \cite{FMS1}:

By using the value in Eq.\ \eq{d=6} with $d=6$, the local part of 
weight up to four is given by
\begin{equation}
S_{\rm loc}[g]=\int d^6x \sqrt{g}\left( -\frac{10}{l}-{l\over 4}R 
+{3l^3\over 320}R^2-{l^3\over 32}R_{ij}R^{ij}\right). 
\label{sloc6;dvv}
\end{equation}
{}From the flow equation of weight $w=6$, 
we thus find
\begin{eqnarray}
 \cW_6&=&-\frac{l}{2\kappa_7^2}\,
  \Bigl[\bigl\{\Sloc,\Sloc\bigr\}\Bigr]_6 \nn
 &=&\frac{l^5}{2\kappa_7^2}\,
  \left({1\over 128}\,RR_{ij}R^{ij}-{3\over 3200}\,R^3
  -{1\over 64}\,R^{ik}R^{jl}R_{ijkl}\right.\nn
 &&~~\left.+\,{1\over 320}\,R^{ij}\nabla_{i}\nabla_{j}R
  -{1\over 128}\,R^{ij}\nabla^2R_{ij}+{1\over 1280}\,
   R\,\nabla^2R\right),
 \label{weyl6}
\end{eqnarray}
which is in perfect agreement with the six-dimensional Weyl anomaly 
given in Ref.\ \cite{HS;weyl}.


\subsection{Callan-Symanzik equation}

Next we derive the Callan-Symanzik equation \cite{dVV}.
Acting on Eq.\ (\ref{cs;functional}) with the functional derivative
\begin{equation}
{\delta\over\delta\phi^{a_1}(x_1)}{\delta\over\delta\phi^{a_2}(x_2)}
\cdots {\delta\over\delta\phi^{a_n}(x_n)}, 
\end{equation}
and then setting $\phi^a=0$, we obtain the relation
\begin{eqnarray}
&&\left[-2g_{ij}(x){\delta\over\delta g_{ij}(x)}
+\beta^a(\phi(x))\,{\delta\over\delta\phi^a(x)} \right]
\bigl\langle \cO_{a_1}(x_1)\cO_{a_2}(x_2)\cdots \cO_{a_n}(x_n) 
 \bigr\rangle \nn
&&~~~~~~+\sum_{k=1}^n\delta(x-x_k)\partial_{a_k}\beta^b(\phi(x))
\bigl\langle \cO_{a_1}(x_1)\cdots \cO_{b}(x_k)
\cdots \cO_{a_n}(x_n) \bigr\rangle=0.
\end{eqnarray}
Recall that $\Gamma$ is the generating functional of correlation
functions with $\phi^a$ regarded as an external field 
coupled to the scaling operator ${\cO_a}(x)$.
By integrating it over ${\bf R}^d$ and considering the finite perturbation 
\begin{equation}
 g_{ij}(x)={1\over a^2}\,\delta_{ij},~~\phi^a(x)=\phi^a,\quad
 (a,\phi^a\!:~{\rm const.})
 \label{finite-perturbation}
\end{equation}
we end up with the Callan-Symanzik equation
\begin{eqnarray}
&&\left[a{\partial\over\partial a}
+\beta^a(\phi)\,{\partial\over\partial\phi^a} \right]
\bigl\langle \cO_{a_1}(x_1)\cO_{a_2}(x_2)\cdots \cO_{a_n}(x_n) \bigr\rangle \nn
&&~~~~~~-\sum_{k=1}^n\gamma_{a_k}^b(\phi)
\bigl\langle \cO_{a_1}(x_1)\cdots \cO_{b}(x_k)
\cdots \cO_{a_n}(x_n) \bigr\rangle=0.
\label{cs}
\end{eqnarray}
Here $\gamma_a^b(\phi)=-\partial_{a}\beta^b(\phi)$ 
is the matrix of anomalous dimensions.

\subsection{Anomalous dimensions}

Here we show that one can generalize 
to arbitrary dimension the argument in Ref.\ \cite{dVV} 
that the scaling dimensions can be calculated directly from the
flow equation \cite{FMS1}. 
{}First, we assume that the bulk scalars are normalized as
$L_{ab}(\phih)=\delta_{ab}$ 
and that the bulk scalar potential $V(\phih)$ has the expansion
\begin{equation}
 V(\phih)=2\Lambda+{1\over 2}\sum_a m_a^2\,\phih_a^2
 +\frac{1}{3!}\,\sum_{a,b,c}g_{abc}\,\phih_a\phih_b\phih_c +\cdots ,
\end{equation}
with $\Lambda=-d(d-1)/2l^2$.
Then it follows from (\ref{hj;potential}) 
that the superpotential $W$ takes the form 
\begin{equation}
 W(\phi)=-\frac{2(d-1)}{l}+{1\over 2}\sum_a \lambda_a\,\phi_a^2
 +\frac{1}{3!}\,\sum_{a,b,c}\lambda_{abc}\,\phi_a\phi_b\phi_c+\cdots,
\end{equation}
with 
\begin{eqnarray}
 &&l\lambda_a={1\over 2}\left( -d + \sqrt{ d^2+4\,m_a^2\,l^2}\right), \\
 &&g_{abc}= \left(\frac{d}{l}
 +\lambda_a+\lambda_b+\lambda_c\right)\lambda_{abc}.
\end{eqnarray}
The beta functions can then be evaluated easily and are found to be 
\begin{equation}
 \beta^a=-\sum_a l\lambda_a\,\phi_a
   -\frac{1}{2}\,\sum_{b,c}\lambda_{abc}\,\phi_b \phi_c+\cdots. 
\end{equation}
Thus, equating the coefficient of the first term 
with $d-\Delta_a$, where $\Delta_a$ is the scaling dimension of the operator 
coupled to $\phi_a$, 
we obtain 
\begin{equation}
 \Delta_a=d+l\lambda_a={1\over 2}
  \left( d + \sqrt{ d^2+4\,m_a^2\,l^2 }\right). 
\end{equation}
This exactly reproduces the result given in Ref.\ 
\cite{M,GKP,W;holography} (see also \S 2.2). 

\subsection{$c$-function revisited}

We here make a comment on how the the holographic $c$-function 
can be formulated within the framework developed in this section. 
For the Euclidean invariant metric 
$\gh_{ij}(x,\tau)=a(\tau)^{-2}\delta_{ij}$, 
the trace of the extrinsic curvature can be written as 
\ba
 \Kh(\tau)=\gh^{ij}\,\frac{1}{2}\,\frac{d}{d\tau}{\gh}_{ij}
  =-d\,\frac{d}{d\tau}\ln a
  =\frac{d}{2(d-1)}\,W\bigl(\phih(\tau)\bigr),
\ea
so that the holographic $c$-function can be rewritten into the following form: 
\ba
 \left(\frac{-1}{\Kh}\right)^{d-1}
  \sim \left(\frac{-1}{W\bigl(\phih(\tau)\bigr)}\right)^{d-1}
  \equiv c(\phih(\tau)). 
 \label{c-function2}
\ea
Thus, by introducing the ``metric'' of the coupling constants as 
\ba
 G_{ab}(\phi)\equiv \frac{1}{2}\,\left(\frac{-1}{W(\phi)}\right)^{d-1}
  L_{ab}(\phi), 
\ea
the beta functions can be expressed as 
\ba
 \beta^a(\phih)\,\left(=a\,\frac{d}{da}\phih^a\right)\,
  =-G^{ab}\bigl(\phih\bigr)\,\partial_b\,c\bigl(\phih\bigr). 
\ea
In this Euclidean setting, the monotonic decreasing of 
the $c$-function can be directly seen 
by assuming that $L_{ab}(\phi)$ (and thus $G_{ab}(\phi)$ also) 
is positive definite: 
\ba
 a\,\frac{d}{da}\,c\bigl(\phih(a)\bigr)=\beta^a\bigl(\phih\bigr)\,
  \partial_a c\bigl(\phih\bigr)
  =-G^{ab}\bigl(\phih\bigr)\,\partial_a c\bigl(\phih\bigr)\,
  \partial_b c\bigl(\phih\bigr)\leq 0. 
\ea
The equality holds when and only when the beta functions vanish.

Let us apply this analysis to the holographic RG flow 
from the $\cN=4$ $SU(N)$ SYM${}_4$ to the $\cN=1$ 
LS fixed point \cite{HRG}, which was mentioned in \S 2.4. 
The vector multiplet of the $\cN=4$ theory can be decomposed 
into a single $\cN=1$ vector multiplet $V=(A_i(x),\lambda(x))$ 
and three $\cN=1$ chiral multiplets $\Phi_I=(\varphi_I(x),\psi_I(x))$ 
$(I=1,2,3)$, each field of which belongs to the adjoint representation 
of $SU(N)$ 
and has the superpotential 
${\cal W}(\Phi)=\tr ([\Phi_1,\Phi_2]\Phi_3)$. 
One can deform the theory by adding to the superpotential 
an $\cN=1$ invariant mass term 
$\delta{\cal W}(\Phi)=(m/2)\,\tr (\Phi_3)^2$. 
This gives rise to an additional term in the potential, 
which can be written schematically as 
${\cal V}=m\,\tr[(\varphi_3)^3\!+\!(\lambda_3)^2]+m^2\,\tr[(\varphi_3)^2]$, 
and the LS fixed point is obtained by taking the limit $m\rightarrow\infty$. 
On the other hand, such deformations have a dual description 
in the $\cN=8$ gauged supergravity theory, 
and in particular, perturbations with 
the operators ${\cal O}_1(x)=\tr[(\varphi_3)^3\!+\!(\lambda_3)^2]$ 
and ${\cal O}_2(x)=\tr[(\varphi_3)^2]$ can be treated 
by considering the time development of two scalar (bulk) fields 
$\phih_a(x,\tau)$ $(a=1,2)$, 
whose superpotential is given by \cite{HRG}
\ba
 W(\phih)=e^{-\phih_2/\sqrt{6}}\left[\cosh\phih_1\cdot
  \left(e^{\sqrt{6}\,\phih_2/2}-2\right)-3\,e^{\sqrt{6}\,\phih_2/2}-2\right]. 
\ea
We here have normalized the scalar fields such that 
they have the kinetic term with $L_{ab}(\phih)=\delta_{ab}$. 
The scalar potential is then given by 
\ba
 V\bigl(\phih\bigr)=\frac{1}{2}\left(\partial_a W\bigl(\phih\bigr)\right)^2
  -\frac{1}{3}\left( W\bigl(\phih\bigr)\right)^2. 
\ea 
The shape of the $W(\phi)$ and $V(\phi)$ is depicted in 
Fig.\ 1 and Fig.\ 2. 
\begin{figure}[ht]
\begin{center}
\rotatebox{270}{
\resizebox{!}{80mm}{
\includegraphics{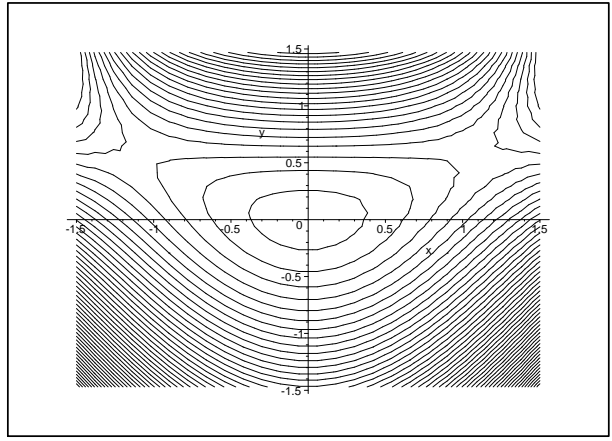}
}
}
\end{center}
\caption{\footnotesize{Superpotential $W(\phi)$. The fixed points 
are at $(\pm\ln 3,(2/\sqrt{6})\ln 2)$. }}
\label{w_fig}
\end{figure}
\begin{figure}[ht]
\begin{center}
\rotatebox{270}{
\resizebox{!}{80mm}{
\includegraphics{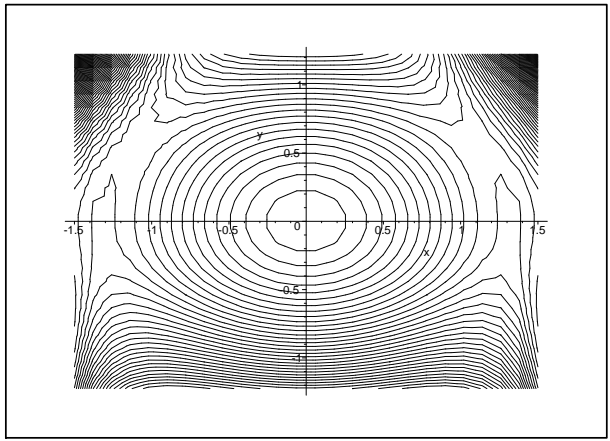}
}
}
\end{center}
\caption{\footnotesize{Scalar potential $V(\phi)$. The fixed points 
$(\pm\ln 3,(2/\sqrt{6})\ln 2)$ are saddle points, 
so that one direction is relevant and the other irrelevant. }}
\label{v_fig}
\end{figure}
The origin $(\phi_a)=(0,0)$ corresponds to the 
UV $\cN=4$ fixed point, 
and, as one can see from the figures, 
there appear another fixed points at 
$(\phi^\ast_a)=(\pm\ln 3,(2/\sqrt{6})\ln 2)$ 
(the two new fixed points are related by $\bZ_2$ transformation 
$\phi_1\rightarrow-\phi_1$), 
which is the LS fixed point. 
Around the origin, the superpotential is expanded as 
\ba
 W=-6-\frac{1}{2}\,(\phi_1)^2-(\phi_2)^2+\cdots, 
\ea
from which one finds that 
\ba
 l=1,\quad\lambda_1=-1,\quad \lambda_2=-2,
\ea
and thus their mass squared in the bulk gravity are calculated to be 
$m_1^2=-3$ and $m_2^2=-4$, respectively.  
The scaling dimensions are then obtained from the standard formula to be 
$\Delta_1=3$ and $\Delta_2=2$, 
which are precisely the scaling dimensions of $\cO_1$ and $\cO_2$ 
in the $\cN=4$ super Yang-Mills theory. 
On the other hand, around the IR fixed point, 
the superpotential is expanded as $W=-4\cdot 2^{2/3}+\cdots$, 
from which one finds that the radius changes from $l=1$ 
to $l^\ast=3\cdot 2^{-5/3}$. 
The mass-squared matrix $\partial_a \partial_b \,V(\phi^\ast)$ 
can be calculated easily as 
\ba
 \bigl(\partial_a\partial_b \,V(\phi^\ast)\bigr)&=&
  \frac{2^{13/4}}{3^2}\left(
   \begin{array}{cc}
   3 & \sqrt{6} \\ \sqrt{6} & 1
   \end{array}
  \right) \nn
 &\rightarrow&
  \frac{2^{13/4}}{3^2}\left(
   \begin{array}{cc}
   2-\sqrt{7} & 0 \\ 0 & 2+\sqrt{7}
   \end{array}
  \right)\quad({\rm diagonalized}),
\ea
so that by using $\Delta^\ast=2+\sqrt{4+m^2\,(l^\ast)^2}$ 
the scaling dimensions are calculated as
$\Delta^\ast_1=1+\sqrt{7}~(<4)$ and $\Delta^\ast_2=3+\sqrt{7}~(>4)$. 
This shows that at the IR fixed point 
the operators acquire large anomalous dimensions 
and one of the two becomes irrelevant.  
The ratio of the central charge can be calculated as before: 
\ba
 \frac{c_{\rm IR}}{c_{\rm UV}}
  =\frac{c\bigl(\phi^\ast)}{c(0)}
  =\left(\frac{-1/\,W(\phi^\ast)}{-1/\,W(0)}\right)^{3}
  =\left(\frac{l^\ast}{l}\right)^{3}
  =\frac{27}{32}, 
\ea
which certainly is less than unity and agrees with the previous result. 
Note that the ridge from the $\cN=4$ fixed point to the $\cN=1$ fixed point 
is given by the curve which has the shape $\phi_2=(\phi_1)^2$ 
around the origin. 
This is an expected result 
since such ridge should preserve the $\cN=1$ symmetry 
and the two scalars are expressed as 
$\phi_1\simeq m$ and $\phi_2\simeq m^2$ around the origin \cite{HRG}.

\subsection{Continuum limit}

In this subsection, we describe a direct prescription for taking 
continuum limits of boundary field theories which is such that 
counterterms can be extracted easily.\footnote{
{}For earlier work on counterterms, see e.g.\ Ref.\ \cite{BK}.
} 
The following argument is based on Ref.\ \cite{FMS1}.

Let $\gb_{ij}(x,\tau)$ and $\phib^a(x,\tau)$
be the classical trajectory of $\gh_{ij}(x,\tau)$ 
and $\phih^a(x,\tau)$ with the boundary condition 
\begin{equation}
 \gb_{ij}(x,\tau_0)=g_{ij}(x),\qquad
 \phib^a(x,\tau_0)=\phi^a(x).
\end{equation}
Recall that the classical action is given as a functional of 
the boundary values $g_{ij}(x)$ and $\phi^a(x)$, 
obtained by substituting these classical solutions into the bulk action:
\begin{equation}
 S\bigl[g_{ij}(x),\phi^a(x)\bigr]=\int d^d x \int_{\tau_0}d\tau\,\sqrt{\gb}\,
 {\cal L}_{d+1}\bigl[\gb(x,\tau),\phib(x,\tau)\bigr]. 
\end{equation}
Also, recall that the fields $g_{ij}(x)$ and $\phi^a(x)$ are 
considered as the bare sources at the cutoff scale corresponding to 
the flow parameter $\tau_0$. 
Although the classical action is actually independent of  
$\tau_0$ due to the Hamilton-Jacobi constraint, 
we still need to tune the fields $g_{ij}(x)$ and $\phi^a(x)$ 
as functions of $\tau_0$ so that 
the low energy physics is fixed and described in terms of finite 
renormalized couplings.

\begin{figure}[htbp]
\begin{center}
\resizebox{!}{70mm}{
 \input{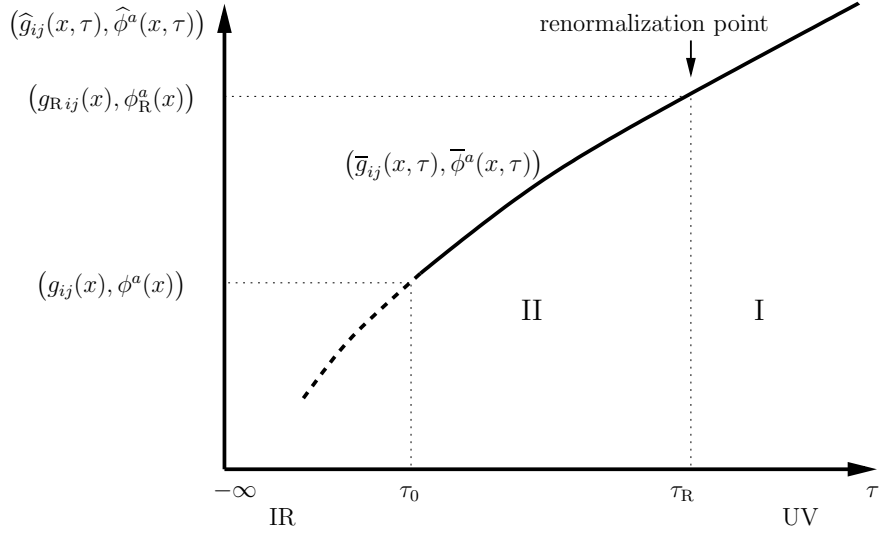}
}
\caption{The evolution of the classical solutions 
$\bigl(\gb_{ij}(x,\tau),\phib^a(x,\tau)\bigr)$ 
along the radial direction $\tau$. 
The region I is defined 
by $\tau\ge \tau_{\rm R}$, and the region II is defined by 
$\tau_0\le \tau < \tau_{\rm R}$. }
\label{renormalize}
\end{center}
\end{figure}

In the holographic RG, such renormalization can be 
easily carried out 
by tuning the bare sources back along the classical trajectory 
in the bulk (see Fig.\ \ref{renormalize}). 
That is, if we would like to fix the couplings at the
``renormalization point'' $\tau=\tau_{\rm R}$ 
to be $\bigl(g_{\rm R}(x),\phi_{\rm R}(x)\bigr)$ 
and to require that physics does not change as the cutoff moves, 
we only need to take the bare sources to be 
\begin{equation}
 g_{ij}(x;\tau_0)=\gb_{ij}(x,\tau_0),\qquad
 \phi^a(x;\tau_0)=\phib^a(x,\tau_0). 
\label{bare}
\end{equation}

The classical action with these running bare sources can be 
easily evaluated by using Eq.\ (\ref{bare}):
\begin{align}
 S\bigl[g_{ij}(x;\tau_0),\,\phi^a(x;\tau_0)\bigr] 
  &= \int d^dx\int_{\tau_0}d\tau 
  \,\sqrt{\gb}\,{\cal L}_{d+1}
  \left[\gb(x,\tau),\phib(x,\tau)\right] \nn
 &= \int d^dx\left( \int_{\tau_{\rm R}}d\tau
   +\int_{\tau_0}^{\tau_{\rm R}}d\tau \right)
  \sqrt{\gb}\,{\cal L}_{d+1} \nn
 &= S_{\rm R}\bigl[g_{\rm R}(x),\phi_{\rm R}(x)\bigr]
  +S_{\rm CT}\bigl[g_{\rm R}(x),\phi_{\rm R}(x);\tau_0,\tau_{\rm R}\bigr].
\label{counter}
\end{align}
Here $S_{\rm R}$ is given by integrating 
$\sqrt{\gb}{\cal L}_{d+1}$ over the region I in Fig.\ 
\ref{renormalize}, and it obeys the Hamiltonian constraint, 
which ensures that $S_{\rm R}$ does not depend on $\tau_{\rm R}$. 
On the other hand, $S_{\rm CT}$ is given by integrating 
$\sqrt{\gb}{\cal L}_{d+1}$ over the region II. 
{ }It also obeys the Hamiltonian constraint 
and thus does not depend on the coordinates of the boundaries 
of integration, $\tau_{\rm R}$ and $\tau_0$, explicitly. 
However, in this case, their dependence implicitly enters  
$S_{\rm CT}$ through the condition that the boundary values at $\tau=\tau_0$
are on the classical trajectory through the renormalization point:
\begin{align}
 S_{\rm CT}&=
  S\bigl[g_{\rm R}(x),\phi_{\rm R}(x);g(x,\tau_0),\phi(x,\tau_0)\bigr] \nn
 &=S\bigl[g_{\rm R}(x),\phi_{\rm R}(x);\gb(x,\tau_0;g_{\rm R},\tau_{\rm R}),
  \phib(x,\tau_0;\phi_{\rm R},\tau_{\rm R})\bigr].
\end{align}
It is thus natural to interpret 
$S_{\rm CT}\bigl[g_{\rm R},\phi_{\rm R};\tau_0,\tau_{\rm R}\bigr]$ as 
the counterterm, and the nonlocal part of 
$S_{\rm R}\bigl[g_{\rm R},\phi_{\rm R}\bigr]$ gives
the renormalized generating functional of the boundary field theory, 
$\Gamma_{\rm R}\bigl[g_{\rm R},\phi_{\rm R}\bigr]$, 
written in terms of the renormalized sources.

Since, as pointed out above, 
$S_{\rm R}\bigl[g_{\rm R},\phi_{\rm R}\bigr]$ also satisfies the Hamiltonian 
constraint, it will yield the same form of the flow equation, 
with all the bare fields replaced by the renormalized fields. 
This suggests that the holographic RG exactly describes 
the so-called renormalized trajectory \cite{WK},    
which is a submanifold in the parameter space, 
consisting of the flows driven by relevant perturbations 
from an RG fixed point at $\tau_0=-\infty$.

There is another scheme of the renormalization 
which was systematically developed by Henningson and Skenderis 
\cite{HS;weyl}.%
\footnote{For a recent discussion based on the Hamilton-Jacobi equation, 
see, {\em e.g.}, Ref.\ \cite{Martelli:2002sp}.}
A detailed comparison of their scheme with that of this subsection 
was given in Ref.\ \cite{FMS1}.%

\resection{Holographic RG and the noncritical string theory}

In this section, we show that the structure of the holographic RG 
can be naturally understood within the framework of noncritical 
string theory. 
In particular, we demonstrate that the Liouville field $\varphi$ 
can be understood to be the $(d+1)$-st coordinate 
appearing in the holographic RG; 
\ba
 \varphi~~(\mbox{Liouville}) ~\longleftrightarrow~ \tau=X^{d+1}.
\ea

\subsection{Noncritical string theory}

We first summarize the basic results on noncritical strings. 
The noncritical string theory \cite{KPZ,DDK} 
is a world-sheet theory where 
only the two-dimensional diffeomorphism ($\Diff$) is imposed 
as a gauge symmetry, 
while the usual critical string theory has the gauge symmetry 
$\Diff\times\Weyl$. 
The nonlinear $\sigma$ model action of the noncritical string theory 
can be written as 
\ba
 S_{{\rm NL}\sigma}[x^i(\xi),\gamma_{ab}(\xi)]&=&\frac{1}{4\pi\alpha'}\,
  \int d^2\xi\,\sqrt{\gamma}\,\left(\gamma^{ab}\,g_{ij}(x(\xi))\,
  \partial_a x^i(\xi)\,\partial_b x^j(\xi)\right.\nn
 &&~~~~\left.+\,T(x(\xi))\,+\,\alpha'\,R_{\gamma}\,\Phi(x(\xi))
  \,+\,\cdots\right).
\ea
Here $\xi=(\xi^a)=(\xi^1,\xi^2)$ are the coordinates of the world-sheet, 
and $\gamma_{ab}(\xi)$ is an intrinsic metric on the world-sheet. 
$x^i~(i=1,2,\cdots,d)$ are the coordinates 
of the $d$-dimensional target space, 
and $g_{ij}(x)$, $T(x)$ and $\Phi(x)$ are, respectively, 
the metric, tachyon and dilaton fields in the target space. 
The partition function is defined as 
\ba
 Z=\int \frac{\cD x^i(\xi)\,\cD\gamma_{ab}(\xi)}{{\rm Vol}(\Diff)}\,
  \exp\left(-S_{{\rm NL}\sigma}[x^i(\xi),\,\gamma_{ab}(\xi)]\right).
\ea
One can see from the above expression that the slope parameter 
$\alpha'$ plays the role of expansion parameter ($\alpha'\sim\hbar$).

The convenient gauge fixing is the conformal gauge for which 
we set the intrinsic metric $\gamma_{ab}(\xi)$ to be 
\ba
 \gamma_{ab}(\xi)=e^{\varphi(\xi)}\cdot\gammah_{ab}(\xi),
\ea
where we have introduced a (fixed) fiducial metric $\gammah_{ab}(\xi)$, 
and the field $\varphi(\xi)$ is called the Liouville field. 
This gauge fixing actually is not complete and leaves 
the residual gauge symmetry consisting of local conformal isometries 
with respect to $\gammah_{ab}$: 
\ba
 \frac{\cD\gamma_{ab}(\xi)}{{\rm Vol}(\Diff)}
  = \frac{\cD\varphi(\xi)}{{\rm Vol}(\Conf)}\,
  e^{-S_{\rm Liouville}[\varphi(\xi),\,\gh_{ab}(\xi)]}, 
\ea
where $S_{\rm Liouville}$ is a local functional 
written with $\varphi(\xi)$ and the fiducial metric $\gh(\xi)$.

As is the case for any scalar fields on the world-sheet, 
the Liouville field $\varphi$ can be regarded as an extra dimensional 
coordinate. 
This interpretation can be pursued further if we change 
the measure of $\varphi$ from the original one 
\ba
 \cD\varphi(\xi)~\leftrightarrow~
  ||\delta\varphi||^2_\gamma\equiv
  \int d^2\xi\sqrt{\gamma}\,(\delta\varphi)^2
  =\int d^2\xi \sqrt{\gammah}\,e^\varphi\,(\delta\varphi)^2
\ea
to the translationally invariant one \cite{DDK}: 
\ba
 \cDh\varphi(\xi)~\leftrightarrow~
  ||\delta\varphi||^2_{\gammah}\equiv
  \int d^2\xi\sqrt{\gammah}\,(\delta\varphi)^2. 
\ea
It will induce a Jacobian factor which can be absorbed 
into the the bare fields $g_{ij}(x)$, $T(x)$ and $\Phi(x)$ 
due to the renormalizability of the NL$\sigma$ model. 
We thus obtain the following expression for the partition function: 
\ba
 Z&=&\int\frac{\cD x^i\,\cD\varphi}{{\rm Vol}(\Conf)}\,
  e^{-S_{{\rm NL}\sigma}}\,e^{-S_{\rm Liouville}} \nn
 &=&\int\frac{\cDh x^i\,\cDh\varphi}{{\rm Vol}(\Conf)}\,
  e^{-\Sh_{{\rm NL}\sigma}[x^i\!,\varphi;\,\gammah_{ab}]},
\ea
where the effective action 
$\Sh_{{\rm NL}\sigma}[x^i\!,\varphi;\,\gammah_{ab}]$ now has the form 
\ba
 \Sh_{{\rm NL}\sigma}&=&\frac{1}{4\pi\alpha'}\int d^2\xi\,\sqrt{\gammah}\,
  \left[\gammah^{ab}\left(\partial_a\varphi\,\partial_b\varphi\,+\,
  \gh_{ij}(x,\phi)\,\partial_a x^i\,\partial_b x^j\right)\right.\nn
 &&~~~~~\left.+\,\Th(x,\varphi)\,
  +\,\alpha'\,R_{\gammah}\cdot\Phih(x,\varphi)\,
  +\,\cdots\right]. 
\ea
Here we have rescaled $\varphi$ such that it has the kinetic term 
in a canonical form. 
The above expression shows that one can introduce a $(d+1)$-dimensional 
space with the coordinates $X^\mu=(x^i,\varphi)~(i=1,...,d)$ 
and the metric 
\ba
 ds^2=\gh_{\mu\nu}(x,\varphi)\,dX^\mu\,dX^\nu
  \equiv(d\varphi)^2+\gh_{ij}(x,\varphi)\,dx^i\,dx^j. 
\ea
Those coefficients cannot take arbitrary values since 
we must impose the conformal symmetry on the effective action, 
which is equivalent to choosing the coefficients such that 
their beta functions vanish. 
One can easily show that the equations $\beta=0$ can be 
derived as the equations of motion of 
the following effective action of the {\em target space}:
\ba
 \bS=\int d^d x\,d\varphi\,\sqrt{\gh}\,e^{-2\Phih}\,
  \left(2\Lambda_0-\widehat{R}-4(\widehat{\nabla}\Phih)^2
  +(\widehat{\nabla}\Th)^2+m_0^2\,\Th^2+O(\alpha')\right)
\ea
with $2\Lambda_0=2(d-25)/3\alpha'$ and $m_0^2=-4/\alpha'$. 
Since the residual conformal isometry can be translated into 
the Weyl symmetry, 
the above discussion shows that 
the $d$-dimensional noncritical string theory is equivalent to 
a $d$-dimensional critical string theory. 

\subsection{Holographic RG in terms of noncritical strings}

As will be further investigated in the following sections, 
one of the basic assumptions in the holographic RG is 
that the (Euclidean) time development should be regular 
interior of the bulk. 
It turns out that this corresponds to the so-called 
Seiberg condition \cite{Seiberg} in the noncritical string theory. 
Let us consider a $(d+1)$-dimensional bosonic string 
theory in the linear dilaton background \cite{Myer}, 
although this does not have asymptotically AdS geometry: 
\ba
 \gh_{ij}=\delta_{ij},\quad \Phih=Q\,\varphi. 
\ea
The coefficient $Q$ is determined from the conformal invariance 
as $Q^2=-\Lambda_0/2=(25-d)/6\alpha'$. 
Then the tachyon vertex with Euclidean momentum 
$k_\mu=(k_i,\alpha)$ is expressed by 
\ba
 \Th&=&e^{i\,k_i x^i+\alpha\,\varphi}\nn
 &=&e^{\Phih}\cdot e^{i\,k_i x^i+(\alpha-Q)\varphi}. 
\ea
Here we extract the factor $e^{\Phih}=e^{Q\,\varphi}$ which comes from 
the curvature arising when an infinitely long cylinder 
is inserted in the world-sheet. 
Thus the momentum along the cylinder is effectively 
$k_\mu|_{\rm cylinder}=(k_i,\,\alpha-Q)$, 
so that the convergence of the wave function inside 
the bulk ($\varphi\rightarrow+\infty$) is equivalent 
to the Seiberg condition $\alpha-Q<0$.

{}Furthermore, the bulk IR cutoff $\tau\geq\tau_0$ 
(or $\varphi\geq\varphi_0$) 
is equivalent to the small-area cutoff of the world-sheet \cite{DW}. 
In fact, when the $(d+1)$-dimensional target space is asymptotically AdS, 
the integration over the zero mode of $\varphi(\xi)$ diverges  
around $\varphi\sim-\infty$. 
This divergence can be regularized by introducing the cutoff $\varphi_0$ 
as we did in the preceding sections: 
\ba
 \int_{-\infty}^\infty d\varphi \int \cDh'\varphi(\xi)\,
  e^{-\Sh_{{\rm NL}\sigma}} ~\Rightarrow~
  \int_{\varphi_0}^\infty d\varphi \int \cDh'\varphi(\xi)\,
  e^{-\Sh_{{\rm NL}\sigma}}.
\ea
On the other hand, the area of the world-sheet can be expressed 
by the zero mode through the volume element 
$\sqrt{\gamma}= e^{\alpha\varphi}$, 
so that this cutoff actually sets a lower bound on the area: 
\ba
 A=\int\sqrt{\gamma}=\int e^{\alpha\varphi} \geq
  \int e^{\alpha\varphi_0}= A_{{\rm min}}. 
\ea
Thus, the holographic RG describes the development 
of string backgrounds as the minimum area of world-sheet is changed, 
which is equivalent, after the Legendre transformation, 
to the development with respect to the two-dimensional 
cosmological constant.

The above two features can be best seen when one sets up 
the holographic RG within the framework of noncritical string theory, 
although it is mathematically equivalent to the critical string theory. 
Taking the translationally invariant measure 
for the Liouville field $\varphi$ is necessary 
in order for $\varphi$ to be interpreted as the RG flow parameter. 
Moreover, those two features are realized automatically 
in (old) matrix models.  
In fact, in such matrix models there exists a bare cosmological term 
which gives rise to the Liouville wall 
so that any physically meaningful wave functions 
are regular inside the bulk of the target space, 
which is nothing but the Seiberg condition. 
{}Furthermore, the continuum limit is obtained 
by fine-tuning couplings such that contributions  
from surfaces with large area survive. 
In fact, the contribution from surfaces with small area 
is always non-universal and discarded in taking the continuum limit, 
and the cutoff on the (physically) small area is naturally set 
by introducing the renormalized cosmological constant term.

The nonlinear $\sigma$ model action 
$S_{{\rm NL}\sigma}[x^i,\gamma_{ab}]$ 
with finitely many ``couplings'' $g_{ij}(x)$, $\Phi(x)$ and $T(x)$ 
gives a renormalizable theory, 
which means that these couplings determine the structure of 
the $(d+1)$-dimensional target space $X^\mu=(x^i,\varphi)$ 
for any value of $\alpha'$. 
Actually the dependence of the renormalized fields 
on $\varphi$ is totally determined by the conformal symmetry 
on the world-sheet. %
This observation implies that the holographic RG structure 
should be preserved for all orders in the $\alpha'$ expansion. 
We will give a few evidences to this expectation.

\resection{Holographic RG for higher-derivative gravity}

In this section, we investigate gravity systems with higher-derivative 
interactions and discuss their relationship 
to the boundary field theories \cite{FMS2,FM}. 
As we show in \S 5.2, 
for a higher-derivative system, 
in order to determine the classical behavior uniquely  
we need more boundary conditions 
than those without higher-derivative interactions. 
Thus, the holographic principle may seem not to work for 
higher-derivative gravity. 
The main aim of this section is to demonstrate that 
the holographic structure still persists for such systems 
by showing that the behavior of bulk fields can be specified 
only by their boundary values. 
This statement is not surprising because 
higher-derivative terms in string theory come from
$\alpha^{\prime}$ corrections;
as we have seen in the the case of non-critical strings,
the renormalizability of the nonlinear $\sigma$ model action 
assures the holographic structure to exist 
for that system.

As a warming-up, we first analyze the system that has 
Euclidean symmetry at each time-slice. 
We introduce a parametrization with which one can 
easily investigate the global structure of the holographic RG 
of the boundary field theories. 
We show that there appear new multicritical fixed points 
in addition to the original conformal fixed points 
existing in the AdS/CFT correspondence.
After grasping basic ideas, 
we then formulate the holographic RG for higher-derivative gravity 
in terms of the Hamilton-Jacobi equation, 
and show that bulk gravity always exhibits
the holographic behavior 
even with higher-derivative interactions. 
We also apply this formulation to a computation of the Weyl anomaly 
and show that the result is consistent 
with a field theoretic calculation.

\subsection{Holographic RG structure in higher-derivative gravity}

In this subsection, we exclusively consider a bulk metric 
with $d$-dimensional Euclidean invariance. 
We introduce a parametrization which allows us to easily investigate 
the global structure of the holographic RG of the boundary field theory.

The bulk metric with $d$-dimensional Euclidean symmetry 
can be written in the following form 
by setting $\gh_{ij} = e^{-2q(\tau)}\,\delta_{ij}$, 
$\widehat{N}=N(\tau)$ and $\lambdah^i=0$ in the ADM decomposition 
\eq{metric;dvv}:%
\footnote{
$q(\tau)$, $N(\tau)$, etc. are bulk fields, 
but in this and the next subsections, 
we do not place the hat (or bar) on (the classical solutions of) 
these bulk fields in order to simplify expressions. 
}
\ba
 ds^2 = N(\tau)^2 d\tau^2 + e^{-2q(\tau)}\,\delta_{ij}\,dx^idx^j.
 \label{metric}
\ea
{}For this metric, the unit length in the $d$-dimensional time-slice 
at $\tau$ is given by $a=e^{q(\tau)}$. 
Since the unit length should grow monotonically 
under the RG flow, 
{\em $dq(\tau)/d\tau$ must be positive 
in order for the bulk metric to have a chance to describe 
the holographic RG flow of the boundary field theories. }

We consider two kinds of gauge fixings (or parametrizations of time). 
One is the temporal gauge which is obtained by setting $N(\tau)=1$: 
\ba
 ds^2=d\tau^2+e^{-2q(\tau)}\delta_{ij}dx^idx^j.
 \label{temporal}
\ea
The other is a gauge fixing that can be made only when 
the above condition 
\ba
 \frac{dq(\tau)}{d\tau} > 0 \qquad (-\infty < \tau < \infty) 
 \label{BScondition}
\ea
is satisfied. 
Then $q$ itself can be regarded as a new time coordinate. 
We call this parametrization the {\it block spin gauge} \cite{FM}.%
\footnote
{
In this gauge, the unit length in the 
$d$-dimensional time slice at $t$ is given by $a(t)=a_0e^{t}$
with a positive constant $a_0$. 
If we consider the time evolution 
$t \to t+\delta t$, the unit length changes as $a \to e^{\delta t}a$. 
In other words, one step of time evolution directly describes 
that of block spin transformation of the $d$-dimensional field theory.
}
By writing $q(\tau)$ as $t$, 
the metric in this gauge is expressed as%
\footnote{
This form of metric sometimes appears in the literature
(see, e.g., Ref.\ \cite{NO2}).}
\ba
 ds^2 = Q(t)^{-2}dt^2 + e^{-2t}\,\delta_{ij}\,dx^idx^j.
 \label{BSmetric}
\ea
Since two parametrizations of time (temporal and block spin) 
are related as 
\ba
 t=q(\tau), 
\ea
together with the condition \eq{BScondition} 
the coefficient $Q(t)$ is given by 
\ba
 Q(t)={dq(\tau)\over d\tau}\bigg|_{\tau=q^{-1}(t)}(>0),  
 \label{BSconditionQ}
\ea 
which we call a ``higher-derivative mode.''%
\footnote{
$Q$ actually appears as a new canonical valuable 
in the Hamiltonian formalism of $R^2$ gravity. 
See the next subsection. 
} 
Note that a constant $Q~(\equiv 1/l)$ gives the AdS metric of radius $l$,
\ba
 ds^2&=&d\tau^2 + e^{-2\tau/l}\,dx_i^2\quad({\rm temporal~gauge})\nn
  &=&l^2 dt^2 + e^{-2t}\,dx_i^2\quad({\rm block~spin~gauge}),
\ea
with the boundary at $\tau=-\infty$ (or $t=-\infty$).

Here we show that the condition (\ref{BScondition}) sets 
a restriction on the possible geometry, 
by solving the Einstein equation both in the temporal and block spin gauges.
In the temporal gauge, the Einstein-Hilbert action
\ba
 \bS_E = \int_{M_{d+1}} d^{d+1}X \sqrt{\hg}\left[2\Lambda-\hR\right] 
 \label{Einstein}
\ea
becomes 
\ba
 \bS_E = -d(d-1){\mathcal{V}}_d\int d\tau e^{-dq(\tau)}
 \left(\dot{q}(\tau)^2+\frac{1}{l^2}\right),
\ea
up to total derivatives.
Here we have parametrized the cosmological constant as 
$\Lambda = -d(d-1)/2l^2$, 
and $\mathcal{V}_d$ is the volume of the $d$-dimensional space.
A general classical solution for this action is given by
\ba
 \frac{dq}{d\tau}= \frac{1}{l}\,\frac{1-Ce^{d\tau/l}}{1+Ce^{d\tau/l}}
 \qquad (C \ge 0).
\ea
This shows that the geometry with a non-vanishing, finite $C$ 
($C \neq 0$ or $\infty$) cannot be described  
in the block spin gauge,  
since $\dot{q}$ vanishes at $\tau=-(l/d)\ln{C}$, 
violating the condition (\ref{BScondition}). 
In fact, in the block spin gauge (\ref{BSmetric}),
the action (\ref{Einstein}) becomes
\ba
 \bS_{E} = -d(d-1){\mathcal{V}}_d\int dt e^{-dt}\left(
 \frac{1}{l^2Q}+Q\right), 
\ea
which readily gives the classical solution as 
\ba
 Q(t) = \frac{1}{l}\quad(>0).
\ea
This actually reproduces only the AdS solution 
among the possible classical solutions obtained in the temporal gauge.

Next we consider a pure $R^2$ gravity theory 
in a $(d+1)$-dimensional manifold $M_{d+1}$ with boundary $\Sigma_d$. 
The action is generally given by
\begin{align}
 \bS = &\int_{M_{d+1}}d^{d+1}X\sqrt{\hg}\left(2\Lambda-\hR
  -a\hR^2 -b\hR_{\mu\nu}^2 -c\hR_{\mu\nu\rho\sigma}^2\right) \nn
 &+\int_{\Sigma_d}d^d x \sqrt{g}\left(
  2K+x_1\,RK+x_2\,R_{ij}K^{ij}+x_3\,K^3 
  +x_4\,KK_{ij}^2+x_5\,K_{ij}^3\right),
 \label{PG:action}
\end{align}
with some given constants $a,b$ and $c$. 
Here $X^\mu=(x^i,t)$ $(i=1,\cdots,d)$ 
and we set the boundary at $t=t_0$. 
$K_{ij}$ and $R_{ijkl}$ are 
the extrinsic curvature and the Riemann tensor on $\Sigma_d$, 
respectively. 
The first term in the boundary terms in (\ref{PG:action}) 
is the Gibbons-Hawking term for Einstein gravity \cite{GH}, 
and  the form of the rest terms are determined by requiring 
that it is invariant under the diffeomorphism 
\ba
 X^{\mu}\rightarrow X^{\prime\mu}=f^{\mu}(X), 
\ea
with the condition 
\ba
 f^{\,t} (x,t\!=\!t_0)=t_0, 
 \label{diffeo;boundary}
\ea
which implies that the diffeomorphism does not change 
the location of the boundary. 
A detailed analysis on this condition is given 
in Appendix \ref{BoundaryTerms}.%
\footnote{
The boundary action in \eq{PG:action}, except for the first term, 
can be interpreted as the generating functional 
of a canonical transformation which shifts the conjugate momentum 
of the higher-derivative mode by a local function. 
}
(Other studies of boundary terms in higher-derivative 
gravity can be found in Refs.\ \cite{Mye} and \cite{NO;boundary}.)

In the block spin gauge,
the equation of motion for $Q$ reads \cite{FM}
\ba
 Q\ddot{Q}+{1\over2}\dot{Q}^2-dQ\dot{Q}={1\over A}\biggl(
 {2\Lambda\over Q^2}+d(d-1)-3BQ^2\biggr), 
 \label{PG:EOM} 
\ea
where $\cdot=d/d t$, and $A$ and $B$ are given by 
\ba
 A=2d\bigl(4da+(d+1)b+4c\bigr), \quad
 B=\frac{d(d-3)}{3}\bigl(d(d+1)a+db+2c\bigr).\label{AB}
\ea
Here we set $t$ to run from $t_0$ to $\infty$.   
The classical action $S$ is obtained by substituting into $\bS$ 
the classical solution $Q(t)$ 
that satisfies the boundary condition $Q(t_0)=Q_0$ 
and has a regular behavior in the limit $t\to+\infty$. 
It is a function of the boundary value, 
$\bS[Q(t)] \equiv S(Q_0,t_0)$.

In the holographic RG, this classical action would be interpreted 
as the bare action of a $d$-dimensional field theory 
with bare coupling $Q_0$ at the UV cutoff $\Lambda_0=\exp (-t_0)$, 
as was discussed in detail in \S 2 and \S 3. 
Our strategy to investigate the global structure of the RG flow 
with respect to $t$ 
is as follows.
We first find the solution that converges to $Q\!=\!{\rm const.}$ 
as $t\to+\infty$ in order to have a finite classical action. 
We next examine the stability of the solution 
by studying a linear perturbation around it.
Since the solution $Q\!=\!{\rm const.}$ gives an AdS geometry, 
the fluctuation of $Q$ around it is regarded 
as the motion of the higher-derivative mode 
in the AdS background, 
which leads to a holographic RG interpretation 
of the higher-derivative mode.

Following the above strategy, 
we first look for AdS solutions ({\em i.e.}, $Q(t)={\rm
const.}$).
By parametrizing the cosmological constant as
\ba
 \Lambda = -{d(d-1)\over 2l^2} + {3B\over 2l^4}, 
\ea
the equation of motion (\ref{PG:EOM}) gives two AdS solutions,
\begin{equation}
 Q^2 = 
  \begin{cases}
  \displaystyle{~~~~~~{1\over l^2}} &\displaystyle{\equiv \,{1\over l_1^2}}
  \,\,, 
   \vspace{2mm}\\
  \displaystyle{{d(d-1)\over 3B}-{1\over l^2}} 
   &\displaystyle{\equiv\, {1\over l_2^2}}\,\,,
   \end{cases}
 \label{AdS_sln}
\end{equation}
where the solution $Q=1/l_2$ exists only when $B>0$.%
\footnote%
{
We consider only the case $Q>0$ because of the condition 
(\ref{BScondition}).
}
We denote by AdS${}^{(i)}\,\,(i=1,2)$ the AdS solution of radius $l_i$.
We assume that we can take the limit $a,b,c\to 0$ smoothly, 
in which the system reduces to Einstein gravity 
on AdS of radius $l=l_1$. 
We also assume that this AdS gravity comes from the low-energy limit 
of a string theory, 
so that its radius $l_1=l$ should be sufficiently larger 
than the string length.
On the other hand, the AdS${^{(2)}}$ solution, if it exists,
appears only when the higher-derivative terms are taken into account. 
As the higher-derivative terms are thought to 
stem from string excitations, their coefficients $a,b$ and $c$ 
(and hence $A$ and $B$) 
are ${\mathcal{O}}(\alpha')$. 
Thus the radius of the AdS${^{(2)}}$ is 
much smaller than that of AdS${^{(1)}}$.

Next, we examine the perturbation of classical solutions 
around (\ref{AdS_sln}), writing $Q(t)$ as
\ba
 Q(t) = {1\over l_i} + X_i(t). 
\ea
The equation of motion (\ref{PG:EOM}) is then linearized as 
\ba
 \ddot{X}_i - d\dot{X}_i -l_i^2m_i^2 X_i =0,
 \label{PG:linear}
\ea
with
\ba
 m_i^2 \equiv -{2\over A}\left(2\Lambda l_i^2+ {3B\over l_i^2}\right).
 \label{mass2}
\ea
The general solution of \eq{PG:linear} 
is given by a linear combination of the functions 
\begin{equation}
 f_i^{\pm}(t) \equiv 
  \exp\left[\left({d\over 2}\pm\sqrt{{d^2\over 4}+l_i^2m_i^2}\right)t
  \right].
 \label{PG:linear_sln} 
\end{equation}
Here $l_i^2m_i^2$ can be easily calculated from (\ref{AdS_sln}) and 
(\ref{mass2}) as
\begin{equation}
 \begin{cases}
  \displaystyle{~l_1^2m_1^2\!} 
   &\displaystyle{=\, {2\over A}\left(d(d-1)l^2-6B\right)}\,\,, 
  \vspace{2mm}\\
  \displaystyle{~l_2^2m_2^2\!} 
  &\displaystyle{=\, -\,{6B\over A}\cdot{d(d-1)l^2-6B \over d(d-1)l^2-3B}}
  \,\,.
 \end{cases}
 \label{l2m2}
\end{equation}

\noindent
\underline{\bf perturbation around AdS${}^{(1)}$}

\noindent
From (\ref{PG:linear_sln}) and (\ref{l2m2}), we see that 
the behavior of $f_1^{\pm}(t)$
depends on the sign of $A$.
For $A>0$, recalling that $A$ is ${\mathcal{O}}(\alpha')$, 
$f_1^+(t)$ grows while $f_1^-(t)$ damps very rapidly. 
On the other hand, for $A<0$, the value in the square root in  
(\ref{PG:linear_sln}) becomes negative,
and thus both $f_1^{\pm}(t)$ 
oscillate rapidly.

\noindent
\underline{\bf perturbation around AdS${}^{(2)}$}

\noindent
We assume $B>0$ because, as mentioned above, 
AdS$^{(2)}$ exists only in this region.
For $A>0$, both of $f_2^{\pm}(t)$ grow exponentially, because 
$l_2^2m_2^2<0$.
On the other hand, for $A<0$, $f_2^{+}(t)$ grows and 
$f_2^{-}(t)$ damps exponentially.

\begin{figure}[ht]
\begin{center}
\includegraphics{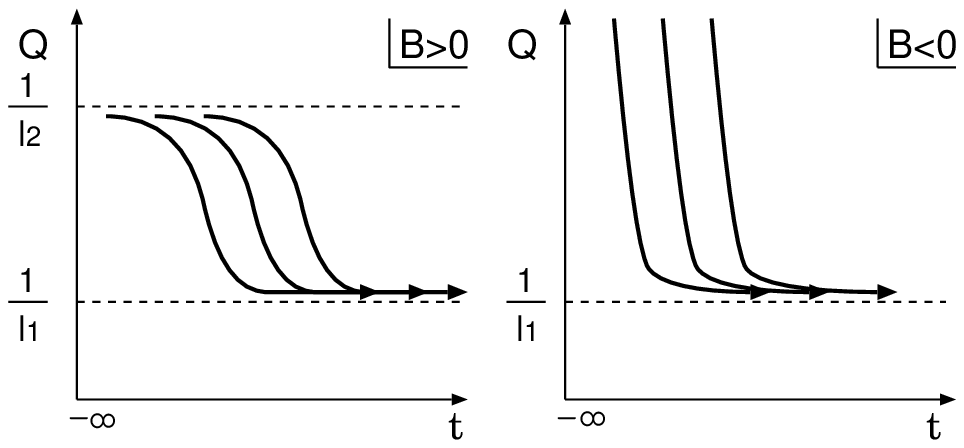}
\end{center}
\caption{\footnotesize{Classical solutions $Q(t)$ for $A>0$.}}
\label{PGflow1}
\end{figure}

\begin{figure}[ht]
\begin{center}
\includegraphics{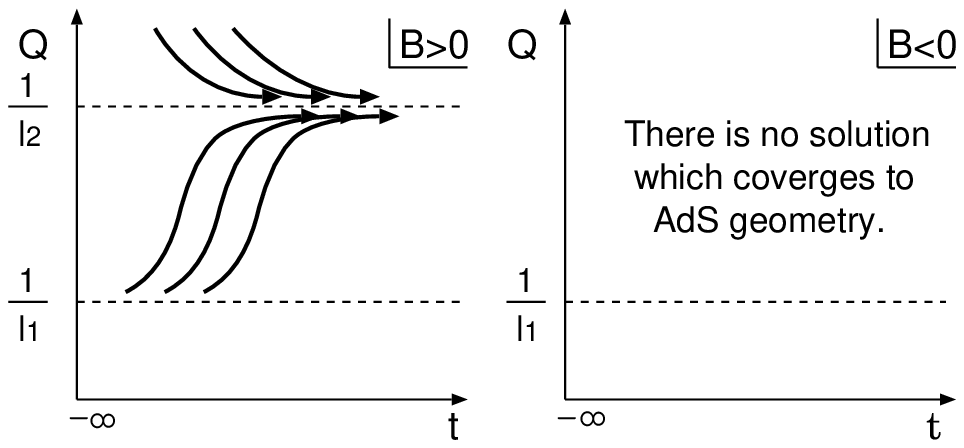}
\end{center}
\caption{\footnotesize{Classical solutions $Q(t)$ for $A<0$.}}
\label{PGflow2}
\end{figure}

\begin{figure}[ht]
\begin{center}
\scalebox{.6}[.6]{\includegraphics{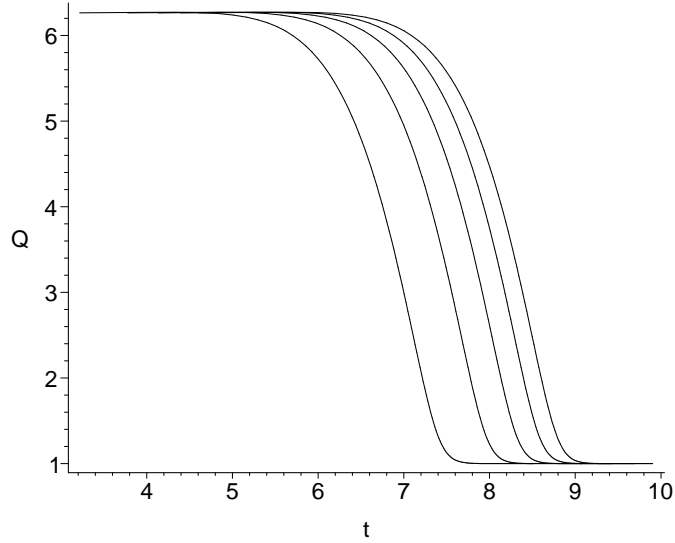}}
\end{center}
\caption{\footnotesize{Result of the numerical calculation of 
classical solutions 
$Q(\tau)$ for the values
$d=4$, $A=0.1$, $B=0.1$ and $l=1$ ($1/l_1=1$ and $1/l_2=6.24$).}} 
\label{numerical}
\end{figure}

Besides, as we explained before, 
the solution of interest to us is the one 
that converges to either AdS$^{(1)}$ or AdS$^{(2)}$ as $t\to+\infty$, 
satisfying the condition that $Q(t)$ be positive for the entire
region of $t$  
[see (\ref{BSconditionQ})]. 
It then turns out that
the classical solutions should behave as in
Figs.\,\ref{PGflow1} and \ref{PGflow2}.
In fact, a numerical analysis with the proper boundary condition 
at $t=+\infty$ 
indicates these behaviors upon choosing the branch $f_i^-(t)$ 
around $Q=1/l_i$. 
The result of the numerical calculation for $A>0$ and $B>0$ is 
shown in Fig.\,{\ref{numerical}}.

Now we give a holographic RG interpretation to the above results. 
We first consider the AdS$^{(1)}$ solution. 
Looking at the equation \eq{eom-exam},  
the equation (\ref{PG:linear}) is 
nothing but the equation of motion of a scalar field 
in the AdS background of radius $l$, with mass squared given by 
\begin{align}
 m_1^2 &= -{2\over A}\left(2\Lambda l^2+ {3B\over l^2}\right) \nn
       &= {2\over A}\left(d(d-1)-{6B\over l^2}\right).
\end{align}
Thus for $A>0$, the higher-derivative mode $Q$ is interpreted as 
a very massive scalar mode, 
and thus it is coupled to a highly irrelevant operator
around the fixed point,  
since its scaling dimension is given by 
\cite{GKP,W;holography}\footnote{
The exponent of the solution $f^-$ in (\ref{PG:linear_sln})
is equal to $d-\Delta$.}
\ba
 \Delta = {d\over2}+\sqrt{{d^4\over4}+l^2m_1^2} \gg d.
\ea
This can also be understood from Fig.\,\ref{PGflow1} 
which depicts a rapid convergence of the RG flow 
to the fixed point $Q(t)=1/l$. 
On the other hand, for $A<0$, the mass squared of the higher-derivative
mode is far below the lower bound for a scalar mode 
in the AdS$^{(1)}$ geometry, $-d^2/4l^2$ \cite{W;holography}, 
and the scaling dimension becomes complex. 
Thus, in this case, the higher-derivative mode makes 
the AdS$^{(1)}$ geometry unstable, 
and a holographic RG interpretation cannot be given to such a solution.

We note here that, 
to obtain the original CFT dual to the AdS$^{(1)}$ 
as the continuum limit is taken, $t\to -\infty$, 
we must fix the higher-derivative mode at the stationary point, 
$Q=1/l_1$. 
Roughly speaking, this is realized by tuning the boundary value 
of the conjugate momentum of the higher-derivative mode to be zero. 
In the next subsection, we adopt this boundary condition 
to derive the flow equation for the $R^2$ gravity theory.

We next consider the AdS$^{(2)}$. 
For $A>0$ and $B>0$ in Fig.\,\ref{PGflow1}, 
it can be seen that classical trajectories 
begin from AdS$^{(2)}$ to AdS$^{(1)}$.
In the context of the holographic RG, 
this means that the AdS$^{(2)}$ solution $Q(t)=1/l_2$ 
corresponds to a multicritical point 
in the phase diagram of the boundary field theory.
From (\ref{AdS_sln}) and (\ref{mass2}), 
the mass squared of the mode $Q$ around the
AdS$^{(2)}$ can be calculated as 
\ba
 m_2^2=-{2\over A}\left(d(d-1)-{6B\over l^2}\right),
\ea
and if this mass squared is above the unitarity bound,
\ba
 l_2^2m_2^2=-{6B\over A}{d(d-1)l^2-6B \over d(d-1)l^2-3B}
 > -{d^2 \over 4},
\ea
the scaling dimension of the corresponding operator is given by 
\begin{align}
 \Delta = {d\over 2}+\sqrt{{d^2\over 4}+l_2^2m_2^2} 
   \,\cong\, {d\over 2}+\sqrt{{d^2\over 4}-{6B\over A}}.
\end{align}
For example, if we consider the case in which $d=4$, $a=b=0$ and $c>0$,%
\footnote
{
This includes IIB supergravity on 
AdS${}_5\times S^5/{\bZ_2}$
which is AdS/CFT dual to ${\mathcal{N}}=2$ $USp(N)$ 
supersymmetric gauge theory
\cite{N=2CFT,BGN}.
} 
we have $A=32c>0$ and $B=8c/3>0$, and thus the scaling
dimension of $Q$ around the AdS$^{(2)}$ is found to be 
$\Delta\cong 2+\sqrt{7/2}$.
It would be interesting to investigate which conformal field theory
describes this fixed point. 

We conclude this subsection with a comment on the $c$-theorem. 
Since the trace of the extrinsic curvature, $\hK$, 
is given by $\hK\sim Q$ in the block spin gauge, 
we see from Eq.\ \eq{c-function} (or Eq.\ \eq{c-function2}) 
that the $c$-function \cite{HRG} is given by $c(Q)=Q^{1-d}$. 
{}Fig.\ 3 shows that it increases when $A>0$, 
but this does not contradict the assertion of the $c$-theorem, 
because in this case, the kinetic term of $Q(t)$ in the bulk action 
has a negative sign. 
This suggests that the obtained multicritical point 
defines a nonunitary theory like a Lee-Yang edge singularity. 


\subsection{Hamilton-Jacobi equation for a  higher-derivative
  Lagrangian}

In the previous subsection, we pointed out that 
the boundary value of the higher-derivative mode must be 
at a stationary point in order to implement
the continuum limit of the boundary field theory. 
To clarify this point further, in this subsection, 
we give a detailed discussion on 
the boundary conditions for higher-derivative modes 
that incorporate the idea of the holographic RG. 
We here discuss a point particle system, 
and will extend our analysis to systems of higher-derivative gravity 
in the next subsection.

A dynamical system with the action%
\footnote{
This $t$ is the coordinate value 
of the boundary and has nothing to do with the time variable 
in the block spin gauge.}
\ba
 \bS\bigl[q(\tau)\bigr]
  =\int_{t'}^{t} d\tau \,L\left(q,\dot{q},\ddot{q}\right) 
 \label{orig_action}
\ea
is described by the equation of motion which is 
a fourth-order differential equation in time $\tau$;
\ba
 {d^2\over d\tau^2}\left({\del L\over\del\ddot{q}}\right)
 -{d\over d\tau}\left({\del L\over\del\dot{q}}\right)
 +{\del L\over\del {q}} = 0. 
\ea
This implies that we need four boundary conditions 
to determine the classical solution uniquely. 
Possible boundary conditions can be found most easily 
by rewriting the system into the Hamiltonian formalism 
with an extra set of canonical variables $(Q,P)$ 
which represents $\dot{q}$ and its canonical momentum.

The Lagrangian in \eq{orig_action} is classically equivalent to 
\ba
 L'(q,Q,\dot{Q};p)=L\left(q,Q,\dot{Q}\right) + p\left(\dot{q}-Q\right),  
\ea
where $p$ is a Lagrange multiplier. 
We then carry out a Legendre transformation from $(Q,\dot{Q})$ 
to $(Q,P)$ through 
\ba
 P={\partial L'\over \partial\dot{Q}}
  \bigl(q,Q,\dot{Q};p\bigr)\,.
 \label{pn}
\ea
Assuming that this equation can be solved with respect to 
$\dot{Q}$ $\left(\equiv \dot{Q}\!\left(q,Q;P\right)\right)$, 
we introduce the Hamiltonian 
\ba
 H(q,Q;\,p,P)\equiv p\,Q+P\dot{Q}\!\left(q,Q;\,P\right)
  -\,L\left(q,Q,\dot{Q}\!\left(q,Q;\,P\right)\right), 
 \label{1st-action1}
\ea
and rewrite the action \eq{orig_action} in the first-order form; 
\ba
 \bS[q,Q;\,p,P]=\int^{t}_{t'} d\tau \left[
  p\,\dot{q}+P\dot{Q}-H(q,Q;\,p,P)\right], 
 \label{1st-action2}
\ea
where $\dot{Q}$ is now 
the time-derivative of the independent variable $Q$. 
The variation of the action \eq{1st-action2} reads
\begin{align}
 \delta\bS=&\int^t_{t^\prime} d\tau \left[\,
  \delta p\left( \dot{q}-{\partial H\over\partial p}\right)
  +\delta P\left(\dot{Q}-{\partial H\over\partial P}\right)
  \right. \nn 
  &\hspace{13mm}\left. 
  -\,\delta q
  \left( \dot{p}+{\partial H\over\partial q}\right)
  -\delta Q\left(\dot{P}
  +{\partial H\over\partial Q}\right)\right] \nn
  &~~+\left( p\,\delta q+ P\,\delta Q\right)\Big|^t_{t^\prime}\, , 
\end{align}
and thus the equation of motion consists of the usual Hamilton equations, 
\ba
 \dot{q}={\partial H\over\partial p},\quad
  \dot{Q}={\partial H\over\partial P},\quad
  \dot{p}=-{\partial H\over\partial q},\quad 
  \dot{P}=-{\partial H\over\partial Q}\,,
 \label{hamiltoneq}
\ea
plus the following constraints which must hold at the boundary, 
$\tau=t$ and $\tau=t^\prime$: 
\ba 
 p\,\delta q+P\,\delta Q =0   \quad (\tau=t, t')\,. 
 \label{bc}
\ea
The latter requirement, \eq{bc}, can be satisfied 
by imposing either Dirichlet boundary conditions, 
\ba
 {\rm \underline{Dirichlet}:}\qquad \delta q=0\,,\quad \delta Q=0\quad 
(\tau=t,t')\,,
\ea
or Neumann boundary conditions, 
\ba
 {\rm \underline{Neumann}:}\qquad p=0\,,\quad P=0\quad (\tau=t,t')\,,
\ea
for each variable $q$ and $Q$.
If, for example, we take the classical solution $(\overline{q},\overline{Q},
\overline{p},\overline{P})$ that satisfies the Dirichlet boundary conditions 
for all $(q,Q)$ with specified boundary values as 
\ba
 \overline{q}(\tau\!=\!t)=q,~\overline{Q}(\tau\!=\!t)
  =Q,\quad~{\rm and}\quad 
  \overline{q}(\tau\!=\!t^\prime)
  =q^\prime,~\overline{Q}(\tau\!=\!t^\prime)=Q^{\prime}\,,
\ea
then after plugging the solution into the action, 
we obtain the classical action that is a function of these boundary values, 
\ba
 S\bigl(t,q,Q;\,t^\prime,q^\prime,Q^{\prime}\bigr)=
  \bS\left[\overline{q}(\tau),\overline{Q}(\tau);
  \,\overline{p}(\tau),\overline{P}(\tau)\right].  
\ea
However, this classical action is not relevant to us 
in the context of the AdS/CFT correspondence, 
since 
we must further set the boundary value $Q$ of the higher-derivative mode 
to a stationary point in order to implement
the continuum limit of the boundary field theory. 
This requirement is equivalent to the condition 
that the higher-derivative mode has vanishing momentum. 
We are thus led to use mixed boundary conditions \cite{FMS2}: 
\ba
 \delta q=0 \quad {\rm and} \quad P=0\qquad (\tau=t,\,t')\,,
\ea
that is, we impose the Dirichlet boundary conditions for $q$ and 
Neumann boundary conditions for $Q$. 
In this case, the classical action 
(to be called the {\em reduced classical action}) 
becomes a function only of the boundary values $q$ and $q^\prime$: 
\ba
 S=S(t,q;\,t^\prime,q^\prime)\,. 
 \label{caction;toy}
\ea
If we further demand the regular behavior in taking $t\rightarrow+\infty$, 
the classical action depends only on the initial value. 
The same argument can be applied to dynamical systems 
of $(d+1)$-dimensional fields with higher-derivative interactions 
of arbitrary order \cite{FMS2}. 
{}Furthermore, the discussion in the previous subsection 
shows that higher-derivative modes should take stationary 
values in order to get a finite result in approaching the boundary. 
This supports our expectation that 
{\em for any bulk system of gravity with higher-derivative interactions, 
if we require the regularity inside the bulk 
and the finiteness near the boundary, 
the Euclidean time development is completely determined 
only by the boundary values of the original fields.} 
That is, the holographic nature still exists for higher-derivative systems.

Now we derive an equation that determines the 
reduced classical action (\ref{caction;toy}). 
This can be derived in two ways, 
and we first explain a more complicated, but straightforward, way 
since this gives us a deeper understanding of the mathematical
structure. 
To this end, we first change the
polarization of the system by performing the canonical 
transformation\footnote
{The following procedure corresponds to a change of representation 
from the $Q$-basis to the $P$-basis in the WKB approximation:
\ba
 \Psi(t,q,Q)=e^{i S(t,q,Q)/\hbar}\rightarrow
  \widehat{\Psi}(t,q,P)=e^{i \widehat{S}(t,q,P)/\hbar}\equiv
  \int dQ \,e^{-i P Q/\hbar}\,\Psi(t,q,Q)\,.\n
\ea
}
\ba
\widehat{\bS}\equiv\bS-\int^t_{t^\prime}d(PQ)\,. 
\ea
Although the Hamilton equation does not change under this transformation, 
the boundary conditions at $\tau=t$ and $\tau=t^\prime$ become 
\ba
 p\,\delta q- Q\delta P=0\quad(\tau=t,t')\,. 
\ea
These boundary conditions can be satisfied by imposing the Dirichlet 
boundary conditions for both $\overline{q}$ and $\overline{P}$: 
\ba
\overline{q}(\tau\!=\!t)=q,~\overline{P}(\tau\!=\!t)
  =P\,,\quad{\rm and}\quad 
\overline{q}(\tau\!=\!t^\prime)
  =q^\prime,~\overline{P}(\tau\!=\!t^\prime)=P^\prime\,. 
\ea
Substituting this solution into $\widehat{\bS}$, 
we obtain a new classical action that is a function 
of these boundary values, 
\ba
 \widehat{S}\left(t,q,P;\,t^\prime,q^\prime,P^\prime\right)
  =\widehat{\bS}\left[\overline{q}(\tau),\overline{Q}(\tau);\,
  \overline{p}(\tau),\overline{P}(\tau)\right]. 
\ea
By taking the variation of $\widehat{\bS}$ and using the equation of 
motion, we can easily show that the new classical action $\widehat{S}$ 
obeys the Hamilton-Jacobi equation: 
\ba
 {\partial\widehat{S}\over\partial t}&=&
  -H\left(q,-{\partial\widehat{S}\over\partial P};\,
 +{\partial\widehat{S}\over\partial q},P\right), \nn
  {\partial\widehat{S}\over\partial t^\prime}&=&
 +H\left(q^\prime,+{\partial\widehat{S}\over\partial P^\prime};\,
  -{\partial\widehat{S}\over\partial q^\prime},P^\prime\right). 
 \label{HJ;after}
\ea
The reduced classical action $S(t,q;t^\prime,q^\prime)$ is then obtained 
by setting $P\!=\!0$ in $\widehat{S}$: 
\ba
 S\left(t,q;t^\prime,q^\prime\right)=
  \widehat{S}\left(t,q,P\!=\!0;\,
  t^\prime,q^\prime,P^\prime\!=\!0\right). 
\ea
Note that the generating function $PQ$ vanishes at the boundary 
when we set $P\!=\!0$. 
In Appendix \ref{HowToSolve},  
we briefly describe how the Hamilton-Jacobi 
equation (\ref{HJ;after}) is solved for a system of a point particle.

In solving the full Hamilton-Jacobi equation, 
we must impose the regularity for $\widehat{S}(t,q,P)$
in the limit $c\!=\!0$ when $P\!=\!0$.  
This is because 
the higher-derivative term is regarded as a perturbation 
and the reduced classical action must have a finite limit 
for $c\!\rightarrow\!0$. 
One can see that the Hamilton-Jacobi equation reduces to an equation
involving the reduced action.
We call it a {\it Hamilton-Jacobi-like} equation.
However, once the regularity condition is imposed, 
we have an alternative way to derive 
the Hamilton-Jacobi-like equation
with greater ease. 
In fact, for any Lagrangian of the form 
\ba
 L(q^i,\dot{q}^i,\ddot{q}^i)=
  L_0(q^i,\dot{q}^i)+c\,L_1(q^i,\dot{q}^i,\ddot{q}^i)\,,
\ea
one can prove the following theorem, 
assuming that the classical solution can be expanded 
around $c\!=\!0$:\footnote
{As long as we think of $L_1$ as a perturbation, 
any classical solution can be expanded as 
\ba
 \bar{q}(\tau)=\bar{q}_0(\tau)+c\,\bar{q}_1(\tau)+{\cal O}(c^2)\,.\n
\ea
Here $\bar{q}_0$ is the classical solution for $L_0$, and 
$\bar{q}_1$ is obtained by solving a second-order differential 
equation. 
Note that we can, in particular, enforce the boundary conditions 
\ba
 \bar{q}_0(\tau\!=\!t)=q,\quad \bar{q}_1(\tau\!=\!t)=0~~{\rm and}~~ 
  \bar{q}_0(\tau\!=\!t^{\prime})=q^{\prime},\quad \bar{q}_1
 (\tau\!=\!t^{\prime})=0\,.
 \label{bc;simple}\n
\ea
In this case, due to the equation of motion for $\bar{q}_0(\tau)$\,, 
the classical action is simply given by 
\ba
 S(q,t;q^{\prime},t^{\prime})
  =\int^t_{t^{\prime}} d\tau\big[ L_0(\bar{q}_0,\dot{\bar{q}}_0)
  +c\,L_1(\bar{q}_0,\dot{\bar{q}}_0,\ddot{\bar{q}}_0)\big]+{\cal O}(c^2)\,.
\n
\ea
This corresponds to the classical action considered in Ref.\ \cite{BGN}.
}

\noindent
\underline{\bf Theorem} \cite{FMS2}\\
{\em Let $H_0(q,p)$ be the Hamiltonian corresponding to $L_0(q,\dot{q})$. 
Then the reduced classical action $S(t,q;\,t',q')
\!=\!S_0(t,q;\,t',q')+c\,S_1(t,q;\,t',q')+{\cal O}(c^2)$ 
satisfies the following equation up to ${\cal O}(c^2)$:}
\ba
 -{\partial S\over\partial t}=\tH(q,p), \quad 
  p_i={\partial S\over\partial q^i}, \quad{\rm and}\quad
  +{\partial S\over\partial t'}=\tH(q',p'), \quad 
  p'_i=-{\partial S\over\partial q^{\prime\,i}},
\label{theorem1}
\ea
{\em where} 
\ba
\tH(q,p)&\equiv&H_0(q,p)-c\,L_1(q,f_1(q,p),f_2(q,p)), \nn
f_1^i(q,p)&\equiv&\left\{H_0,q^i\right\}
  ={\partial H_0\over\partial p_i}, \nn
f_2^i(q,p)&\equiv&\left\{H_0,\left\{H_0,q^i\right\}\right\}
 ={\partial^2 H_0\over\partial p_i\partial q^j}
 {\partial H_0\over\partial p_j}
 -{\partial^2 H_0\over\partial p_i\partial p_j}
 {\partial H_0\over\partial q^j}.
\nn
 &&\left(\left\{F(q,p),G(q,p)\right\}\equiv
  \frac{\partial F}{\partial p_i}\frac{\partial G}{\partial q^i}-
  \frac{\partial G}{\partial p_i}\frac{\partial F}{\partial q^i}\right)
\label{theorem2}
\ea
We call $\tH$ a {\em pseudo-Hamiltonian}.

\noindent
A proof of this theorem is given in Appendix \ref{ProofofTheorem}. 
One can see easily that this correctly reproduces 
(\ref{HJ;c1}) and (\ref{HJ;c2}) for the Lagrangian given in
\eq{lag}--\eq{lag2}.

\subsection{Application to higher-derivative gravity}

Here we apply the formalism developed in the previous subsection 
to a system of higher-derivative gravity with the action \eq{PG:action}. 
We first derive the Hamilton-Jacobi-like equation of the system.
We also show that the coefficients $x_1,\cdots,x_5$ 
must obey some relations so that we can impose 
the mixed boundary condition consistently.

The action (\ref{PG:action}) 
is expressed in terms of the ADM parametrization as 
\ba
 \bS
 =\int_{\tau_0}^\infty d\tau \int d^d x \sqrt{\hg}\,
  \left[\bcL_0\bigl(\hg,\hK;\,N,\lambdah\bigr)
  +\bcL_1\bigl(\hg,\hK,\dot{\hK};\,\widehat{N},\lambdah\bigr)
  \right],
 \label{action}
\ea
where\footnote
{We here use the following abbreviated notation: 
 $\hK^n_{ij}\equiv \hK_{i_1}^{i_2}\hK_{i_2}^{i_3}\cdots \hK_{i_n}^{i_1},\,
 (\hK^2)_{ij}\equiv \hK_{ik}\hK^k_j$.
}
\ba
 \frac{1}{\widehat{N}}\,\bcL_0&=&2\Lambda-\hR
+\hK_{ij}^2-\hK^2, 
\ea
\ba
 \frac{1}{\widehat{N}}\,\bcL_1&=&-a\hR^2-b\hR_{ij}^2-c\hR_{ijkl}^2 
+\Big[(-6a+2x_1)\hK_{ij}^2+(2a-x_1)\hK^2\Big]\hR \nn
&&+\Big[-2(2b+4c-x_2)(\hK^2)_{ij}+(2b+2x_1-x_2)\hK\hK_{ij}\Big]\hR^{ij} \nn
&&+\,2(6c+x_2)\hK_{ik}\hK_{jl}\hR^{ijkl} \nn
&&-\,2(2b+c-3x_5)\hK_{ij}^4+(4b+4x_4-x_5)\hK\hK_{ij}^3 \nn
&&-\,(9a+b+2c-2x_4)\left(\hK_{ij}^2\right)^2
+(6a-b+6x_3-x_4)\hK^2\hK_{ij}^2 \nn
&&-\,(a+x_3)\hK^4 \nn 
&&-\,(4b+2x_1-x_2)\hK_{ij}\hnab^i\hnab^j\hK
+2(b-4c+x_2)\hK_{ij}\hnab^j\hnab_k\hK^{ki} \nn
&&+\,(8c+x_2)\hK_{ij}\hnab^2\hK^{ij}+2(b+x_1)\hK\hnab^2\hK \nn
&&-\Big[(4a+b)\hg^{ij}\hg^{kl}+(b+4c)\hg^{ik}\hg^{jl}\Big]\hL_{ij}\hL_{kl} \nn
&&+\bigg[\Big\{(4a-x_1)\hR+(12a+2b-x_4)\hK_{kl}^2-(4a+3x_3)\hK^2\Big\}\hg^{ij} 
\nn
&&~~~+(2b-x_2)\hR^{ij}+(4b+8c-3x_5)(\hK^2)^{ij}
-2(b+x_4)\hK\hK^{ij}\bigg]\hL_{ij}, 
\ea
with
\begin{equation}
  \hK_{ij}=\frac{1}{2\widehat{N}}
 \left(\dot{\hg}_{ij}-\hnab_i\lambdah_j-\hnab_j\lambdah_i\right),
 \label{kij}
\end{equation}
and 
\begin{equation}
 \hL_{ij}=\frac{1}{\widehat{N}}\left( \dot{\hK}_{ij}
  -\lambdah^k\,\hnab_k\hK_{ij}
  -\hnab_i\lambdah^k\,\hK_{kj}
  -\hnab_{j}\lambdah^k\,\hK_{ik}
  +\hnab_i\hnab_{j}\widehat{N}\right)\,. 
 \label{lij}
\end{equation}
For details of the ADM decomposition, see Appendix \ref{ADM-review}.

We now derive the Hamilton-Jacobi-like equation of $R^2$ gravity
by using the Theorem, (\ref{theorem1}) and (\ref{theorem2}).
We first rewrite the Lagrangian density of zero-th order, $\bcL_0$, 
into the first-order form 
\ba
 \bcL_0 \rightarrow 
  \hpi^{ij}\dot{\hg}_{ij}-\bcH_0\,, 
 \label{L0toH0}
\ea
where the zero-th order Hamiltonian density $\bcH_0$ is given by 
\ba
 \bcH_0\bigl(\hg,\hpi;\,\widehat{N},\lambdah\bigr)
  =\widehat{N}\,\left(\hpi_{ij}^2-{1\over d-1}\hpi^2-2\Lambda+\hR\right) 
  -2\lambdah_i\,\hnab_j\hpi^{ij}\,. 
\ea
Then by using the Theorem, the pseudo-Hamiltonian density 
is given by 
\ba
 \widetilde{\bcH}\bigl(\hg,\hpi;\,\widehat{N},\lambdah\bigr)
  =\bcH_0\bigl(\hg,\hpi;\,\widehat{N},\lambdah\bigr)
  -\bcL_1\bigl(\hg,\hK^0(g,\pi),
  \hK^1(\hg,\hpi);\,\widehat{N},\lambdah\bigr)\,. 
 \label{hj-like}
\ea
Here $\hK^0_{ij}(\hg,\hpi)$ is obtained by replacing 
$\dot{\hg}_{ij}(x)$ in \eq{kij}
with 
$\left\{\int d^d y \sqrt{\hg}\,\bcH_0(y),\,\hg_{ij}(x)\right\}$, 
and it is calculated to be 
\ba
  \hK^0_{ij}=\hpi_{ij}-{1\over d-1}\hpi\,\hg_{ij}\,. 
\ea
On the other hand, 
$\hK^1_{ij}\equiv\left\{\int d^d y \sqrt{\hg}\,\bcH_0(y),\,\hK^0_{ij}\right\}$ 
is found to be equivalent to replacing $\hL_{ij}$ in $\bcL_1$ 
by 
\begin{align}
 \hL^0_{ij} =&\, -{1\over 2(d-1)^2}\Bigl[ 
  2(d-1)\Lambda+(d-1)\hR+(d-1)\hpi_{kl}^2-3\hpi^2\Bigr] \hg_{ij} \nn 
  &~+\hR_{ij}+2(\hpi^2)_{ij}-{3\over d-1}\hpi\hpi_{ij}\,.
 \label{eqlast}
\end{align}
Using Eqs.\ \eq{L0toH0}--\eq{eqlast}, 
we obtain the following Hamilton-Jacobi-like equation 
for the reduced classical action \cite{FMS2}: 
\ba
 0&\!=\!&\int d^d x \sqrt{g}\,
  \widetilde{\bcH}\left(g(x),\pi(x);N(x),\lambda^i(x)\right) \nn
 &\!=\!&\int d^d x \sqrt{g}\,
  \left[N(x)\,\widetilde{\cH}(g(x),\pi(x))
  +\lambda^i(x)\,\widetilde{\cP}_i(g(x),\pi(x))\right]\,,
 \label{hj-like2a}
\ea
\ba
 \pi^{ij}(x) =\frac{-1}{\sqrt{g}}\frac{\delta S}{\delta g_{ij}(x)}\,, 
 \label{hj-like2b}
\ea
where\footnote
{We have ignored those terms in $\widetilde{\cH}$ 
that contain the covariant derivative $\nabla$. 
This is justified when we consider the holographic Weyl anomaly 
in four dimensions. 
Actually, it turns out that they give only total derivative terms 
in the Weyl anomaly. 
}
$g_{ij}$ and $\pi^{ij}$ are the boundary values of 
$\hg_{ij}$ and $\hpi^{ij}$, respectively, and 
\ba
 \widetilde{\cH}(g,\pi)
  &\equiv&\pi_{ij}^2-{1\over d-1}\pi^2-2\Lambda+R\nn
 &&+\,\alpha_1\,\pi_{ij}^4+\alpha_2\,\pi\pi_{ij}^3
  +\alpha_3\left(\pi_{ij}^2\right)^2+\alpha_4\,\pi^2\pi_{ij}^2
  +\alpha_5\,\pi^4 \nn
 &&+\,\beta_1\,\Lambda\pi_{ij}^2+\beta_2\,\Lambda\pi^2+\beta_3\,R\,\pi_{ij}^2
  +\beta_4\,R\,\pi^2\nn
 &&+\,\beta_5\,R_{ij}(\pi^2)^{ij}
  +\beta_6\,R_{ij}\,\pi\pi^{ij}
  +\beta_7\,R_{ijkl}\,\pi^{ik}\pi^{jl} \nn
 &&+\,\gamma_1\,\Lambda^2+\gamma_2\,\Lambda R+\gamma_3\,R^2+\gamma_4\,R_{ij}^2
  +\gamma_5\,R_{ijkl}^2\,,
 \label{hj-like2c}\\
 \widetilde{\cP}_i(g,\pi)&\equiv&-2\nabla^j\pi_{ij}\,,
 \label{hj-like2d}
\ea
with 
\begin{align}
\alpha_1 = &\, 2c, \quad \alpha_2={2x_5 \over (d-1)}, \nn
\alpha_3 =&\, {1\over 4(d-1)^2}
\Big[ 4a+(d^2-3d+4)b+4(d-2)(2d-3)c \nn
&\qquad\qquad~~ -2(d-1)(dx_4+3x_5) \Big], \nn
\alpha_4 =&\, {1\over 2(d-1)^3}\left[ 
-4a-(d^2-3d+4)b-4(2d^2-5d+4)c \right. \nn
&\left.\qquad\qquad~~ -3dx_3+(2d^2-7d+2)x_4-3(2d-1)x_5\right],\nn
\alpha_5 =&\, {1\over 4(d-1)^4}\left[ 
4a+(d^2-3d+4)b+4(2d^2-5d+4)c \right. \nn
&\left.\qquad\qquad~~ +2(3d-4)x_3-2(d^2-6d+6)x_4+2(5d-6)x_5\right], 
\end{align}
\begin{align}
\beta_1= &\, {1\over (d-1)^2}\Big[ 4da-d(d-3)b-4(d-2)c 
-(d-1)(dx_4+3x_5)\Big],\nn
\beta_2 =&\, {1\over (d-1)^3}\Big[-4da+d(d-3)b+4(d-2)c \nn
&\qquad\qquad~~-3dx_3+(d^2-2d-2)x_4+3(d-2)x_5 \Big], \nn
\beta_3 =&\, {1\over 2(d-1)^2}\Big[4a+(d^2-3d+4)b-4(3d-4)c \nn
&\qquad\qquad~~-(d-1)(dx_1+x_2-(d-2)x_4+3x_5)\Big], \nn
\beta_4 =&\, {1\over 2(d-1)^3}\Big[-4a-(d^2-3d+4)b+4(d-2)c \nn 
&\qquad\qquad~~-(d-1)(d-4)x_1-3(d-1)x_2+3(d-2)x_3\nn
&\qquad\qquad~~-(d^2-8d+10)x_4 +3(3d-4)x_5\Big], \nn
\beta_5 =&\, 16c+3x_5,\qquad 
\beta_6={2(x_1+2x_2-x_4-3x_5)\over d-1},\qquad
\beta_7=-12c-2x_2,
\end{align}
\ba
\gamma_1&=&{d\over (d-1)^2}\Big[ 4da+(d+1)b+4c \Big],\nn
\gamma_2&=&{1\over (d-1)^2}\Big[ 4da-d(d-3)b-4(d-2)c 
-(d-1)(dx_1+x_2)\Big],\nn
\gamma_3&=&{1\over 4(d-1)^2}\Big[ 4a+(d^2-3d+4)b-4(3d-4)c 
+2(d-1)((d-2)x_1-x_2)\Big],\nn
\gamma_4&=&4c+x_2,\qquad \gamma_5=c. 
\ea
Here $R_{ijkl}$ is the Riemann tensor made 
of the metric tensor of the $d$-dimensional boundary 
$\tau=\tau_0$.
Since the (true) classical action $\hS[g(x),P(x)]$ 
is independent of the choice of $N$ and $\lambda^i$ 
(and thus, so is $S[g(x)]$), 
from Eqs.\ \eq{hj-like2a}--\eq{hj-like2d} 
we finally obtain the following equation that determines the reduced 
classical action: 
\ba
 \widetilde{\cH}\bigl(g_{ij}(x),\pi^{ij}(x)\bigr)=0\,,\quad 
  \widetilde{\cP}_i\bigl(g_{ij}(x),\pi^{ij}(x)\bigr)=0\,,\quad 
  \pi^{ij}(x)=\frac{-1}{\sqrt{g}}\frac{\delta S}{\delta g_{ij}(x)}\,.
 \label{basic}
\ea

We make a few comments 
on the possible form of the boundary action $\bS_b$ 
and the cosmological constant $\Lambda$. 
As discussed above, in order that the boundary field theory 
has a continuum limit, 
the geometry must be asymptotically AdS: 
\ba
 ds^2\rightarrow d\tau^2+e^{-2\tau/l}\eta_{ij}(x)dx^idx^j
\quad {\rm for}~~\tau\rightarrow -\infty. 
\label{asymp;ads}
\ea
This should be consistent with our boundary condition 
$P^{ij}\!=\!0$. 
Explicitly investigating the equations of motion 
derived from the action \eq{action}, 
we can show that this compatibility gives rise to the relation 
\ba
 d^2\,x_3+d\,x_4+x_5&=&-{4\over 3}\Big( d(d+1)a+db+2c\Big). 
 \label{coef}
\ea
It can also be shown that the asymptotic behavior (\ref{asymp;ads}) 
determines the cosmological constant $\Lambda$ as
\ba
 \Lambda=-{d(d-1)\over 2l^2}+{d(d-3)\over2l^4}
 \Big[ d(d+1)a+db+2c \Big]. 
 \label{cc}
\ea

\subsection{Solution to the flow equation and the Weyl anomaly}

We first note that the basic equation, \eq{basic}, 
can be rewritten as a flow equation of the form \cite{FMS2}
\ba
 \{ S, S \}+\{ S,S,S,S \} ={\cal L}_d, 
\ea
with 
\begin{align}
 \left( \sqrt{g}\right)^2\,\{ S, S \} \equiv &\, \left[  
  \left({\delta S\over\delta g_{ij}}\right)^2
  -{1\over d-1}\left( g_{ij}{\delta S\over \delta g_{ij}} \right)^2 
  \right. \nn
 &\left.
  ~+ \beta_1\,\Lambda \left( {\delta S \over \delta g_{ij}} \right)^2
  +\beta_2\,\Lambda \left( g_{ij}{ \delta S \over \delta g_{ij}} \right)^2
  +\beta_3\, R \left( { \delta S \over \delta g_{ij}} \right)^2
\right.\nn
&\left.
  ~+ \beta_4\, R \left( g_{ij}{ \delta S \over \delta g_{ij}} \right)^2
  +\beta_5\, R_{ij}g_{kl} 
   { \delta S \over \delta g_{ik}}{ \delta S \over \delta g_{jl}} 
\right.\nn
&\left.
  ~+ \beta_6\, R_{ij} { \delta S \over \delta g_{ij}}
  \,g_{kl}{ \delta S \over \delta g_{kl}} 
  +\beta_7\, R_{ijkl} { \delta S \over \delta g_{ik}}
  \,{ \delta S \over \delta g_{jl}} \right], 
\end{align}
\ba
 \left( \sqrt{g}\right)^4 \{ S,S,S,S \}&\equiv&\left[ 
  \alpha_1 \left({\delta S\over\delta g_{ij}}\right)^4
  +\alpha_2 \left(g_{kl}{\delta S\over\delta g_{kl}}\right)
  \left({\delta S\over\delta g_{ij}}\right)^3
  +\alpha_3 \left(\left({\delta S\over\delta g_{ij}}\right)^2\right)^2 
  \right.\nn
 &&\left. ~+ \alpha_4 
   \left(g_{kl}{\delta S\over\delta g_{kl}}\right)^2
   \left({\delta S\over\delta g_{ij}}\right)^2
  +\alpha_5 \left(g_{ij}{\delta S\over\delta g_{ij}}\right)^4 \right], 
\ea
\ba
 {\cal L}_d&\equiv&2\Lambda-R-\gamma_1\Lambda^2-\gamma_2\Lambda 
R
  -\gamma_3 R^2-\gamma_4R_{ij}^2-\gamma_5R_{ijkl}^2. 
\ea
As in \S 3, we decompose the reduced classical action 
into the local part and the non-local part, 
\begin{equation}
 \frac{1}{2\kappa_{d+1}^2}S[g(x)]
  =\frac{1}{2\kappa_{d+1}^2}S_{\rm loc}[g(x)]-\Gamma[g(x)]\,.
\end{equation}
Following the prescription given in \S 3, 
we first determine the weight $0$ and $2$ parts of the $S_{\rm loc}$, 
\ba
  \left[\cLloc\right]_0= W\,, \qquad
  \left[\cLloc\right]_2= -\Phi\,R,
\ea
\begin{align}
 W =&\, -{2(d-1)\over l}+{1\over l^3}\Big[
  -4d(d+1)a-4db-8c+d(d^2x_3+dx_4+x_5)\Big], \nn
 \Phi =&\, {l\over d-2}-
  {2\over (d-1)(d-2)\,l}\Big[ d(d+1)a+d\,b+2c\Big]\nn
 &+{1\over l}\left[d\,x_1+x_2
  +{3(d^2x_3+d\,x_4+x_5)\over 2(d-1)}\right], 
 \label{wphi}
\end{align}
where (\ref{cc}) has been used.

For $d=4$, the weight $4$ part of the flow equation is 
an equation that the generating functional $\Gamma$ obeys, 
\begin{eqnarray}
&&2\Big[\{ \Sloc,\,\Gamma\}\Big]_4 
+4\Big[\{ \Sloc,\,\Sloc,\,\Sloc,\,\Gamma\}\Big]_4\nn
&&=\frac{1}{2\kappa_5^2}\left(
 \Big[\{ \Sloc,\,\Sloc\}\Big]_4
 +\Big[\{ \Sloc,\,\Sloc,\,\Sloc,\,\Sloc\}\Big]_4 
\right.\nn
&&\hspace{35mm}\hspace{-15mm}+\,\gamma_3R^2+\gamma_4R_{ij}^2
+\gamma_5R_{ijkl}^2\Bigr). 
\end{eqnarray}
{}From this, we can evaluate the trace of the stress tensor 
for the boundary field theory: 
\ba
 \langle T^i_i \rangle_g\equiv 
  {2\over\sqrt{g}}\,g_{ij}{\delta\Gamma\over\delta g_{ij}}.
\ea
In fact, using the values in (\ref{wphi}), 
we can show that the trace is given by \cite{FMS2}
\begin{align}
 \langle T^i_i\rangle_g=
  {2l^3\over 2\kappa_5^2}\Biggl[&
  \left({-1\over 24}+{5a\over 3l^2}+{b\over 3l^2}
  +{c\over 3l^2}\right)R^2  +\left( {1 \over 8}-{5a\over l^2}
  -{b\over l^2}-{3c\over 2l^2}\right)R_{ij}^2  \nn 
  &+{c\over 2l^2}R_{ijkl}^2 
  \Biggr]\,.
 \label{anomaly;bulk}
\end{align}
This correctly reproduces the result\footnote
{The authors of Refs.\ \cite{BGN} and \cite{NO} parametrized 
the cosmological constant $\Lambda$ as 
\ba
 \Lambda=-{d(d-1)\over 2 L^2}\,, \n
\ea
so that their $L$ is related to our $l$, the radius of asymptotic AdS, as 
\ba
 l^2=L^2\left[ 
  1-{(d-3) \over (d-1)L^2}\big( d(d+1)a+db+2c\big)\right].\n
\ea
}
obtained in Refs.\ \cite{BGN} and \cite{NO}, 
where the Weyl anomaly was calculated by perturbatively solving 
the equation of motion near the boundary and by looking at 
the logarithmically divergent term, as in Ref.\ \cite{HS;weyl}.

For the case of ${\cal N}\!=\!2$ superconformal $USp(N)$ gauge theory 
in four dimensions, 
we choose $2\kappa_5^2$ such that 
\begin{equation}
 {1\over 2\kappa_5^2}=
  {{\rm Vol}(S^5/\bZ_2)\,({\rm radius~of~}S^5/\bZ_2)^5\over 2\kappa^2},
 \label{5newton}
\end{equation}
where $2\kappa^2=(2\pi)^7g_s^2$ is the ten-dimensional Newton constant 
\cite{Pol}, 
and the radius of $S^5/\bZ_2$ could be set to 
$(8\pi g_sN)^{1/4}$ \cite{APTY}. 
In this relation, we note the replacement $N \to 2N$ as compared to
the $AdS_5\times S^5$ case. 
This is because here we must quantize 
the RR $5$-form flux over $S_5/\bZ_2$ instead of over $S^5$ \cite{N=2CFT}. 
For the $AdS_5$ radius $l$, 
we may also set $l=(8\pi g_s N)^{1/4}$. 
Setting the values $a=b=0$ and 
$c/2l^2=1/32N+{\cal O}(1/N^2)$, as determined in Ref.\ \cite{BGN}, 
we find that the Weyl anomaly (\ref{anomaly;bulk}) takes the form 
\ba
 \langle T^i_i\rangle_g&=&
  {N^2\over 2\pi^2}\left[ 
  \left({-1\over 24}+{1\over 48N}\right)R^2
  +\left({1\over 8}-{3\over 32N}\right)R_{ij}^2
  +{1\over 32N}R_{ijkl}^2\right] 
 +{\cal O}(N^0)\,. \nn
\ea
This is different from the field theoretical result \cite{Duff;Weyl}, 
\ba
 \langle T^i_i\rangle_g&=&
  {N^2\over 2\pi^2}\left[ 
  \left({-1\over 24}-{1\over 32N}\right)R^2
  +\left({1\over 8}+{1\over 16N}\right)R_{ij}^2
  +{1\over 32N}R_{ijkl}^2\right] 
 +{\cal O}(N^0)\,. \nn
\label{anomaly;bgn}
\ea
As was pointed out in Ref.\ \cite{BGN}, 
the discrepancy could be accounted for 
by possible corrections to the radius $l$ 
as well as to the five-dimensional Newton constant. 
In fact, if these corrections are 
\ba
 l=(8\pi g_s N)^{1/4}\left(1+\frac{\xi}{N}\right)\,,\quad
 \frac{1}{2\kappa_5^2}= 
  \frac{{\rm Vol}(S^5/\bZ_2)\,(8\pi g_s N)^{5/4}}{2\kappa^2}
  \left( 1+{\eta\over N}\right)\,,
\ea
then the field theoretical result is correctly reproduced 
for $3\xi+\eta=5/4$. 

\resection{Conclusion}

In this article, we have investigated various aspects of 
the AdS/CFT correspondence and the holographic renormalization group 
(RG). 

In \S 2,
we gave a review of the basic idea of the AdS/CFT 
correspondence and the holographic RG, 
and calculated the scaling dimensions of the scaling operators 
which are dual to bulk scalar fields in the AdS background. 
As a typical example of the AdS/CFT correspondence, 
we considered the duality between 
the $\cN=4$ $SU(N)$ SYM$_4$ 
and Type IIB supergravity on AdS$_5\times S^5$.
As a consistency check for the duality,
we showed the one-to-one 
correspondence between 
the short chiral primary multiplets of the CFT 
and the Kaluza-Klein spectra of supergravity. 
We also demonstrated the holographic description of 
RG flows that interpolate between a UV and an IR fixed points, 
by considering the example of an RG flow 
from the $\cN=4$ $SU(N)$ SYM$_4$ to 
the $\cN=1$ Leigh-Strassler fixed point.
The ``c-function'' was defined from the view point 
of the holographic RG, and shown to obey an analog of
Zamolodchikov's c-theorem.

In \S 3, 
we explored the formulation of the holographic RG 
based on the Hamilton-Jacobi equation of bulk gravity 
given by de Boer, Verlinde and Verlinde. 
A systematic prescription for calculating the Weyl anomaly of the 
boundary CFT was proposed. 
We also derived the Callan-Symanzik equation for $n$-point functions 
in the boundary field theory. 
We calculated the scaling dimensions of scaling operators 
from the  coefficients of the RG beta functions, 
and showed that they are in precise agreement with known results 
in the AdS/CFT correspondence. 
We explained how we take the continuum limit of the boundary field
theory, 
and concluded that the holographic RG describes 
the so-called renormalized trajectory.

We discussed the holographic RG in the framework of the 
noncritical string theory in \S 4. 
In the holographic RG, 
we must introduce an IR cutoff to regularize the 
infinite volume of the bulk space-time, and 
the (Euclidean) time development 
of fields in the gravity theory is required to be regular 
interior of the bulk.  
We demonstrated that this basic requirement in the holographic RG can be 
understood naturally in the context of noncritical strings.

In \S 5, 
the holographic RG for $R^2$ gravity was investigated. 
In general, when we work
in the Hamiltonian formalism, 
we must introduce new valuables 
which we call the ``{\em higher-derivative modes}.'' 
We introduced a parametrization of the metric 
in which the Euclidean time evolution of the system can be 
directly interpreted as an RG transformation 
of the boundary field theory. 
We examined classical solutions of the system 
under this parametrization. 
We found that the stability of an AdS solution depends on 
the coefficients of the curvature squared terms, and  
the fluctuation of the higher-derivative mode around 
a stable AdS solution is interpreted as a very massive scalar field 
in the background of the AdS space-time. 
In the AdS/CFT correspondence, 
this means that 
the fluctuation of the higher-derivative 
mode corresponds to a highly irrelevant operator of the 
boundary CFT.   
Thus, we must fix the boundary values of higher-derivative 
modes at stationary values 
in order to implement the continuum limit of the boundary field theory.   
We discussed that the condition is automatically satisfied by 
adopting the mixed boundary condition, that is, 
the Dirichlet boundary condition for the usual valuables and 
the Neumann boundary condition for the higher-derivative modes. 
We also discussed that when the coefficients of the curvature squared
terms satisfy an appropriate condition, 
there appears another conformal fixed point in the parameter space 
of the boundary field theories. 

Using the prescription with the mixed boundary conditions, 
we derived a Hamilton-Jacobi-like equation for $R^2$ gravity 
which describes RG flows of the dual field theory. 
As an application, we calculated the $1/N$ correction of the 
Weyl anomaly of $\cN=2$ $USp(N)$ supersymmetric gauge theory
in four dimensions.
We found that the result is consistent with a field theoretical calculation.

We here make a comment on 
field redefinitions of bulk gravity in the context of the 
AdS/CFT correspondence \cite{FM2}. 
The AdS/CFT correspondence should have the property that 
any physical quantities of the $d$-dimensional boundary field theory 
calculated from $(d+1)$-dimensional bulk gravity are invariant 
under field redefinitions of the fields 
in ten-dimensional supergravity. 
This is because ten-dimensional 
classical supergravity represents the on-shell 
structure of massless modes of superstrings, 
and the on-shell amplitudes (more precisely, the residues of 
one-particle poles of correlation functions 
for external momenta) should be invariant under 
redefinitions of fields \cite{KOS} (see also Ref.\ \cite{GW} 
for discussions in the context of string theory).%
\footnote{See also Ref.\ \cite{LM} for recent 
discussion about scheme independence in the renormalization group structure.}  

As an example, let us show \cite{FM2} that 
the holographic Weyl anomaly of the $\cN=4$ $SU(N)$ SYM$_4$ 
does not change under the field redefinition of the ten-dimensional 
metric of the form 
\ba
 \bG_{MN} \to \bG_{MN}' \equiv  \bG_{MN}+\alpha \bR\bG_{MN}+\beta\bR_{MN}. 
  \label{redef}
\ea
The bosonic part of the ten-dimensional Type IIB supergravity action 
is given by 
\ba
\bS_{10}=\frac{1}{2\kappa_{10}^2}\int d^{10}X \sqrt{-\bG}
\left[e^{-2\phi}\left(\bR+4\left|d\phi\right|^2\right)
-{1\over4}\left|F_5\right|^2\right].
\label{IIB-1}
\ea
In the context of the AdS$_5$/CFT$_4$ correspondence, 
we are interested in the AdS$_5$$\times$$S^5$ 
solution that is realized 
as the near horizon limit of the black 3-brane solution:  
\begin{gather}
 ds^2 = \frac{l^2}{r^2}\,dr^2
        +\frac{r^2}{l^2}\,\eta_{ij}\,dx^idx^j
        +l^2 \,d\Omega_5^2, \nn
 (F_5)_{r0123}= -{4\over g_s}\,{r^3\over l^4}, \quad
 (F_5)_{y^1\cdots y^5}={4\over g_s}\,l^4, \nn
 e^{\phi}=g_s. 
  \label{sln-1}
\end{gather}
Here, $d\Omega_5^2$ is 
the metric of the unit five-sphere and $i,j\in\{0,1,2,3\}$.
In this case,
the AdS$_5$ and $S^5$ have the same radius, $l$, 
whose value is determined by the D3-brane charge as 
\ba
 l=(4\pi g_s N)^{1/4}, 
\label{radius}
\ea
where $N$ is the number of the coincident D3-branes, and we have set 
the string length $l_s$ to $1$.
The action of the effective five-dimensional gravity is given by 
compactifying the ten-dimensional action \eq{IIB-1} on $S^5$: 
\ba
 \bS_5 = {\pi^3l^5\over 2\kappa_{10}^2g_s^2}\int d^5x
 \sqrt{-\hg}\left({12\over l^2}+\hR \right). 
\ea
The holographic Weyl anomaly calculated from this action is 
given in \eq{result;weyl}, which reproduces the Weyl anomaly of 
the $\cN=4$ $SU(N)$ SYM$_4$ as mentioned in \S 3.3.

On the other hand, if we make the field redefinition \eq{redef}, 
the obtained new ten-dimensional gravity action is 
\begin{align}
 \widetilde{\bS}_{10}[&\bG_{MN}] \equiv 
 \bS_{10}[\bG_{MN}+\alpha \bR\bG_{MN}+\beta\bR_{MN}] \nn
 &=\frac{1}{2\kappa_{10}^2}\int d^{10}X \sqrt{-\bG}
 \Biggl\{
  e^{-2\phi}\biggl[\bR+4\left|d\phi\right|^2
 +a\bR^2 +b\bR_{MN}^2 \nn
 &\qquad\qquad\qquad\qquad\qquad\qquad\quad
 +a\bR\left|d\phi\right|^2 
  +b\,\bR^{MN}\partial_M\phi\,\partial_N\phi\biggr] \nn
  &\qquad\qquad-{1\over4}\left|F_5\right|^2
  +{b\over8}\,\bR\left|F_5\right|^2
  -{b\over4}\,{1\over 4!}\,\bR_{MN}(F_5)^{MPQRS}(F_5)^{N}_{\ PQRS}
 \Biggr\}.
 \label{IIB-2}
\end{align}
Here $a$ and $b$ are defined as
\ba
 a=4\alpha+{1\over2}\beta,\quad
 b=-\beta.
\ea
The AdS$_5\times S^5$ solution for the action \eq{IIB-2} is 
given by 
\begin{gather}
 ds^2 = \left(1-{8b\over {l'}^2}\right){{l'}^2\over r^2}\,dr^2
 +{r^2\over {l'}^2}\,\eta_{ij}\,dx^idx^j
 +{l'}^2 d\Omega_5^2, \nn
 (F_5)_{r0123} = {4\over g_s}\left(1+{8b\over {l'}^2}\right){r^3\over {l'}^4},
 \quad 
 (F_5)_{y^1\cdots y^5} = 
 {4\over g_s}\left(1-{8b\over {l'}^2}\right){l'}^4, \nn 
 e^{\phi} = g_s, 
\label{sln-2}
\end{gather}
where the new radius of the $S^5$ is related to $l$ by 
\ba
 l'=\left(1+\frac{2b}{l^2}\right)l.
\ea 
Note that after the field redefinition, the radius of $S^5$, $l'$,  
differs from that of AdS$_5$, $L$, which is expressed as  
\ba
 L \equiv \left(1-{4b\over {l'}^2}\right)l'
  =\left(1-{2b\over l^2}\right)l.
\ea
From the solution (\ref{sln-2}), 
we compactify ten-dimensional spacetime on $S^5$ of radius $l'$. 
Then, the (dimensionally reduced) five-dimensional action is obtained as  
\begin{align} 
 \widetilde{\bS}_5 = \frac{\pi^3{l'}^5}{2\kappa_{10}^2g_s^2}
 &\left(1+{40a+4b\over {l'}^2} \right) \times \nn
 &\int d^5x \sqrt{-\hg} 
 \left[\left({12\over {l'}^2}-{80a-80b\over {l'}^4}\right)
 +\hR+a\hR^2+b\hR_{\mu\nu}^2\right].
\end{align}
This action has an AdS$_5$ solution with radius 
$\left(1-4b/{l'}^2\right)l'$, which is consistent with the 
AdS$_5\times S^5$ solution (\ref{sln-2}). 
The corresponding Weyl anomaly is calculated 
by using the formula \eq{anomaly;bulk} as
\begin{align}
 \langle T_i^i \rangle &= 
 {2L^3\over 2\kappa_5^2}\left(1-{40a+8b\over {l'}^2}\right)
 \left(-{1\over24}R^2+{1\over8}R_{ij}^2\right) \nn
 &= \frac{2\pi^3{l'}^8}{2\kappa_{10}^2g_s^2}
 \left(1-{16b\over {l'}^2}\right)
 \left(-{1\over24}R^2+{1\over8}R_{ij}^2\right) \nn
 &=\frac{2\pi^3l^8}{2\kappa_{10}^2g_s^2}
 \left(-{1\over24}R^2+{1\over8}R_{ij}^2\right) \nn
 &= {N^2\over 4\pi^2}\left(-{1\over24}R^2+{1\over8}R_{ij}^2\right).
\end{align}
This is identical to the result (\ref{result;weyl}).

We conclude this article by making a few comments on future directions
in the AdS/CFT correspondence and the holographic RG.

Once we start with AdS$_{d+1}$ gravity with $d\ge 4$, the dual 
$d$-dimensional conformal field theory is in general at a 
non-trivial fixed point, 
because operators of dual CFT coupled to bulk modes 
have non-trivial anomalous dimensions. 
It is thus natural to conjecture that any CFT in higher dimensions
which has an AdS dual is a non-abelian gauge theory.%
\footnote{
The situation is different when $d\le 3$. 
Actually, an AdS$_4$ dual of the the critical $O(N)$ 
vector model in three dimension is proposed in Ref.\ \cite{KP;vec}.}
In fact, all the known examples of the AdS/CFT correspondence involve
non-abelian gauge theories. 
Furthermore, a non-trivial fixed point for $d\ge 4$ seems unlikely
besides non-abelian gauge theories because of triviality.
It would be nice to study the conjecture in more detail.
In particular, it is interesting to investigate 
if there is a chance to gain information on 
the gauge symmetry of the boundary theory only from bulk supergravity.

The equation \eq{hj;potential} seems to imply some hidden
symmetry in bulk.
In fact, the form of \eq{hj;potential} is reminiscent of a scalar
potential of supergravity with $W(\phi)$ a ``superpotential.''
Moreover, as pointed out in Ref.\ \cite{HRG}, holographic RG flows can be
described by first-order differential equations via the superpotential.
These facts might suggest that bulk gravity has 
a hidden supersymmetry or some novel symmetry.

To show the gauge/string duality from the loop equations of 
the Yang-Mills theory \cite{Pol;loop,MM;loop} 
is an old but fascinating idea \cite{Pol;96}-\cite{Pol;01}. 
A strong coupling analysis in lattice gauge theory 
\cite{wilson;74,Kogut:ag} shows that
elementary excitations in gauge theory are strings of color flux,
and the interaction of strings would be suppressed in the large $N$ limit, 
as mentioned in Introduction. 
It is thus reasonable that 
we can describe a gauge theory in terms of strings of 
color flux.
In this framework, a gauge theory would be described by 
the Wilson loop; 
\begin{equation}
 W[C(s)] = \biggl\langle 
  {\rm Tr}{\mathcal{P}}\exp\left(i\oint_C dx^i A_i\right)
   \biggr\rangle,  \qquad \left(0\le s \le 2\pi\right)
   \label{Wloop}
\end{equation}
where $s$ parametrizes the contour $C$. 
The Wilson loop \eq{Wloop} has a reparametrization invariance 
$s\to s'(s)$. 
Here we can allow for the $s'(s)$ to ``go backward'' 
on the way of $s\in\left[0,2\pi\right]$, that is, 
$ds'(s)/ds$ can vanish at some $s$. 
This characteristic symmetry of the Wilson loop is called 
the {\em zigzag symmetry} \cite{Pol;zigzag}.  
{}Fundamental equations that characterize the Wilson loops are 
the loop equations, and written schematically as  
\begin{equation}
 \hat{L}(s)W[C] = W*W, 
\label{loopeq}
\end{equation} 
where $\hat{L}$ is the loop Laplacian and the right-hand side 
represents the interaction of two loops 
(or intersection of a single loop) at a single point. 
For an accurate definition of the loop equations, see the literature 
\cite{Pol;loop,MM;loop}.

The equivalence between gauge theory and string theory 
means that there is an open string with its ends on the loop $C$ 
such that the functional $W[C]$ defined by 
\begin{equation}
 W[C]=\int {\mathcal{D}}x^i \,{\mathcal{D}}\varphi \,
  e^{-\Sh\left[x^i,\varphi\right]}
  \qquad \left(i=1,\cdots,4\right) 
\end{equation}
satisfies the loop equation \eq{loopeq} and has the zigzag symmetry. 
Here $\varphi$ and $x^i$ express the Liouville field and 
matter fields on the string world-sheet, respectively. 
So far, lots of efforts have been made 
to find the duality.  
For example, in Ref.\,\cite{Pol;zigzag}, it is argued that 
world sheet supersymmetry eliminates boundary tachyonic modes  
and the zigzag symmetry is to be expected.%
\footnote{ 
We expect that this world-sheet supersymmetry might be enhanced to  
the space-time hidden supersymmetry mentioned above. 
}
It would be nice to pursue these ideas to gain a deeper
insight into the gauge/string correspondence.

As discussed in \S 2.4, the Penrose limit of AdS$_5\times S^5$
leads us to the maximally supersymmetric pp-wave background, 
on which string theory is exactly solvable in the light-cone gauge.
From the exact result of the string spectra, 
Berenstein, Maldacena and Nastase made a prediction 
about the anomalous dimensions of
${\cal N}=4$ SYM composite operators for $N,J \gg 1$ with $N/J^2$ fixed, 
expressed as exact functions of $\lambda=4\pi g_sN=g_{\rm YM}^2N$.
In order to confirm this pp-wave/CFT correspondence, we have to compute
the exact anomalous dimensions from the field theory side.
That computation was done in Ref.\ \cite{SZ;02}, reproducing the exact
anomalous dimensions. (For a related work, see 
Ref.\ \cite{OS;quiver}).
So the pp-wave/CFT correspondence is justified beyond the supergravity
approximation. 
One of the problems there, however, 
is that the holography is not manifest in the pp-wave backgrounds.
Since a Penrose limit zooms in the local geometry near a null
geodesic of a given background, the resulting background has a totally
different boundary compared to the original one.
Thus the holographic rules in the AdS/CFT correspondence are no longer
valid in the pp-wave backgrounds.
Although several attempts have been made to understand 
how the holography works in the pp-wave backgrounds 
\cite{Das;02,LOR;02,BN;02}, 
there still remain a lot of issues to be clarified. 
In particular, it might be possible to formulate 
the holographic principle on a pp-wave background 
beyond the supergravity approximation 
because the string theory on it is simple enough.

\section*{Acknowledgements}
The authors would like to thank 
T.\ Asakawa, T.\ Fukuda, K.\ Hosomichi, Y.\ Hyakutake, H.\ Kawai, 
T.\ Kobayashi, T.\ Kubota, H.\ Kunitomo, A.\ Miwa, S.\ Nakamura, 
M.\ Ninomiya, S.\ Nojiri, S.\ Ogushi, K.\ Ohashi, Y.\ Oz,
N.\ Sasakura, J.\ Sonnenschein and H.\ Sonoda  
for useful discussions.

\appendix

\resection{Variations of curvature}

In this appendix, we list the variations of the curvature tensor, 
Ricci tensor and Ricci scalar with respect to the metric.

Our convention is\footnote{
The sign is opposite to that adopted in Ref.\ \cite{HS;weyl}.
}
\begin{eqnarray}
 R^{\mu}_{~\,\nu\lambda\sigma} &\equiv& 
 \partial_{\lambda}\Gamma_{\sigma\nu}^{\mu}
  +\Gamma_{\lambda\rho}^{\mu}\Gamma_{\sigma\nu}^{\rho}
  - (\lambda \leftrightarrow \sigma), \nn
 R_{\mu\nu} &\equiv& R^{\rho}_{~\mu\rho\nu},
 \qquad R\,\equiv\,G^{\mu\nu}\,R_{\mu\nu}. 
\end{eqnarray}
The fundamental formula is 
\ba
\delta\Gamma^\kappa_{\mu\nu}
=\frac{1}{2}\,G^{\kappa\lambda}
 \,\left(\nabla_{\mu}\,\delta G_{\nu\lambda}
 + \nabla_{\nu}\,\delta G_{\mu\lambda} 
 - \nabla_{\lambda}\,\delta G_{\mu\nu}
 \right),
\ea
from which one can calculate the variations of curvatures:
\begin{eqnarray}
 \delta R^\mu_{~\,\nu\lambda\sigma}&=&
  \nabla_\lambda\,\delta\Gamma^\mu_{\sigma\nu}
  -\nabla_\sigma\,\delta\Gamma^\mu_{\lambda\nu},\\
 \delta R_{\mu\nu\lambda\sigma} &=&
  \frac{1}{2}\Bigl[
  \nabla_{\lambda}\nabla_{\nu}\delta G_{\sigma\mu}
  -\nabla_{\lambda}\nabla_{\mu}\delta G_{\sigma\nu} 
  -\nabla_{\sigma}\nabla_{\nu}\delta G_{\lambda\mu}
  +\nabla_{\sigma}\nabla_{\mu}\delta G_{\lambda\nu}\nn
 &&~~ +\delta G_{\mu\rho}\,R^{\rho}_{~\nu\lambda\sigma} 
  -\delta G_{\nu\rho}\,R^{\rho}_{~\mu\lambda\sigma} 
  \Bigr],\\
 \delta R_{\mu\nu} &=& \frac{1}{2}\left[\nabla^{\rho}
  \left(\nabla_{\mu}\delta G_{\nu\rho} + \nabla_{\nu}\delta 
  G_{\mu\rho}\right) - \nabla^2 \delta G_{\mu\nu} - 
  \nabla_{\mu}\nabla_{\nu}
   \left(G^{\rho\lambda}\delta G_{\rho\lambda}\right)\right], \nonumber \\
 \\
 \delta R &=& -\delta G_{\mu\nu}\,
  R^{\mu\nu}+\nabla^{\mu}\nabla^{\nu}
  \delta G_{\mu\nu} - \nabla^2\left(G^{\mu\nu}\delta G_{\mu\nu}\right).
\end{eqnarray}
Here note that 
\ba
 \Bigl[\nabla_\mu,\nabla_\nu\Bigr]\,\delta G_{\lambda\sigma}
  =-\delta G_{\rho\sigma}\,R^\rho_{~\lambda\mu\nu}
  -\delta G_{\lambda\rho}\,R^\rho_{~\sigma\mu\nu}.
\ea

%
\resection{Variations of $S_{\rm loc}[g(x),\phi(x)]$}

In this appendix, we list the variations of $S_{\rm loc}[g(x),\phi(x)]$.

\noindent\underline{\bf Pure gravity}:

If we only consider terms with weight $w\leq4$ of the form 
\begin{equation}
 S_{\rm loc}[g] = \int d^d x\sqrt{g}\left(
 W - \Phi R + XR^2 + YR_{ij}R^{ij} 
 + ZR_{ijkl}R^{ijkl}
\right),
\end{equation}
then we have 
\begin{eqnarray}
\frac{1}{\sqrt{g}}\frac{\delta S_{\rm loc}}{\delta g_{ij}}
 \!&=&\!\frac{1}{2}\Bigl(
 W - \Phi R + XR^2 + YR_{ij}R^{ij} 
 + ZR_{ijkl}R^{ijkl} \Bigr)
  g^{ij} \nonumber \\
 &&+\,\Phi R^{ij} -2X \Bigl(RR^{ij}-\nabla^{i}
 \nabla^{j}R\Bigr) 
 -Y\Bigl(2R^{i}_{\,~k}R^{jk}
 -2\nabla_{k}\nabla^{\left(i\right.}R^{\left.j\right)k}
 +\nabla^2R^{ij}\Bigr) 
 \nonumber \\
 && -2Z\Bigl(R^{i}_{~\,klm}
 R^{jklm}-2\nabla^{k}\nabla^{l}
 R^{\left(i~~j\right)}_{~~kl~}\Bigr)
 -\left(2X+\frac{1}{2}\,Y\right)g^{ij}\,\nabla^2R, 
\end{eqnarray}
and thus
\begin{eqnarray}
 \frac{1}{\sqrt{g}}\,g_{ij}\,
  \frac{\delta S_{\rm loc}}{\delta g_{ij}}
 &=&\frac{d}{2}\,W - \frac{d-2}{2}\,\Phi\,R 
 +\frac{d-4}{2}\left(XR^2+YR_{ij}R^{ij}+ZR_{ijkl}
 R^{ijkl}\right) \nonumber \\
 &&~- \left(2(d-1)X+\frac{d}{2}\,Y+2Z\right)\nabla^2R.
\end{eqnarray}
In the last expression, we have used the Bianchi identity: 
$\nabla^i R_{ij}=(1/2)\nabla_j R$.

\noindent\underline{\bf Gravity coupled to scalars}:

{}For $S_{\rm loc}[g,\phi]$ of the form
\begin{equation}
 S_{\rm loc}[g,\phi] = \int d^d x \sqrt{g}\left(
 W(\phi) - \Phi(\phi)R + \frac{1}{2}M_{ab}(\phi)g^{ij}
 \partial_{i}\phi^a \partial_{j}\phi^b 
\right),
\end{equation}
we have
\begin{eqnarray}
\frac{1}{\sqrt{g}}\frac{\delta\Sloc}{\delta g_{ij}} &=&
 \frac{1}{2}\left(W-\Phi R + \frac{1}{2}\,M_{ab}\,
 \partial_{k}\phi^a\,\partial^{k}\phi^b \right)g^{ij} \nn
 &&+\,\Phi \,R^{ij}
 +g^{ij}\,\nabla^2\Phi - \nabla^{i}
  \nabla^{j}\Phi-\frac{1}{2}\,M_{ab}\,
  \partial^{i}\phi^a\,\partial^{j}\phi^b, \\
\frac{1}{\sqrt{g}}\frac{\delta\Sloc}{\delta \phi^a} &=&
 \partial_a W - \partial_a \Phi\, R -M_{ab}\,\nabla^2\phi^b
 - \Gamma_{a;bc}^{(M)}\,\partial_{i}\phi^b\,\partial^{i}\phi^c,
\end{eqnarray}
where $\Gamma_{bc}^{(M)a}(\phi)\equiv
M^{ad}(\phi)\,\Gamma_{d;bc}^{(M)}(\phi)$ is the Christoffel symbol 
constructed from $M_{ab}(\phi)$.

%
\resection{ADM decomposition}
\label{ADM-review}
\setcounter{equation}{0}

In this appendix, we summarize the components of the Riemann tensor, 
Ricci tensor and scalar curvature written in terms of the ADM 
decomposition.%
\footnote{
In this appendix, we use a different convention from 
that we have used this article; 
that is, 
quantities in the $(d+1)$-dimensional manifold wear a hat $\hat{\ }$ while 
quantities in the $d$-dimensional equal-time slice do not. 
}

In the ADM decomposition, the metric takes the form 
\ba
 ds^2&=&\hg_{\mu\nu}\,dX^\mu dX^\nu \nn
 &=&N(x,{\tau})^2 d{\tau}^2+g_{ij}(x,{\tau})
 \Bigl(dx^{i}+\lambda^i(x,{\tau})d{\tau}\Bigr) 
 \Bigl(dx^{j}+\lambda^{j}(x,{\tau})d{\tau}\Bigr). 
\ea
Here we use the following basis instead of the 
coordinate basis $\partial_{\mu}$: 
\ba
 \widehat{e}_{\widehat{n}}={1\over N}
 (\partial_{\tau}-\lambda^i\partial_i,),\qquad
 \widehat{e}_i=\partial_i. 
\ea
In this basis, the components of the metric are given by 
\ba
\left( 
\begin{array}{cc}
 {\hg}(\widehat{e}_{\widehat{n}},\widehat{e}_{\widehat{n}}) & 
 {\hg}(\widehat{e}_{\widehat{n}},\widehat{e}_j) \\
 {\hg}(\widehat{e}_j,\widehat{e}_{\widehat{n}}) & 
 {\hg}(\widehat{e}_i,\widehat{e}_j) 
 \end{array}
 \right)
=
\left( 
\begin{array}{cc}
1 & 0 \\
0 & g_{ij}
\end{array}
\right).
\ea
{}For the purpose of computing the Riemann tensor in this basis, it is 
useful to start with the formula 
\ba
\hR^{\sigma}_{\,\,\,\rho\mu\nu}\,\widehat{e}_{\sigma}&=&
\hR(\widehat{e}_{\mu},\widehat{e}_{\nu})\widehat{e}_{\rho} \nn
&=&\left[ \hnab_{\widehat{e}_\mu},
\hnab_{\widehat{e}_\nu} \right]\widehat{e}_{\rho}
-\hnab_{[\widehat{e}_{\mu},\widehat{e}_{\nu} ]}\,\widehat{e}_{\rho}. 
\ea
Each component can be calculated explicitly 
by using the equations 
\ba
\hnab_{\widehat{e}_i}\widehat{e}_j&=&
-K_{ij}\widehat{e}_{\widehat{n}}+\Gamma^k_{ij}\,\widehat{e}_k, \nn
\hnab_{\widehat{e}_i}\widehat{e}_{\widehat{n}}&=&
K_i^k\,\widehat{e}_k, \nn
\hnab_{\widehat{e}_{\widehat{n}}}
\widehat{e}_j&=&
{1\over N}\,\partial_jN\,\widehat{e}_{\widehat{n}}+
\left(K^k_j+{1\over N}
  \,\partial_j\lambda^k\right)\widehat{e}_k, \nn
\hnab_{\widehat{e}_{\widehat{n}}}
\widehat{e}_{\widehat{n}}&=&
-{1\over N}\,g^{kl}\,\partial_k N\,\widehat{e}_l, \nn
\left[ \widehat{e}_{\widehat{n}}, \widehat{e}_{i}\right]&=&
{1\over N}\,\partial_iN\,\widehat{e}_{\widehat{n}}+
{1\over N}\,\partial_i\lambda^k\,\widehat{e}_{k}, 
\ea
where $K_{ij}$ is the extrinsic curvature and $\Gamma^i_{jk}$ is 
the affine connection with respect to $g_{ij}$. 
We thus obtain 
\ba
\hR_{ijkl}&=&R_{ijkl}-K_{ik}K_{jl}+K_{il}K_{jk}, \nn
\hR_{\widehat{n}jkl}&=&\nabla_lK_{jk}-\nabla_kK_{jl}, \nn
\hR_{\widehat{n}j\widehat{n}l}&=&(K^2)_{jl}-L_{jl},
\ea
with
\begin{align}
K_{ij}&={1\over 2N}\left(\dot{g}_{ij}
+\nabla_i\lambda_j+\nabla_j\lambda_i\right), \\
L_{ij}&={1\over N}\left( 
\dot{K}_{ij}-\lambda^k\,\nabla_k K_{ij}-\nabla_i\lambda^k\,K_{kj}
-\nabla_j\lambda^k\,K_{kj}+\nabla_i\nabla_j N\right). 
\end{align}
The components of the Ricci tensor 
$\hR_{\mu\nu}\equiv {\hR}^{\rho}_{\,\,\,\mu\rho\nu}
={\hR}_{\nu\mu}$ are given by 
\ba
\hR_{ij}&=&R_{ij}+2(K^2)_{ij}-KK_{ij}-L_{ij}, \nn
\hR_{i\widehat{n}}&=&\nabla^kK_{ki}-\nabla_iK, \nn
\hR_{\widehat{n}\widehat{n}}&=&K_{ij}^2-g^{ij}L_{ij}\,,
\ea
and the scalar curvature is 
\begin{align}
\hR&=R+3K_{ij}^2-K^2-2g^{ij}L_{ij} \nn
&=R-K_{ij}^2+K^2-{2\over N}\left(\dot{K}
+\lambda_k\left(\nabla^kN-\lambda^kK\right)\right), 
\label{scalar;ap}
\end{align}
where we use the fact 
\ba
 g^{ij}L_{ij}={1\over N}\left[
 \dot{K}+\nabla_k\left(\nabla^kN-\lambda^kK\right)\right]
 +\left(2K_{ij}^2-K^2\right). 
\ea

%
\resection{Boundary terms}
\label{BoundaryTerms}
\setcounter{equation}{0}

In this appendix, we supplement the discussion of
the possible boundary terms in \eq{PG:action}. 
In this appendix we omit the hat on the bulk fields. 

We first consider the infinitesimal transformation 
\ba
 x^i\rightarrow x^{\prime i}=x^{i}+\epsilon^i(x,\tau),\qquad 
  \tau\rightarrow \tau^{\prime}=\tau+\epsilon(x,\tau). 
 \label{diffeo;inf}
\ea
Under this transformation, $N,\lambda_i$ and $g_{ij}$ are found 
to transform as 
\ba
{1\over N^{\prime}}&=&{1\over N}(1+\dot{\epsilon}
-\lambda^i\partial_i\epsilon), \nn
\lambda^{\prime}_i&=&\lambda_i-\partial_i\epsilon^j\lambda_j
-\dot{\epsilon}\lambda_i-\partial_i\epsilon\,(N^2+\lambda^2)
-g_{ij}\dot{\epsilon}^j, \nn
g^{\prime}_{ij}&=&g_{ij}-\partial_i\epsilon^kg_{kj}
-\partial_j\epsilon^kg_{ik}-\partial_i\epsilon\,\lambda_j
-\partial_j\epsilon\,\lambda_i. 
\ea
{}Furthermore, $\Gamma^i_{jk}$, the affine connection defined by $g_{ij}$, 
transforms under the diffeomorphism (\ref{diffeo;inf}) as 
\ba
\Gamma^{\prime i}_{jk}=\Gamma^i_{jk}-\partial_j\,\partial_k\epsilon^i
+\Gamma^m_{jk}\,\partial_m\epsilon^i
-\Gamma^i_{mk}\,\partial_j\epsilon^m
-\Gamma^i_{jm}\partial_k\epsilon^m+\tilde{\delta}\Gamma^i_{jk}, 
\ea
with 
\ba
\tilde{\delta}\Gamma^i_{jk}=-\lambda^i\nabla_j\nabla_k\epsilon 
-\partial_j\epsilon\nabla_k\lambda^i-\partial_k\epsilon\nabla_j\lambda^i
-Ng^{il}(\partial_j\epsilon\,K_{lk}+\partial_k\epsilon\,K_{lj}
-\partial_l\epsilon\,K_{jk}).  \nn
\ea
Note that $\tilde{\delta}\Gamma^i_{jk}$ does not contain $\epsilon^i$. 
{}From these relations, it is straightforward to verify that 
the extrinsic curvature transforms as 
\begin{align}
K^{\prime}_{ij}=&\,K_{ij}
-\partial_i\epsilon^l\,K_{lj}-\partial_k\epsilon^l\,K_{jl} \nn
&+N\nabla_i\nabla_j\epsilon
+\partial_i\epsilon\,(\partial_jN-\lambda^lK_{jl})
+\partial_j\epsilon\,(\partial_iN-\lambda^lK_{lj}). 
\end{align}
We can also show that the Riemann curvature $R^i_{\,\,jkl}$ 
transforms under (\ref{diffeo;inf}) as 
\ba
R^{\prime i}_{\,\,\,jkl}&=&R^i_{\,\,jkl}
+\partial_m\epsilon^i\,R^m_{\,\,jkl}
-\partial_j\epsilon^m\,R^i_{\,\,mkl}
-\partial_k\epsilon^m\,R^i_{\,\,jml}
-\partial_l\epsilon^m\,R^i_{\,\,jkm} \nn
&&-\partial_k\epsilon\,\dot{\Gamma}^i_{lj}
+\partial_l\epsilon\,\dot{\Gamma}^i_{kj}
+\nabla_k\tilde{\delta}\Gamma^i_{lj}
-\nabla_l\tilde{\delta}\Gamma^i_{kj}. 
\ea

As argued in \S $5$, 
we focus on the diffeomorphism that obeys the condition 
(\ref{diffeo;boundary}). 
This is equivalent to the following relation in an infinitesimal form: 
\ba
 \partial_i\epsilon(\tau\!=\!\tau_0)=0. 
 \label{r0;ap}
\ea
Therefore, we find that the boundary action in \eq{PG:action} is invariant 
under this diffeomorphism.

We remark that in the above, we have discarded boundary terms of the form 
\ba
 \bS'_b=\int_{\Sigma_d}d^d x \sqrt{g}\left(K^{ij}L_{ij}
  +K g^{ij}L_{ij}\right),
\ea
although these are allowed by the diffeomorphism.\footnote
{By definition, the $(d+1)$-dimensional scalar curvature 
$\widehat{R}$ is a scalar. 
It thus follows from (\ref{scalar;ap}) that $L_{ij}(\tau\!=\!\tau_0)$ 
transforms as a tensor under the diffeomorphism with (\ref{r0;ap}). 
}
The reason is that if there were such boundary terms, 
they would require us to further introduce an extra boundary condition, 
since  
\ba
 \delta \bS'_b = \int_{\Sigma_d}d^d x \sqrt{g}
  \left[\cdots +\delta\dot{K}_{ij}P_2^{ij}(g_{kl},K_{kl})\right].
\ea

\resection{Example of derivation of the Hamilton-Jacobi-like equation}
\label{HowToSolve}
\setcounter{equation}{0}

We briefly describe how the Hamilton-Jacobi equation (\ref{HJ;after}) 
is solved. 
{}For simplicity, we consider the case $N\!=\!1$ 
and focus only on the upper boundary at $\tau\!=\!t$. 
Motivated by the gravitational system 
considered in the next section, 
we assume that the Lagrangian takes the form 
\ba
L(q,\dot{q},\ddot{q})=L_0(q,\dot{q})+c L_1(q,\dot{q},\ddot{q}), 
\label{lag}
\ea
where 
\ba
 L_0(q,\dot{q})&=&{1\over 2}
  m_{ij}(q)\dot{q}^i\dot{q}^j-V(q), \nn
 L_1(q,\dot{q},\ddot{q})&=&
  {1\over 2}n_{ij}(q)\ddot{q}^i\ddot{q}^j
  -A_i(q,\dot{q})\ddot{q}^i-\phi(q,\dot{q}), 
 \label{lag1}
\ea
with
\ba
 A_i(q,\dot{q})&=&a_{ijk}^{(2)}(q)\dot{q}^j\dot{q}^k
  +a_{i}^{(0)}(q), \nn
 \phi(q,\dot{q})&=&
  \phi^{(4)}_{ijkl}(q)\dot{q}^i\dot{q}^j\dot{q}^k\dot{q}^l
  +\phi^{(2)}_{ij}(q)\dot{q}^i\dot{q}^j+\phi^{(0)}(q). 
 \label{lag2}
\ea
{}We further assume that the determinants of the 
matrices $m_{ij}(q)$ and $n_{ij}(q)$ have the same 
signature.
Following the procedure discussed in \S 5, 
this Lagrangian can be rewritten into the first-order form 
\ba
 L=p\,\dot{q}+P\dot{Q}-H(q,Q;\,p,P)\,,
\ea
with the Hamiltonian 
\begin{align}
 H(q,Q;p,P) =&\,\,p_iQ^i-{1\over 2}m_{ij}(q)Q^iQ^j+V(q) \nn
 &+{1\over 2c}
  n^{ij}(q)\Big(P_i+cA_i(q,Q)\Big)\Big(P_j+cA_j(q,Q)\Big)
  +c\,\phi(q,Q), 
 \label{trueH}
\end{align}
where $(n^{ij})=(n_{ij})^{-1}$. 
The Hamilton-Jacobi equation \eq{HJ;after} is solved 
as a double expansion with respect to $c$ and $P$ 
by assuming that the classical action takes the form 
\ba
 \widehat{S}(t,q,P)&=&
  {1\over\sqrt{c}}\,\widehat{S}_{-1/2}(t,q,P)+\widehat{S}_0(t,q,P)
  +\sqrt{c}\,\widehat{S}_{1/2}(t,q,P)+c\,\widehat{S}_1(t,q,P) \nn
 &&+{\cal O}(c^{3/2}). 
 \label{step_first}
\ea
After some simple algebra, the coefficients are found to be 
\ba
 \widehat{S}_{-1/2}&=&
  {1\over 2}u^{ij}(q)P_iP_j+{\cal O}(P^3),\nn
  \widehat{S}_{0}&=&S_0(t,q)-P_i\,\partial^iS_0+{\cal O}(P^2),
\nn
 \widehat{S}_{1/2}&=&P_i\,u^{ij}(q)n_{jk}(q)
  \left[ \bGam^k_{lm}\,\partial^lS_0\,\partial^mS_0+\partial^kV(q)
  +n^{kl}(q)A_l\!\left( q,{\partial S_0\over\partial q}\right)\right] \nn
 &&+{\cal O}(P^2).
\ea
Here, 
\ba
\partial_i\equiv {\partial\over\partial q^i},\qquad 
\partial^i\equiv m^{ij}\partial_i, 
\ea
and $\bGam^i_{jk}$ is the affine connection defined by $m_{ij}$. 
Also $u^{ij}$ is defined by the relation 
\ba
u^{ik}(q)u^{jl}(q)m_{kl}(q)=n^{ij}(q). 
\ea
Furthermore, $S_{0}(t,q)=\widehat{S}_{0}(t,q,P\!=\!0)$ and 
$S_{1}(t,q)=\widehat{S}_{1}(t,q,P\!=\!0)$
satisfy the equations 
\begin{align} 
 -{\partial S_0\over\partial t} =&\, {1\over 2}m_{ij}(q)
  {\partial S_0\over\partial q^i}{\partial S_0\over\partial q^j}+V(q), \nn
 -{\partial S_1\over\partial t} =&\, m_{ij}(q)
  {\partial S_1\over\partial q^i}{\partial S_0\over\partial q^j} \nn
  &-{1\over 2}n_{ij}(q)
  \left(\bGam^i_{kl}\,\partial^kS_0\,\partial^lS_0+\partial^iV(q)\right)
  \left(\bGam^j_{mn}\,\partial^mS_0\,\partial^nS_0+\partial^jV(q)\right) \nn
  &-A_i\!\left(q,{\partial S_0\over\partial q}\right)
  \left(\bGam^i_{kl}\,\partial^kS_0\,\partial^lS_0+\partial^iV(q)\right)
  +\phi\!\left( q,{\partial S_0\over\partial q}\right)\,,
 \label{sol}
\end{align}
which can be expressed as a Hamilton-Jacobi-like equation 
for the reduced classical action 
$S(t,q)\!=\!S_0(t,q)+c\,S_1(t,q)+{\cal O}(c^2)$: 
\ba
 -{\partial S\over\partial t}=\tH(q,p),\qquad
  p_i={\partial S\over\partial q^i}\,,
 \label{HJ;c1}
\ea
where 
\ba
\tH(q,p)&=&{1\over 2}m^{ij}(q)p_ip_j+V(q) \nn
&&+\,c\left[ 
-{1\over 2}n_{ij}(q)
\left(\bGam^i_{kl}\,p^kp^l+\partial^iV(q)\right)
\left(\bGam^j_{mn}\,p^mp^n+\partial^jV(q)\right) \right.\nn
&&~~~~~~-
A_i(q,p)\left(\bGam^i_{kl}\,p^kp^l+\partial^iV(q)\right)
+\phi(q,p)\Bigl].
\label{HJ;c2}\label{step_last}
\ea
It is important to note that $\tH$ is not the Hamiltonian. 
In fact, the Hamilton equation for $\tH$ does not 
coincide with that obtained from (\ref{trueH}).

\resection{Proof of Theorem}
\label{ProofofTheorem}
\setcounter{equation}{0}

In this appendix, we give a detailed proof of 
Theorem, \eq{theorem1} and \eq{theorem2}, 
for the action  
\ba
 \bS=\int^t_{t'} d\tau \Big[ L_0(q^i,\dot{q}^i)
  +c\,L_1(q^i,\dot{q}^i,\ddot{q}^i)\Big], 
\ea
where $i$ runs over some values.  
In the following discussion, 
we focus only on the upper boundary, for simplicity.

We first rewrite the zero-th order Lagrangian $L_0$ 
into the first-order form 
by introducing the conjugate momentum $p_{0i}$ of $q^i$ as  
\ba
 \bS[q(\tau),p_0(\tau)]
  =\int^t d\tau \Big[ p_{0i}\dot{q}^i-H_0(q,p_0)+c\,L_1(q,\dot{q},\ddot{q})
\Big], 
\ea
through the Legendre transformation from $(q,\dot{q})$ to 
$(q,p_0)$ defined by 
\ba
 p_{0i}={\partial L_0\over\partial\dot{q}^i}(q,\dot{q})\,. 
\ea
{}From this, the equation of motion for $p_{0i}$ and $q^i$ is given
by 
\ba
\dot{q}^i&=&{\partial H_0\over\partial p_{0i}}, 
\label{h1;ap} \\
\dot{p_{0i}}&=&-{\partial H_0\over\partial q^i}
+c\left[{\partial L_1\over\partial q^i}
-{d\over d\tau}\left({\partial L_1\over\partial\dot{q}^i}\right)
+{d^2\over d\tau^2}\left({\partial L_1\over\partial\ddot{q}^i}\right)
\right]. 
\label{h2;ap}
\ea
Let $\bar{q}(\tau),\,\bar{p}_0(\tau)$  be the solution to this equation 
of motion that satisfies the boundary condition 
\ba
 \bar{q}^i(\tau\!=\!t)=q^i\,. 
\ea
Since this condition determines the classical trajectory uniquely 
[together with the lower boundary values $\bar{q}^i(\tau\!=\!t')=q^{\prime\,i}
$ 
that we have not written here explicitly], 
the boundary value of $\bar{p}_{0}$ is completely specified by $t$ and $q$: 
$\bar{p}_0(\tau\!=\!t)\!=\!p_0(t,q)$. 
By plugging the classical solution into the action $\bS$, the 
classical action is obtained as a function of the boundary value $q^i$ 
and $t$: 
\ba
 S(t,q)=\bS[\bar{q}(\tau),\bar{p}_0(\tau)].
\ea
In order to derive a differential equation that determines $S(t,q)$, 
we then take the variation of $S(t,q)$. 
Using (\ref{h1;ap}) and (\ref{h2;ap}), this is easily evaluated to be 
\ba
\delta S&=&
\delta t\Big[p_{0i}\dot{q}^i-H_0(q,p_0)+c\,L_1(q,\dot{q},\ddot{q})\Big] \nn
&&+\,\delta\bar{q}^i(t)\left[ p_{0i}+c\left( 
{\partial L_1 \over\partial\dot{q}^i}(q,\dot{q},\ddot{q})-
{d\over d\tau}\!\left.\left( {\partial L_1 \over\partial\ddot{q}^i}
 (\bar{q},\dot{\bar{q}},\ddot{\bar{q}})\right)\right|_{\tau=t}
  \right)\right] \nn
&&
 +\,c\,\delta\dot{\bar{q}}^i(t)\,{\partial L_1\over\partial\ddot{q}^i}
  (q,\dot{q},\ddot{q}), 
\label{var}
\ea
where 
\ba
 \dot{q}^i\equiv{d\bar{q}^i\over d\tau}(\tau\!=\!t),\qquad 
 \ddot{q}^i\equiv{d^2\bar{q}^i\over d\tau^2}(\tau\!=\!t)\,,
\ea
and $\delta\bar{q}^i(t)$ and $\delta\dot{\bar{q}}^i(t)$ are understood 
to be $\delta\bar{q}^i(\tau)|_{\tau=t}$ 
and $d\,\delta\bar{q}^i(\tau)/d\tau|_{\tau=t}$, respectively. 
By expanding the classical solution $\bar{q}^i(\tau)$ around $\tau\!=\!t$, 
we find that the variations $\delta \bar{q}^i(t)$ 
and $\delta \dot{\bar{q}}^i(t)$ are given by 
\ba
 \delta\bar{q}^i(t)=\delta q^i-\dot{q}^i\,\delta t, \qquad
  \delta\dot{\bar{q}}^i(t)=\delta \dot{q}^i-\ddot{q}^i\,\delta t. 
\ea
Here it is important to note that 
$\dot{q}$ can be written in terms of $q$ and $t$, since 
the classical solution is determined uniquely by the boundary value
$q$. 
Actually it can be shown that 
\ba
\delta\dot{q}^i&=&
{\partial^2H_0\over\partial q^j\partial p_{0i}}\,\delta q^j
+{\partial^2 H_0\over\partial p_{0i}p_{0j}}\,\delta p_{0j} \nn
&=&
{\partial^2H_0\over\partial q^j\partial p_{0i}}\,\delta q^j
+{\partial^2 H_0\over\partial p_{0i}p_{0j}}\,\left( 
{\partial p_{0j}\over\partial t}\delta t
+{\partial p_{0j}\over\partial q^k}\delta q^k\right), 
\ea
where we have used (\ref{h1;ap}) as well as the fact that 
$p_{0}=p_{0}(t,q)$. 
From these relations, 
the variation (\ref{var}) is found to be 
\ba
\delta S=p_i\,\delta q^i-\widetilde{H}(q,p)\,\delta t, 
\ea
with
\ba
 p_i&=&p_{0i} 
 +c\left[ {\partial L_1\over\partial\dot{q}^i}(q,\dot{q},\ddot{q})
  -{d\over d\tau}\!
  \left.\left( {\partial L_1\over\partial\ddot{q}^i}
  (\bar{q},\dot{\bar{q}},\ddot{\bar{q}})\right)\right|_{\tau=t}\right.\nn
 &&\left.
  +\,{\partial L_1\over\partial\ddot{q}^j}\left(
  {\partial^2H_0\over\partial q^i\partial p_{0j}}
  +{\partial^2H_0\over\partial p_{0j}\partial p_{0k}}
  {\partial p_{0k}\over\partial q^i}\right)\right], 
 \label{p;ap}\\
 \tH(q,p)&=&H_0(q,p_0) \nn
  &&+\,c\left[ 
  -L_1(q,\dot{q},\ddot{q})
  +\dot{q}^i\left(
  {\partial L_1\over\partial\dot{q}^i}(q,\dot{q},\ddot{q})
  -{d\over d\tau}\!\left.\left(
   {\partial L_1\over\partial\ddot{q}^i}(\bar{q},\dot{\bar{q}},\ddot{\bar
  {q}})
   \right)\right|_{\tau=t}\right)
  \right. \nn
&&~~~~~\,\left. +\,
{\partial L_1\over\partial\ddot{q}^i}\left(
\ddot{q}^i
-{\partial^2H_0\over\partial p_{0i}\partial p_{0j}}
{\partial p_{0j}\over\partial t}
\right)\right].
\ea
In order to compute $\tH(q,p)$, we first note that the Hamilton
equation appearing in (\ref{h1;ap}) and (\ref{h2;ap}) gives the relation 
\ba
\ddot{q}^i=
{\partial^2H_0\over\partial p_{0i}\partial q^j}
{\partial H_0\over\partial p_{0j}}
+{\partial^2 H_0\over\partial p_{0i}\partial p_{0j}}
\left( {\partial p_{0j}\over\partial q^k}
{\partial H_0\over\partial p_{0k}}
+{\partial p_{0k}\over\partial t}\right)\,.
\ea
It is then easy to verify that $\tH(q,p)$ takes the form 
\ba
\tH(q,p)=H_0(q,p)-c\,L_1(q,\dot{q},\ddot{q})+{\cal O}(c^2). 
\ea
Here $\dot{q}^i$ and $\ddot{q}^i$ in $L_1$ can be replaced by 
\ba
 f_1^i(q,p)\equiv\left\{H_0(q,p),q^i\right\} 
  =\frac{\partial H_0}{\partial p_i}(q,p) 
\ea
and
\ba
 f_2^i(q,p)&\equiv&\left\{H_0(q,p),\left\{H_0(q,p),q^i\right\}\right\}\nn
 &=&{\partial^2 H_0\over\partial p_i\partial q^j}(q,p)
  {\partial H_0\over\partial p_j}(q,p)
  -{\partial^2 H_0\over\partial p_i\partial p_j}(q,p)
  {\partial H_0\over\partial q^j}(q,p)\,,
\ea
respectively, up to ${\cal O}(c^2)$. 
This completes the proof of (\ref{theorem1}) and (\ref{theorem2}).


\end{document}